%% file: VC_preprint.tex
\documentclass[final,10pt]{elsarticle}

\usepackage{array}
\usepackage{amssymb}

\usepackage{natbib}
\bibpunct{(}{)}{;}{a}{,}{,}

\usepackage{graphicx}
\usepackage{epstopdf}
\usepackage{lineno}
\usepackage{rotating}
\usepackage{booktabs}
\usepackage{multirow}
\usepackage{rotating}

\usepackage{longtable}

\begin{document}

\begin{frontmatter}

\title{Effects of a radially varying electrical conductivity on 3D numerical dynamos}

\journal{Physics of the Earth and planetary  Interiors}

\author[Nat]{Natalia G\'{o}mez-P\'{e}rez\corref{cor1}} 
\cortext[cor1]{Corresponding author}

\ead{ngomezperez@ciw.edu}

\author[M]{Moritz Heimpel}
\author[Johannes]{Johannes Wicht}

\address[Nat]{Department of Terrestrial Magnetism,
Carnegie Institution of Washington,
5241 Broad Branch Road N.W., Washington, DC 20015-1305}

\address[M]{Department of Physics,
Room 238 CEB, 11322 - 89 Av., 
University of Alberta, 
Edmonton, Alberta, Canada T6G 2G7}

\address[Johannes]{	
Max Planck Institute for Solar System Research, 
Max-Planck-Str. 2,
37191 Katlenburg-Lindau}

\begin{abstract}

The transition from liquid metal to silicate rock in the cores of the
terrestrial planets is likely to be accompanied by a gradient in the
composition of the outer core liquid.  The electrical conductivity of
a volatile enriched liquid alloy can be substantially lower than a
light-element-depleted fluid found close to the inner core boundary.
In this paper, we investigate the effect of
radially variable electrical conductivity on planetary dynamo action
using an electrical conductivity that decreases exponentially as a
function of radius.  We find that numerical solutions with continuous,
radially outward decreasing electrical conductivity profiles result in
strongly modified flow and magnetic field dynamics, compared to
solutions with homogeneous electrical conductivity.  
The force balances at the top of the simulated fluid determine the
overall character of the flow. 
The relationship between 
Coriolis and Lorentz forces near the outer boundary
controls the flow and magnetic field intensity and morphology of the system.
Our results imply that a low conductivity layer near the top of Mercury's
liquid outer core is consistent with its weak magnetic field.

\end{abstract}

\end{frontmatter}


\section{Introduction}

Variations in the physical properties of fluids in 
planetary dynamos define the character of the observed intrinsic
magnetic field (e.g., strength, geometry and time variability).  
Changes in the electrical
conductivity of the fluid as a function of depth may become relevant
in the context of terrestrial and gas giant planets.  In this paper we
explore (with a focus on the terrestrial planets) how the radial
variation of electrical conductivity in planetary cores may result in
changes to dynamo-generated magnetic fields.

\subsection{Terrestrial planets}
The cores of the terrestrial planets are composed principally of iron,
with minor but significant amounts of nickel and lighter elements.  It
has long been known that an  iron-nickel core would have too high a
density to be compatible with Earth's moment of inertia and seismic
data \citep[e.g.,][]{Birch1952,Poirier1994}.  A compatible Earth core
density model can result from the inclusion of about 8\% by weight of
one or more light elements.  Detailed models of core composition are
based primarily on the constraints of seismology, mineral physics,
geochemistry, metallurgy, and cosmochemisty. Silicon, sulphur and
oxygen are the primary candidates for the light elements.  Sulphur is
likely to be a significant component but the depletion of light
elements in the process of accretion in the inner solar system limits
sulphur to about 2~wt\% in the core.

Recent reviews of core differentiation and composition distinguish
between models considering Silicon versus those considering Oxygen as the
primary light element. Composition models give weight percents of Fe
$\simeq$ 85\%-88\%, Ni $\simeq$ 5\% Si $\simeq$ 0-7\%, O $\simeq$
0-4\% and S $\simeq$ 2\% \citep[e.g][]{McDonough2003,Wood2006}.
Solidification of a more or less pure iron-nickel inner core may
exclude the lighter elements, which would then be enriched in the
outer core. The ratio of the inner core radius (1221~km) to the core
radius (3480~km) is 0.35 and the mass of the inner core is only about
5\% of the total core mass. So the bulk composition of the outer core
is only slightly different than that of the whole core.

Convection in the liquid outer core is driven by a combination of
compositional and/or thermal buoyancy. Thermal buoyancy available to
drive convection and dynamo action originates primarily from the
latent heat of solidification at the inner core boundary (ICB) and
possibly from cooling at the core-mantle boundary (CMB). Secular
cooling at the CMB does not guarantee a source of convective
instability since heat can be conducted through a stably stratified
layer (a super-adiabatic heat flux is needed). Compositional buoyancy
originates at the ICB due to light element enrichment in the residual
liquid associated with inner core solidification.  A source of
compositional buoyancy near the CMB could come from precipitation from
a silicate enriched layer at the top of the core, leaving a light
element depleted, heavy residual liquid and a silicate sediment layer
at the CMB \citep{Buffett2000}.

Temperature and pressure dependence of electrical conductivity 
may lead to sizable variations with depth.
\citep[][]{Stacey2001}.  Assuming a well-mixed outer core and a Si
concentration of $X_{S_i}=0.25$ in a dual species alloy, the authors
found the most extreme variation in the fluid core of Earth results in
a factor of 1/2 difference for the electrical conductivity and about
3/4 for the thermal conductivity from the ICB to the CMB.  Although
the effects of different impurities in the alloy have not been
explored experimentally, \citet{Stacey2001} argue that the effects of
impurities other than Si do not significantly change their estimates.
A later revision \citep{Stacey2007} found that the variation in
electrical and thermal conductivities had been overestimated due to
the assumption of treating Fe as an electronically simple metal.
Their later results predict a difference in the electrical
conductivity of a factor of 0.78 between that of the CMB when compared
to the ICB's. Those estimates were made under the assumption of a
well-mixed outer core. However gradients in material properties would
be amplified in a compositionally stratified layer.

The existence of a thermally and/or compositionally stably stratified
layer near the top of the Earth's core has been both, suggested
and argued \citep{Jacobs1975,Fearn1981, Lister1998, Braginsky1993,
  Braginsky2007}.  Here we will focus on the possibility of a
compositionally stratified layer near the top of the core since the
electrical conductivity in such a layer would decrease with radius
(due to the higher light-element concentration when compared to the
bottom of the fluid core).  Two mechanisms by which a compositionally
stratified layer can grow are by chemical diffusion from the mantle
directly to the top of the core \citep{Lister1998} and by buoyant
transport of light element enriched residual liquid from inner core
solidification \citep{Moffatt1994}.  Chemical diffusion is a slow
process and would not result in a layer thickness greater than about
10~km \citep{Lister1998}. On the other hand, buoyant rise of light
element rich residual liquid could be an efficient process to build a
layer of greater thickness. A thick layer with a significantly
anomalous density would likely be detected as a seismically fast
layer. Seismological constraints have been so far
inconclusive. However, recent seismological results using outermost
core waveforms seem to be consistent with the existence of a low
density layer of about 100 km thickness
\citep{Alexandrakis2007,Tanaka2007}

It has been confirmed that Mercury has a liquid iron core
\citep{Margot2007}.  Furthermore, it is likely that Mercury's weak
intrinsic magnetic field is generated by a dynamo in its liquid outer
core. Neither the details of Mercury's core composition, nor the size
of its inner core is known \citep{Hauck2004,Solomon2008,Heimpel2008}.
Observations of Mercury's contractional lobate crustal faults imply a
planetary radial contraction of about 3~km, small compared to an
estimated 17~km of contraction that would result from a completely
solidified core.  This implies that Mercury may have a relatively
large outer core, perhaps earth-like proportion. For a thick-shell
outer core to exist in Mercury, given its small size,
\citet{Hauck2004} estimates that the light elements (such as S or Si)
in the core have a relatively high concentration. This suggests the
possibility of a large and stratified Hermean outer core with a thick,
and stably stratified outermost layer. Such a layer would be
compositionally stratified \citep[in contrast to thermally stratified
  models previously proposed, e.g.,][]{Christensen2006a} with a radial
increase in the proportion of a light elements, and a radial decrease
in electrical conductivity.

\subsection{Gas giant planets}
The magnetic field in the gas giants is generated
within a metallic hydrogen region.  Experimental results have found
that the transition from metallic to molecular hydrogen yields a wide
range of pressures where the electrical conductivity is non-negligible
thus varying slowly with depth in planetary interiors
\citep{Nellis2000}.  
The internal structure has been deduced to first order based on
measurements of the moment of inertia and total mass of each planet
\citep[e.g.,][]{Guillot2005}.  However, constraints on hydrogen and
helium mixtures at high pressures need to be found in order to better
determine the internal structure of the gas giants.  The additional
effect of helium may complicate further the underdetermined internal
layering of gas giants \citep[e.g.,][]{Stevenson2008}.  The metallic to
molecular hydrogen transition is of particular interest from the point
of view of the internal dynamics.
The depth and radial extent of this transition are important for magnetic 
field generation and in understanding the observable magnetic field morphology. 
This is a task for future investigation.

 In this paper, we present results from a numerical dynamo model with 
 radially varying electrical conductivity
 \citep{Gomez2007a}.  
 To implement the radially variable conductivity, we
modified an existent numerical code
 \citep[originally MagIC 2.0,][]{Wicht2002}, that uses the Boussinesq
 approximation.  We focus this study on the effect of the varying
 conductivity on the dynamo action, and on the generated magnetic
 field. With this new implementation we performed a set of tests to
 analyze its consistency with previously published work.  In
 section~\ref{sec:strat} we include a review of numerical models that
 worked with stratified liquids. In section~\ref{sec:method} we
 present the necessary modification to the dynamo equations, and to
 the numerical code. The parameters explored for twenty runs studied
 in this paper are included in section~\ref{sec:param}.  We present
 the results in section~\ref{sec:result}, the discussion and conclusions are found
 in section~\ref{sec:diss} and~\ref{sec:con}, respectively. 
 We also included a table of symbols in 
 appendix~\ref{sec:symbols}.

\section{Stratified planetary interiors in numerical simulations}\label{sec:strat}

The study of planetary and solar dynamos has evolved rapidly in the
past two decades due to code development and the accessibility to
powerful computers used to solve numerically the equations of motion
of rotating fluids.  Solutions of numerical dynamos share
characteristics with Earth's dynamo with respect to temporal
variability, evolution and mean geometry
\citep[e.g.,][]{Kageyama1993,Glatzmaier1995a,Kuang1997,Christensen1998}.  More
recently, numerical models have been successful in reproducing  
non-dipolar and weak fields such as those observed for the ice
giants \citep[e.g.,][]{Gomez2007b,Stanley2006}.  Nevertheless, these
models of planetary dynamos rely on strong assumptions that do not
necessarily represent realistic physical conditions expected for
planetary interiors. For example, due to hardware limitations, numerical 
models cannot simultaneously resolve planetary- and small-scale flow, both of  
which are prevalent for low viscosity fluids such as liquid iron or metallic 
hydrogen in a planetary setting.

\subsection{Buoyancy stratification}
\citet{Christensen2008}, and \citet{Christensen2006a} implemented a
stable stratified (non-convecting) layer at the top of the electrical
conductive fluid in planetary dynamo models.  They showed that small
scale magnetic fields, may be generated in the deep interior of the
electrically conductive fluid, shielded by a stably stratified layer.
This outer layer acts as a filter that mainly damps the rapidly
varying small scale components by the magnetic skin effect.  They
conclude that the axisymmetric field of Saturn and the relatively weak
field of Mercury may be explained by a stably stratified layer at the
top of the dynamo region.  
Similarly, \citet{Stanley2008} using
buoyantly stratified fluids, found that zonal winds developed in a
stable layer can affect the magnetic field time evolution rendering a
secular variation of thermally stratified models
inconsistent with Earth's observed field.

\subsection{Density stratification}

We discuss briefly the effects of density variation (anelastic versus
Boussinesq approximations) in this section, since they are closely
linked to those of the electrical conductivity.  In order to solve the
anelastic dynamo equations numerically, \citet{Glatzmaier1984} used a
poloidal and toroidal decomposition for the velocity and magnetic
fields \citep[for the decomposition of solenoidal vectors the reader
  may refer to][]{Chandrasekhar1961}. This allows for the use of a
pseudo-spectral algorithm, which has been extensively used in
planetary and solar dynamo simulations
\citep[e.g.,][]{Glatzmaier1984,Dormy1998,Christensen1999,Sakuraba1999,Clune1999}.
This mathematical approach requires field vectors to be solenoidal
(i.e.,\ divergence free) and thus, approximations are used for the
continuity equation, $\nabla\cdot(\rho\mathbf{u})+\partial
\rho/\partial t =0 $, where $\mathbf{u}$ is the velocity, $\rho$ the
density of the fluid and $t$ is time.  
If the time scale of convection
is large compared to the time for sound waves to travel across the
depth of the shell,
then $\partial\rho / \partial t \ll
\nabla\cdot(\rho\mathbf{u})$ and the anelastic approximation, $\nabla
\cdot(\rho\mathbf{u})=0$, is obtained. In addition, if the density
changes in the fluid are much smaller than the averaged density, one
obtains the Boussinesq approximation, $\nabla\cdot\mathbf{u}=0$, where
the buoyancy is determined by temperature changes exclusively, but the
density perturbations with respect to the mean are neglected.

Differences between the results obtained using Boussinesq and
anelastic approximations have been compared in the context of magnetic
convection in 2D models \citep{Evonuk2004}. They find significant
effects due to strong density stratification 
(about three orders of
magnitude variation in density). The length scales of the top and
bottom thermal plumes differ due to the compressibility of the liquid.
3D effects on the flow are very significant as well, and although the
effect on the density variation is likely to be important in
three-dimensional systems, a detailed analysis of density variations
in 3D flows needs to be completed.

\citet{Anufriev2005} compared Boussinesq and anelastic approximations
analytically and found that they differ in the thermodynamic
equations.  In the latter, the rate of work to expand or contract the
fluid is included in the energy balance, while the Boussinesq treats a
thermal balance as the total energy balance.  The anelastic
approximation takes into account a more complete picture by including
cooling or heating done by the flow.  They propose a change in the
thermal boundary conditions in Boussinesq models in order to allow for
a realistic energy balance in the system.  This suggestion should be
seriously considered in future numerical simulations that use the
Boussinesq approximation.

We are not aware of previous numerical simulations that studied the
effects of varying electrical conductivities in the context of
planetary interiors.  Anelastic models of the interior of stars
include variations on magnetic and viscous diffusivities, and thermal
expansion coefficient changing as a function of background density,
$\propto\bar\rho^{-1/2}$ \citep[][]{Featherstone2007,Clune1999}.
Their numerical models include high resolution runs, specially
targeted for stellar dynamos 
(e.g., the heat transport throughout the
simulated volume is set up to follow hydrogen burning at the deep
interior).  Here, we study the effects of radially varying electrical
conductivity on the flow and on the magnetic field generation in the
liquid cores of terrestrial planets.

\section{Methodology}\label{sec:method}

\subsection{Governing equations}

To describe the behavior of an electrically conductive fluid bounded
by co-rotating spherical shells, we use the magnetohydrodynamic
equations in the frame of the rotating fluid. Consider a spherical
shell with inner and outer boundary radii $r_i$ and $r_o$
respectively, rotating with an angular velocity
$\mathbf{\Omega}$. Using the Boussinesq approximation, these equations
are written as:

\begin{small}
\begin{eqnarray}
E\left(\frac{\partial \mathbf{u}}{\partial {t}} + (\mathbf{u} \cdot \nabla) \mathbf{u} 
- \nabla^2 \mathbf{u} 
\right)&+&2
\mathbf{\hat z} \times \mathbf{u} \nonumber\\
 & = & -\nabla {P} + \frac{Ra E}{Pr} \frac {g} {g_o} \mathbf{\hat r} {T} 
+ \frac1 
{Pm^*}\left(\nabla \times \mathbf{B}\right) \times \mathbf{B}, 
\label{eq:Navier-Stokes_CH1}\\
\nabla \cdot \mathbf{u} = 0, &  & \nabla \cdot \mathbf{B} = 
0,\label{eq:ZeroDivergence_CH1}\\
\frac{\partial {T} }{\partial {t}} + \mathbf{u} \cdot \nabla {T} & = & \frac{1}{Pr} 
\nabla^2 {T},
\label{eq:Heat_CH1}\\
\frac{\partial \mathbf{B}}{\partial {t}} & = & \nabla \times \left( \mathbf{u} \times 
\mathbf{B} \right) - 
\frac{1}{Pm^*} 
\nabla\times\left( \tilde\lambda(\nabla\times\mathbf{B})\right), 
\label{eq:Maxwell_CH1}
\end{eqnarray}
\end{small}
where $\mathbf{u}$\label{in:u} and $\mathbf{B}$\label{in:B} are the
velocity and magnetic induction vectors, respectively; $t$\label{in:t}
is time; $ T$\label{in:T} and $ P$\label{in:P} are the temperature and
pressure scalars, respectively; $g$\label{in:g} is the acceleration of gravity
which changes linearly with depth, and $g_o$ is the value of
this acceleration at the outer boundary;
$\mathbf{\hat{r}}$ is the radial unit vector, and $\mathbf{\hat{z}}$
is the unit vector in the direction of the angular momentum of the
rotating frame. 
$\tilde{\lambda}(r)=\frac{\lambda(r)}{\lambda_i}$ is the
normalized magnetic diffusivity, where $\lambda(r)$ and $\lambda_i=\lambda(r_i)$ are
the magnetic diffusivities, in units of m$^2$s$^{-1}$, of the liquid
core as a function of $r$ and of the solid inner core respectively.

Equations~\ref{eq:Navier-Stokes_CH1} to \ref{eq:Maxwell_CH1} are 
expressed in terms of the following 
non-dimensional 
parameters: 
the Rayleigh number,
\begin{center}
\begin{equation}
Ra=\frac{\alpha g_o \,\Delta T\, D^3} {\kappa \nu},\label{in:Ra}
\end{equation} 
\end{center}
where $\alpha$ is the thermal expansion coefficient and $\kappa$ is the 
thermal diffusivity.
The Ekman number, 
\begin{center}
\begin{equation}
E=\frac{\nu}{\Omega\,D^2}, \label{in:E}
\end{equation} 
\end{center}
which is the ratio between viscous and Coriolis forces in the system.
The Prandtl number,
\begin{center}
\begin{equation}
Pr=\frac{\nu}{\kappa},\label{in:Pr}
\end{equation} 
\end{center}
which is the ratio between the viscous and the thermal diffusivities.
And a modified magnetic Prandtl number,
\begin{center}
\begin{equation}
Pm^*=\frac{\nu}{\lambda_i},\label{in:Pm}
\end{equation} 
\end{center}
which is the ratio between the viscous diffusivity of the fluid and
the magnetic diffusivity at the inner boundary,
$\lambda(r_i)=\lambda_i$.  Note that the
magnetic Prandtl number changes as a function of radius,
we used the maximum magnetic Prandtl number for the 
definition of the non-dimensional units used in our numerical model.

The second term in the right hand side of  equation~\ref{eq:Maxwell_CH1} may be written as: 
\begin{equation}
\frac{1}{Pm^*} 
\nabla\times\left( \tilde\lambda(\nabla\times\mathbf{B})\right)
=\frac{1}{Pm^*} 
\left[
- \tilde\lambda \, \nabla^2\mathbf{B}
+
\nabla  \tilde\lambda \times \left(\nabla\times\mathbf{B}\right)
\right],
\label{eq:non-advTerm}
\end{equation}
in which the variable conductivity modifies the term 
$\nabla  \tilde\lambda \times \left(\nabla\times\mathbf{B}\right)$
exclusively.

The magnetic induction equation is modified by the spatial variability of $\lambda(r)$.
Since $\mathbf{B}$ is solenoidal, we define the
magnetic field as a function of two scalar potentials \citep[see]
[Appendix III]{Chandrasekhar1961}:
 \begin{equation}
 \mathbf{B}=\nabla\times\nabla\times {\Phi} \mathbf{\hat{r}} + \nabla\times {\Psi}
\mathbf{\hat{r}},
 \end{equation}
 where $\Phi$ is the toroidal and $\Psi$ the poloidal potential. 
 From this decomposition the time variation of the potentials \citep{Gomez2007a}:
 \begin{eqnarray}
 \frac{L_H}{r^2}\frac{\partial \Phi}{\partial t} &=& \mathbf{\hat{r}}\cdot (\nabla\times 
(\mathbf{u}\times\mathbf{B}))  + 
\frac{\tilde\lambda}{Pm^*}\frac{L_H}{r^2}\left[ \frac{\partial^2}{\partial r^2} - 
\frac{L_H}{r^2}\right]\Phi.\\
\frac{L_H}{r^2}\frac{\partial \Psi}{\partial t} &=& \mathbf{\hat{r}}\cdot \nabla
\times(\nabla\times (\mathbf{u}\times
\mathbf{B}))  
\nonumber\\&& \quad +
\frac{\tilde\lambda}{Pm^*}\frac{L_H}{r^2}\left[ \frac{\partial^2}{\partial r^2} - 
\frac{L_H}{r^2}\right]\Psi
+
\frac{\nabla_r \tilde\lambda}{Pm^*} \frac{L_H}{r^2} \frac{\partial \Psi }{\partial r},
\label{eq:toroidal_time}
 \end{eqnarray}
 where $\nabla_r=\mathbf{\hat{r}}\cdot\nabla$.  It is convenient to
 write the horizontal component of the angular momentum operator as
 $L_H=-r^2\nabla_H$, with $\nabla_H=\nabla-\nabla_r$.  The term
 $-\lambda\nabla^2\mathbf{B}$ in equation~\ref{eq:non-advTerm}
 introduces a zero order effect of a varying electrical diffusivity,
 it results on a decrease of the ohmic dissipation with depth, for
 both, poloidal and toroidal components of the magnetic field.  The
 radial derivative of the diffusivity has a first order effect,
 involving exclusively the poloidal component.  This is controlled by
 the term $\nabla \tilde\lambda \times
 \left(\nabla\times\mathbf{B}\right)$ in
 equation~\ref{eq:non-advTerm}.

\subsection{Radially variable electrical conductivity}

\begin{figure}[h!]
   \centering
   \includegraphics[width=4in]{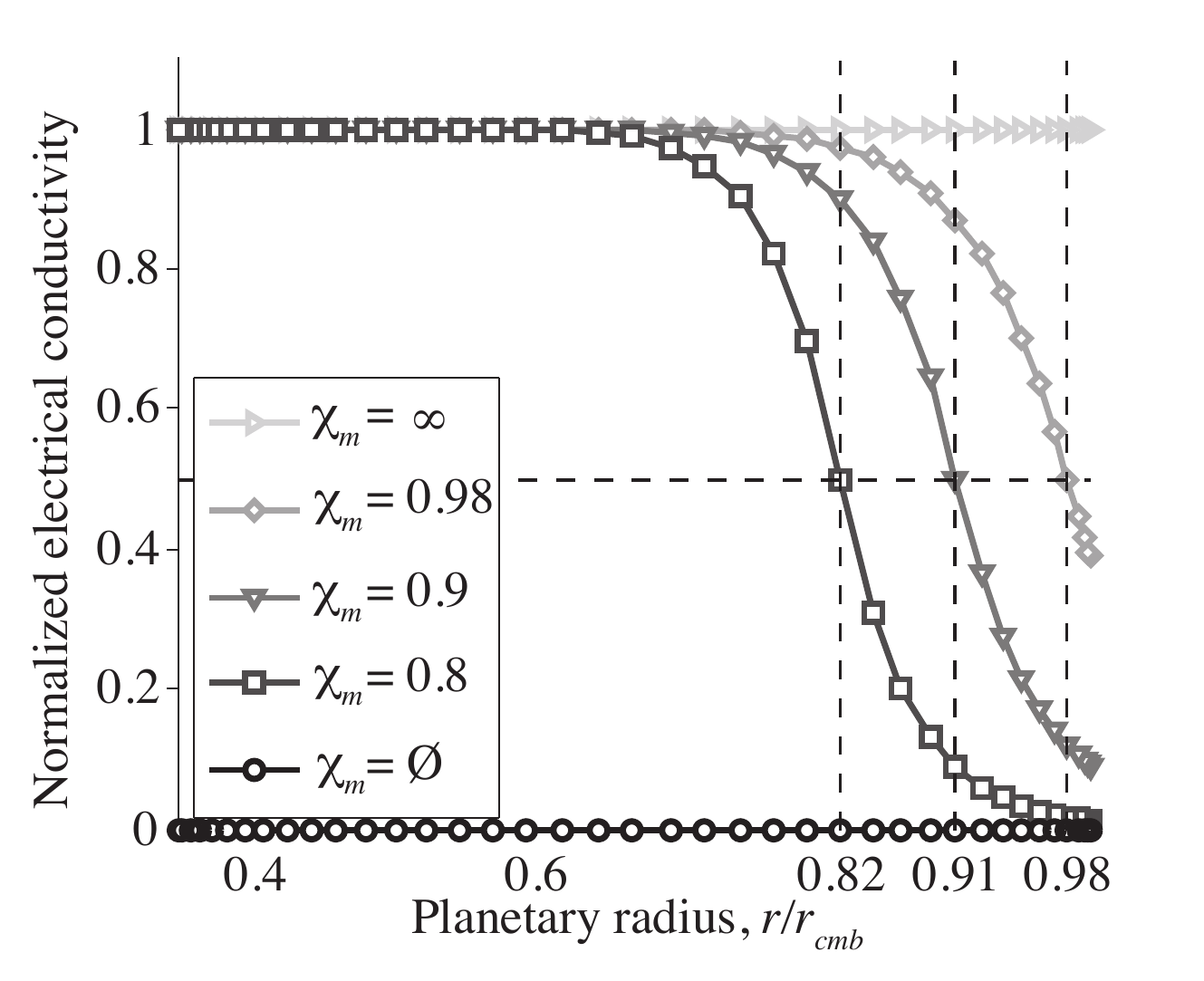}
   \caption{Normalized electrical conductivity as a function of radius.
    The electrical conductivity is normalized
     with the conductivity of the inner core. $r_m$ was chosen to be a
     grid point and thus $\chi_m=0.91\approx 0.9$ and $
     \chi_m=0.82\approx 0.8$ are labelled in the text as $\chi_m=0.9$
     and $ \chi_m=0.8$ respectively.}
   \label{fig:sigma_r}
\end{figure}

The radial change in electrical conductivity, $\sigma(r)=\frac{1}{\mu_0\,\lambda(r)}$,
where $\mu_0$ is the magnetic permeability of vacuum,
 is modeled with  a
piecewise continuous function whose derivative is also continuous.
The outermost part decreases exponentially
while the interior changes as a power of the radius.  One may write
the normalized electrical conductivity as:
\begin{equation}
\tilde\sigma(r)= \left\{
  \begin{array}{l@{\quad,\quad}l}
     1+(\tilde{\sigma}_m-1)\left( \frac{r-r_i}{r_m-r_i} \right)^a & r< r_m \\
     \tilde{\sigma}_m \exp\left[ \frac{k}{\tilde{\sigma}_m}(r-r_m)\right] & r\geq r_m
  \end{array}
  \right. 
\label{eq:sigma}
\end{equation}
where $r_i$ is the inner boundary, $r_m$ 
is the radius where
were the electrical conductivity changes from
a polynomial 
 to an exponential function, $a$ determines how fast the
exponential function decreases outside the conductive volume,
$\tilde{\sigma}_m$ is chosen to be the value of the normalized conductivity
where the functions match, $\tilde\sigma(r_m) = \tilde{\sigma}_m$, and $k=
a(\tilde{\sigma}_m -1)(r_m -r_i)^{-1}$ (see figure~\ref{fig:sigma_r}).
We choose to implement a continuous differentiable function in order
to have an analytical solution for $\lambda$ and $\nabla\lambda$.
This function for the electrical conductivity is chosen to model a
layering between an light-element-enriched fluid with a
light-element-poor fluid. During the solidification of the core, heavy
elements are preferentially deposited on the inner core. As a result,
light element enriched liquid rises buoyantly to the top of the outer
core, just below the CMB.  A more realistic representation of the
electrical conductivity variability, and of miscibility of iron and
light element enriched iron at high pressures, needs to be studied.
From an experimental point of view such measurements are very
challenging.  
First principles numerical simulations
\citep[e.g.,][]{Caracas2009} may lead to some useful results and a more
realistic characterization of $\lambda(r)$ may be implemented in the
future.

\section{Dynamo regimes studied}\label{sec:param}

We study self-sustained dynamos in two main regimes. First, we include
two sets of cases with relatively low $Ra$, where the
resultant fields are non-reversing and dipole dominated.  Second, we
analyzed strongly forced and time variable cases where dynamos result
in dipolar magnetic fields with strong multipolar components.  
We chose a values for
$Ra=3.4Ra_{c}=2.5\times10^6$ (set A), $Ra=7.8Ra_{c}=8.5\times10^7$
(set B) and $Ra=16.8Ra_{c}=1.3\times10^7$ (sets C and C'), where the value
of $Ra_{c}$ was calculated using the empirical formula for the onset
of convection in rotating spherical shells from
\citet{Al-Shamali2004}.  
For sets A, C
and C', $E=10^{-4}$ and $E=10^{-5}$ for set B.  Common parameters used
for all sets are a Prandtl number $Pr=1$, and a radius ratio
$\chi=\frac{r_i}{r_o}=0.35$.  For sets A, B and C non-slip and fixed
temperature boundary conditions 
are used at both, bottom and top
boundaries.  We also include cases with top and bottom free-slip boundaries,
i.e., set C'. For this set, all other parameters identical to those of set C.
The models account for an electrically conductive inner core where
electrical conductivity 
is homogeneous in the inner core, continuous 
through the ICB, and may vary radially in the outer core.  The electrical
conductivity for the mantle was chosen to be zero.  All simulations in
sets 
A, C and C' have the same resolution, with 61 radial levels for the
outer core and 17 for the inner core; 288 ($l_{max}=120$) grid points
in the azimuthal direction and half of that in the latitudinal
direction.  For set B the latitudinal and azimuthal resolution was
increased to 256 and 512 respectively ($l_{max}=170$) and 65 radial
levels were used.  Hyperdyffusivities were not used for any of the
runs presented in this paper.  For the purpose of normalization only,
the magnetic Prandtl number, $Pm^*=5$, is defined by the ratio
between the viscous diffusivity of the fluid, $\nu$, and electrical
diffusivity of the inner core, $\lambda_i$ (which is the minimum
diffusivity in the simulated volume).

For each set A, B, C and C' we compare two cases with homogeneous
electrical conductivity in the fluid with cases with electrical
conductivity varying as a function of radius.  To label these cases
$\chi_m={r_m}/{r_o}$ is used, where $r_m$ is the same used in
equation~\ref{eq:sigma}.  Cases with homogeneous conductivity are:
non-magnetic case (from here on referred to as $\chi_m=\emptyset$) and
a case with constant electrical conductivity ($\chi_m=\infty$).  The
radially variable electrical conductivity cases A, B, C and C' have
$\chi_m\approx0.8$, $\chi_m\approx0.9$ and $\chi_m\approx0.98$, 
see figure~\ref{fig:sigma_r}.
 The shape of the electrical conductivity function is defined 
 by equation~\ref{eq:sigma}.  For all cases $a=10$, and
$\tilde{\sigma}_m=\frac{1}{2}$.

\section{Results}\label{sec:result}

\begin{figure}[h!]
   \centering
   \includegraphics[width=5in]{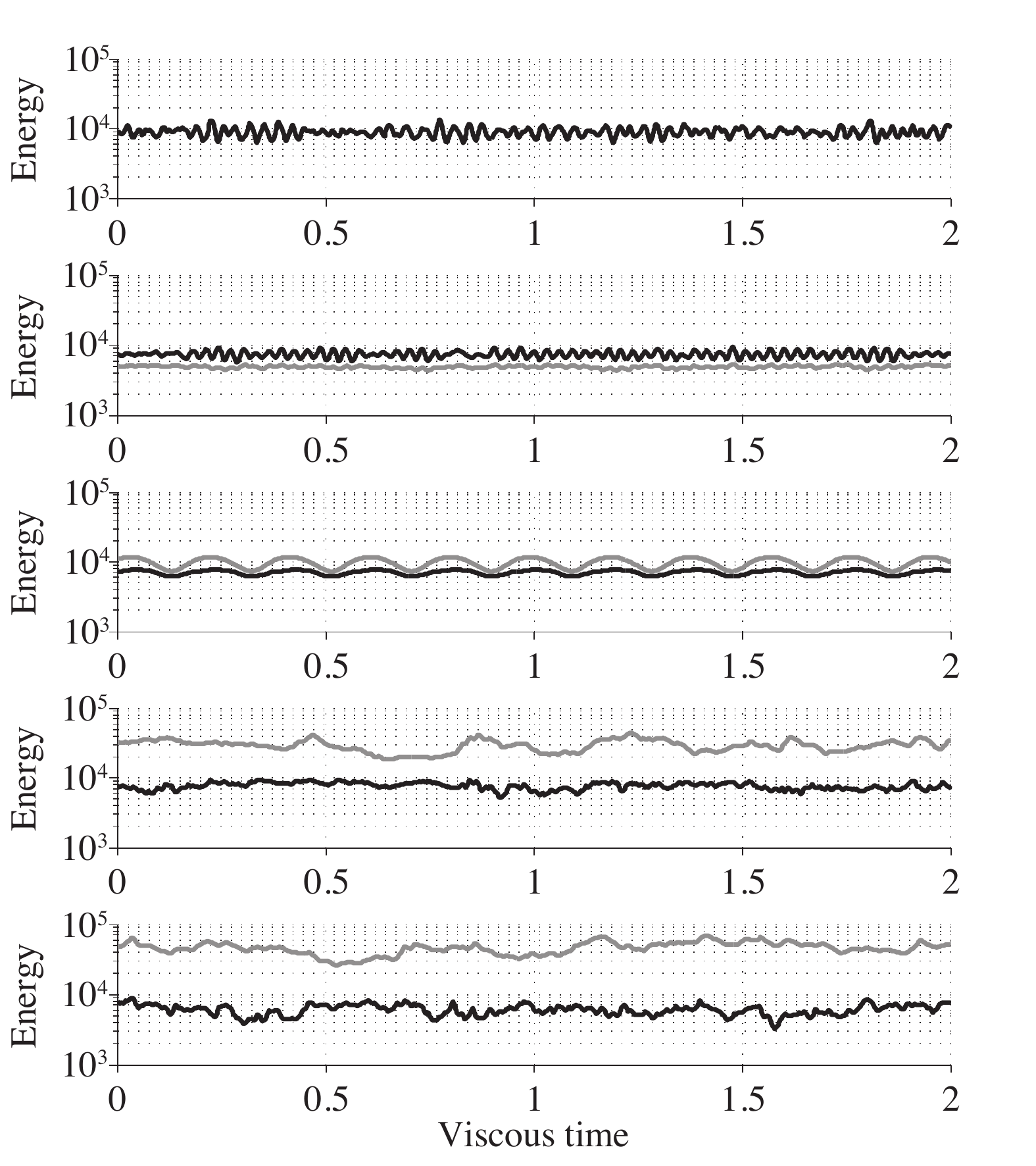}
   \caption{For set A, 
     time series of the total kinetic energy in the
     fluid (in back) and the total magnetic energy (in grey). From top
     to bottom increasing values of $\chi_m$; $ \chi_m=\emptyset$,
     $\chi_m=0.8$, $\chi_m=0.9$, $\chi_m=0.98$, and $\chi_m=\infty$ in
     rows one to five.}
   \label{fig:TS_en_3Rac}
\end{figure}

\begin{figure}[h!]
   \centering
   \includegraphics[width=5in]{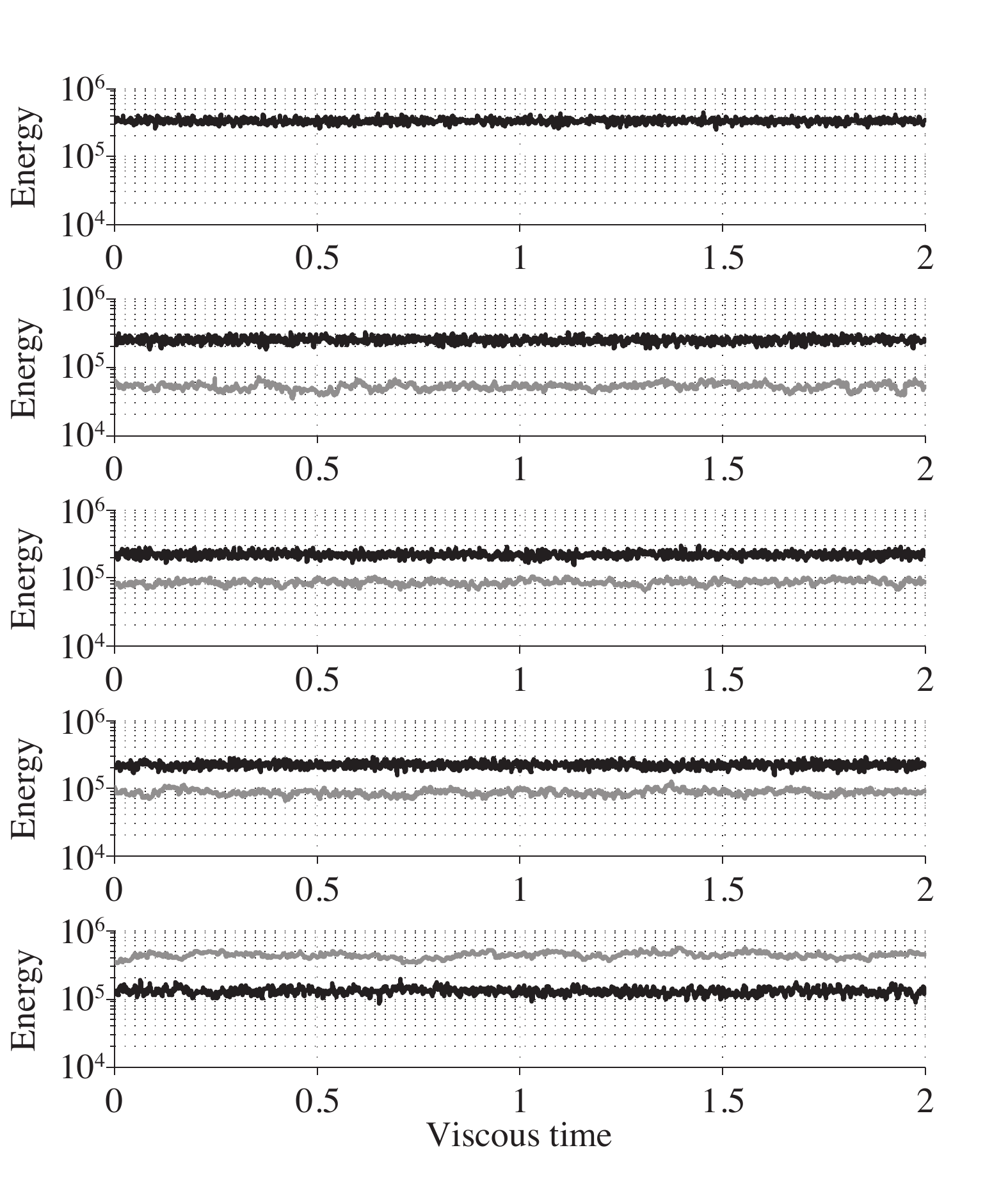}
   \caption{Similar to figure~\ref{fig:TS_en_3Rac} but for set C.
$\chi_m=\emptyset$, $\chi_m=0.8$, $\chi_m=0.9$, $
\chi_m=0.98$, and $\chi_m=\infty$ in rows one to five. }
   \label{fig:TS_en_16Rac}
\end{figure}

Significant changes in the internal dynamics of the simulations are
obtained by varying the electrical conductivity profile. 
Energy time series for sets A and C are included in 
figures~\ref{fig:TS_en_3Rac}~and~\ref{fig:TS_en_16Rac}.
The time variability of the solution varies with $\chi_m$. We find 
quasi-stationary solutions for set A for $\chi_m=0.8$ and $\chi_m=0.9$.
The solutions for sets B, C and C' did not result in quasi-stationary solutions,
as may be seen for set C in figure~\ref{fig:TS_en_16Rac}.

We include
Table~\ref{tab:averages} with time-averaged values, where the average
has been taken over two viscous diffusion times.  The standard
deviation from the time-averaged value is also included.
\begin{sidewaystable}[htbp]
  \begin{center}
      \caption{ Time-averaged values for all cases studied in this
        paper. In the first and second columns the ratios
        $Ra/Ra_{c}$ and $\chi_m=r_m/r_o$.  Columns three and four,
        the time-averaged mean Reynolds, $Re$, and Elsasser $\Lambda$
        numbers.  In column five, the mean Elsasser, $\Lambda^{CMB}$,
        number at the top boundary.  In columns six and seven
        the ratios of axial dipole energy to total field energy at the
        CMB ($Q^a_o$), and for the fluid outer core ($Q^a$)
        respectively. Note that the values presented for the magnetic
        field at $r=r_o$ are the sum of both, poloidal and toroidal
        components, as opposed to only poloidal (corresponding to the
        external field).  Lastly, in columns eight and nine, the time- and space-averaged
        Hartmann and thermal boundary layer thicknesses, 
        $\delta_H$ and $\delta_k$.}
        \input{tab_1.txt}
   \label{tab:averages}
   \end{center}
\end{sidewaystable}

The Reynolds number, $Re=\frac{\left<|\mathbf{u}|\right>D}{\nu}$, for
$\chi_m=\emptyset$ is always greater than for the corresponding
$\chi_m=\infty$.  In particular, the toroidal kinetic energy is
decreased due to the presence of Lorentz forces.

Values of the averaged Elsasser number,
$\Lambda=\frac{\left<\mathbf{B}^2\right>}{\rho\mu_0\lambda_i\Omega}$
in the fluid outer core and at the top of the simulated fluid are
included in columns four and five in Table~\ref{tab:averages}.
It is important to note that the definition of $\Lambda$ uses 
$\lambda_i$ instead of $\lambda(r)$ to be consistent with the 
normalization and the non-dimensional parameters. 
Using the total magnetic energy, $M$, the axisymmetric dipolar magnetic energy, $M^a$, 
and the total and axisymmetric  energy at the CMB, $M_o$, and $M^a_o$, respectively; we calculate
the ratio of the axisymmetric dipole to the whole magnetic energy 
for the whole fluid ($Q^a=M^a/M$) and at the CMB ($Q^a_o=M^a_o/M_o$).
These values are shown in columns 7 and 6, respectively.

The 
axisymmetric dipole component is dominant
for all runs in set A, see $Q^a_o$ in column six in Table~\ref{tab:averages}.  
In contrast, the magnetic field energy is distributed over higher
multipoles in sets B, C, and C', due to the higher $Ra/Ra_{c}$ value. 
For these three sets, the variable conductivity cases exhibit broader
length scale features and weaker fields at the CMB than case
$\chi_m=\infty$, this is due to the diffusion of magnetic field
through the low electrical conductivity volume.

\begin{figure}[h!]
   \centering
   \includegraphics[width=4in]{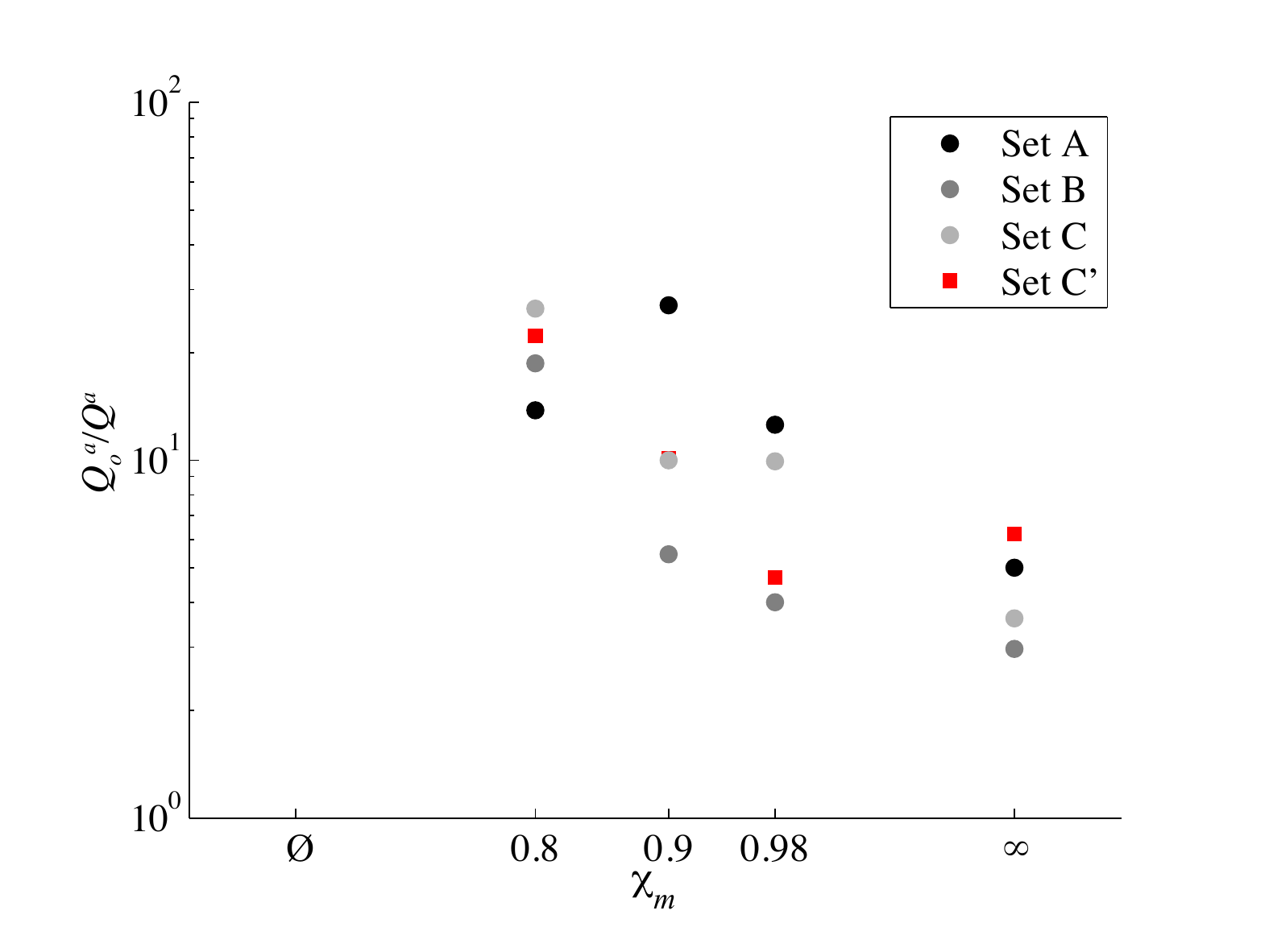}
   \caption{Time average of the ratio of the relative axisymmetric
     dipole energy at the CMB ($Q^a_o$) and the relative
     axisymmetric dipole energy over the fluid core ($Q^a$).  The
     ratio has been calculated for all runs where magnetic fields are
     non-zero and the color refers to sets A (Black), B (dark grey),
     and C (light grey). Values of simulations in set C' are
     shown with red squares.}
   \label{fig:RatioRelat}
\end{figure}

Observable characteristics of dynamos are limited to the radial
component of the magnetic field at the CMB and its temporal
variability.  It is  important to understand how the flow dynamics
is correlated to the surface field.  Quantities such as $Q^a_o$ in
Table~\ref{tab:averages} serve as an indication of the observable
magnetic field. In contrast, $Q^a$ characterizes how the magnetic
field being carried by the core flow behaves.  We compare how much of
the fluid core axisymmetry is carried to the CMB surface field.  To do
this, we take the ratio of the normalized surface axisymmetric dipole
energy and the normalized fluid core axisymmetric dipole energy, this
quantity is shown as a function of $\chi_m$ in figure \ref{fig:RatioRelat}.
With the exception of set A a tendency for decreasing $Q^a_o/Q^a$
is found for increasing $\chi_m$. This is reasonable since higher
harmonics are preferentially damped by the diffusion through the
non-conducting volume.  Thus, models with larger non-conductive
volumes result in stronger axisymmetric dipoles at the top surface.
The exception to this in set A is given by the average of
$\chi_m=0.9$, the quasi-stationary solution results in an increased
axisymmetry when compared to $\chi_m=0.8$.
It is possible that this is a
consequence of the periodic variation (see panel 3 in
figure~\ref{fig:TS_en_3Rac}) reinforcing and stabilizing the
axisymmetric components of the magnetic field.

\begin{sidewaysfigure}
   \centering
   \begin{tabular}{m{0.1in}|m{1.5in}|m{1.5in}|m{1.5in}|m{1.5in}|m{1.5in}}
    &\multicolumn{1}{c|}{$\chi_m=\emptyset$}  
    &\multicolumn{1}{c|}{ $\chi_m=0.8$} 
    &\multicolumn{1}{c|}{ $\chi_m=0.9$ }
    &\multicolumn{1}{c|}{ $\chi_m=0.98$} 
    &\multicolumn{1}{c}{$\chi_m=\infty$} \\

   A&
   \includegraphics[width=1.5in]{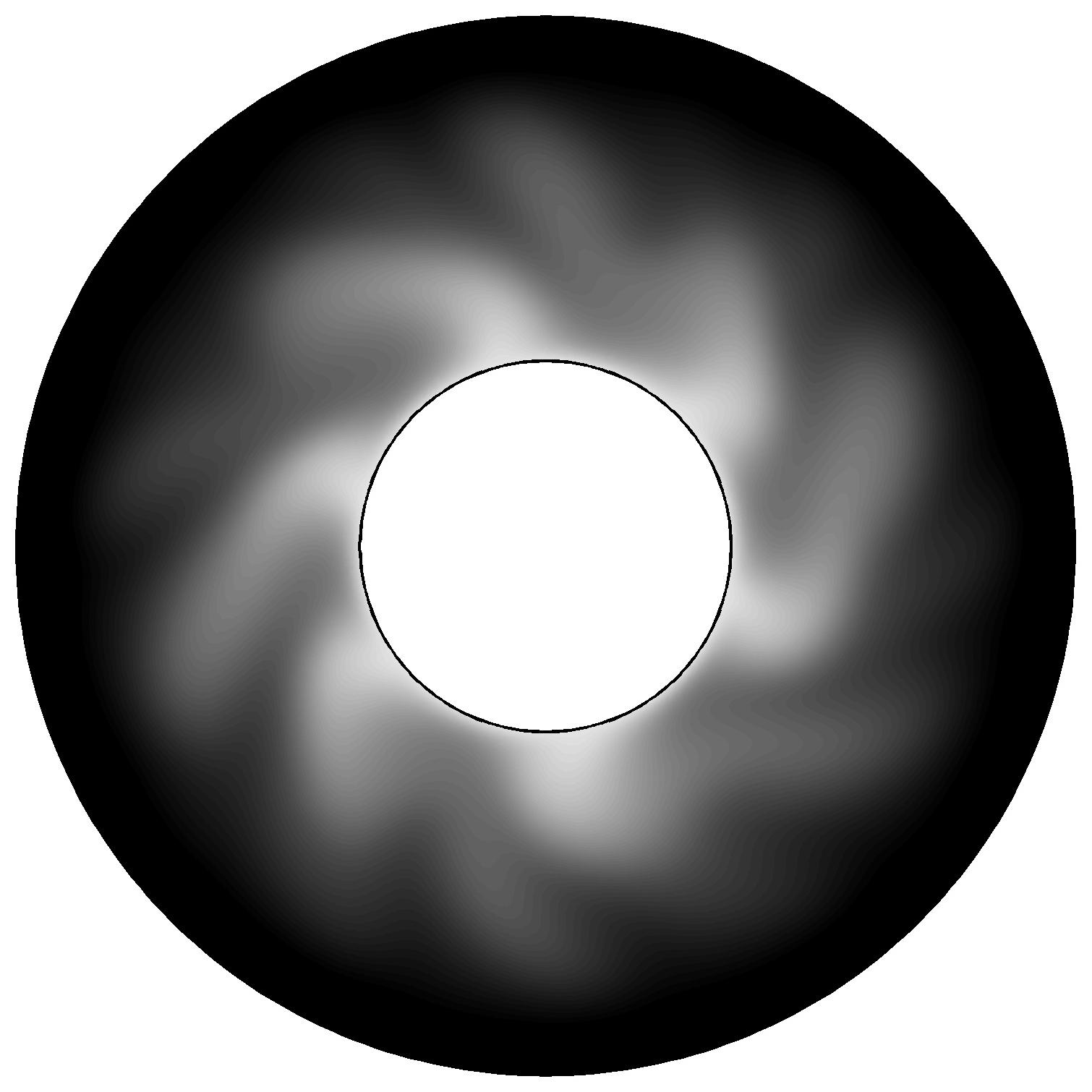}&
   \includegraphics[width=1.5in]{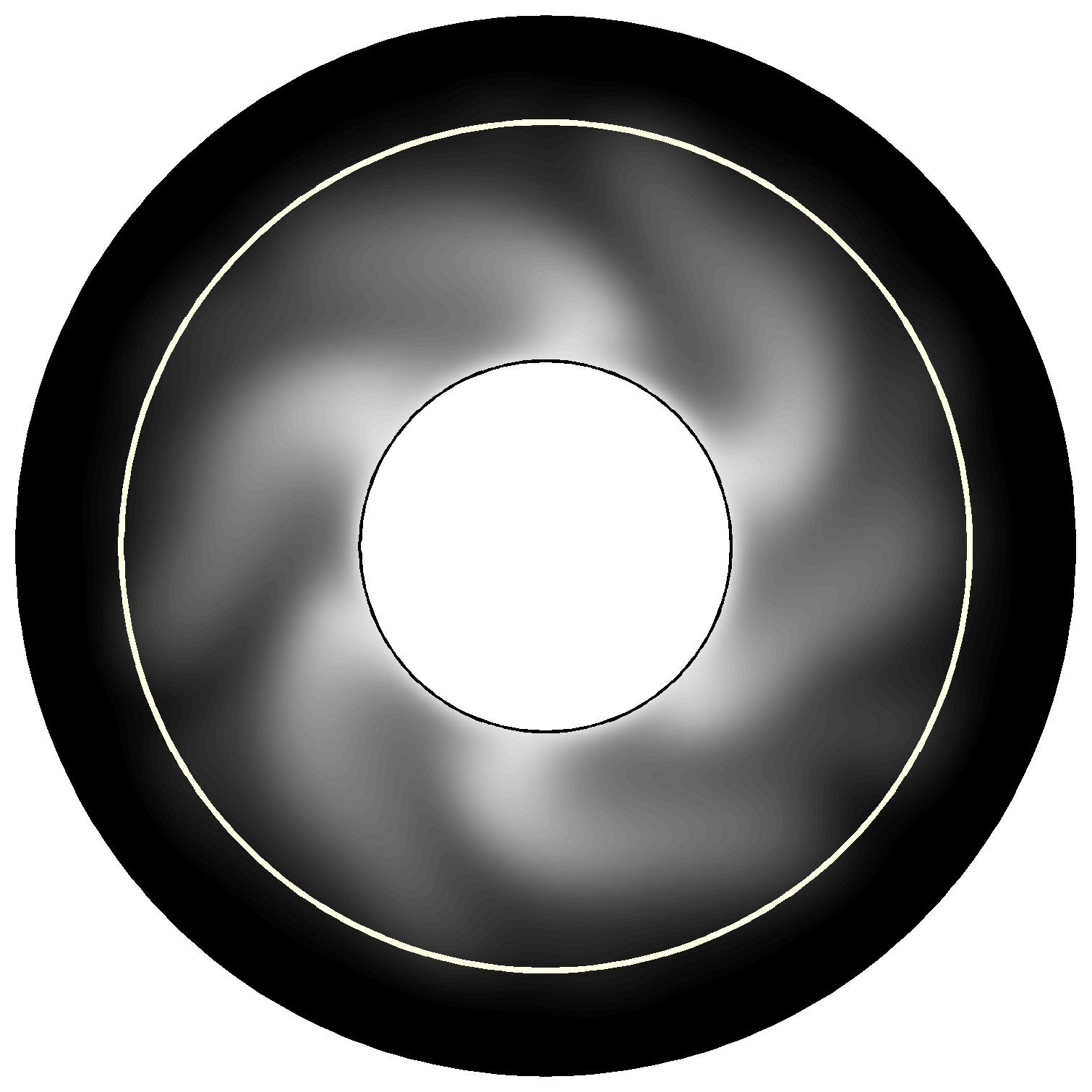}&
   \includegraphics[width=1.5in]{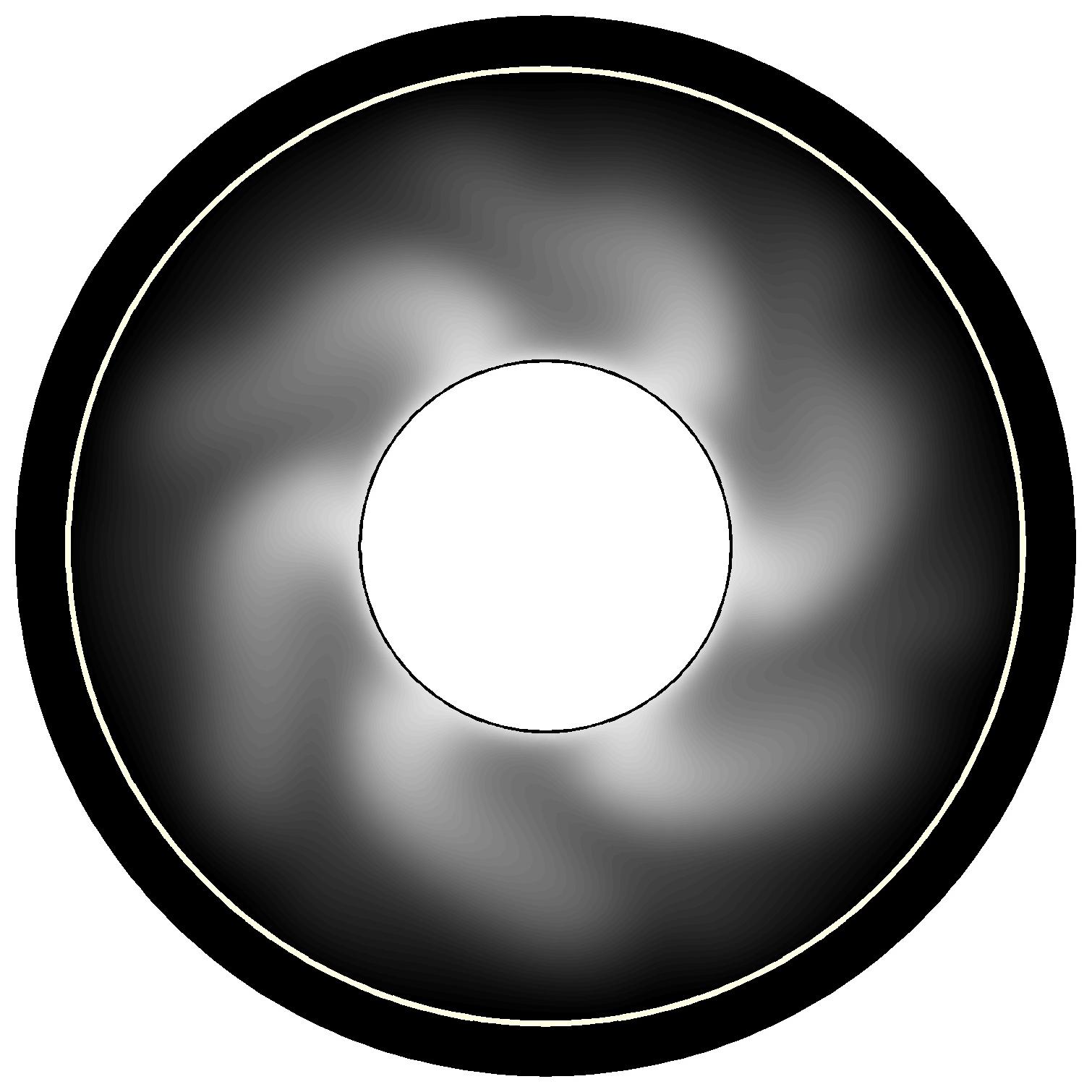}&
   \includegraphics[width=1.5in]{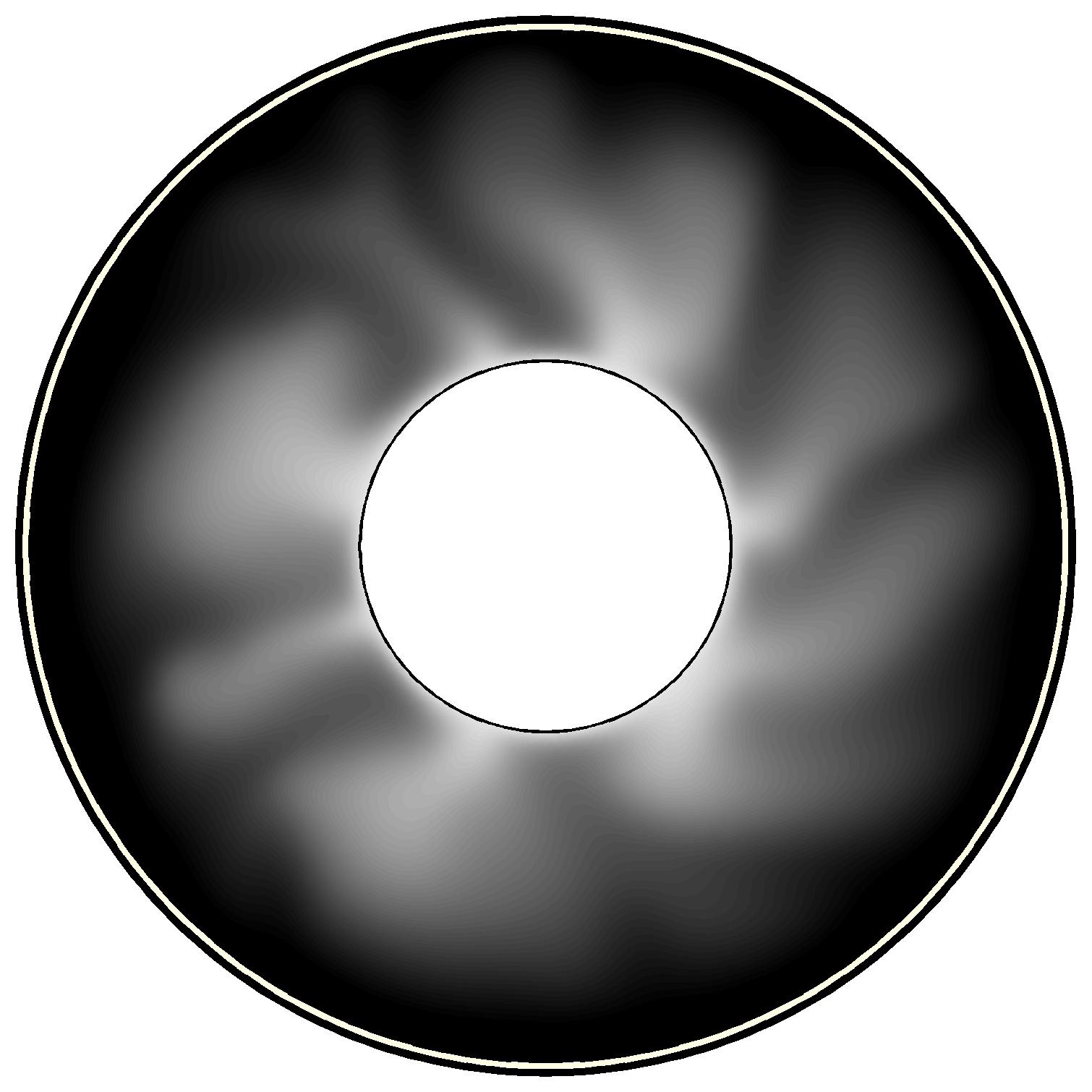}&
   \includegraphics[width=1.5in]{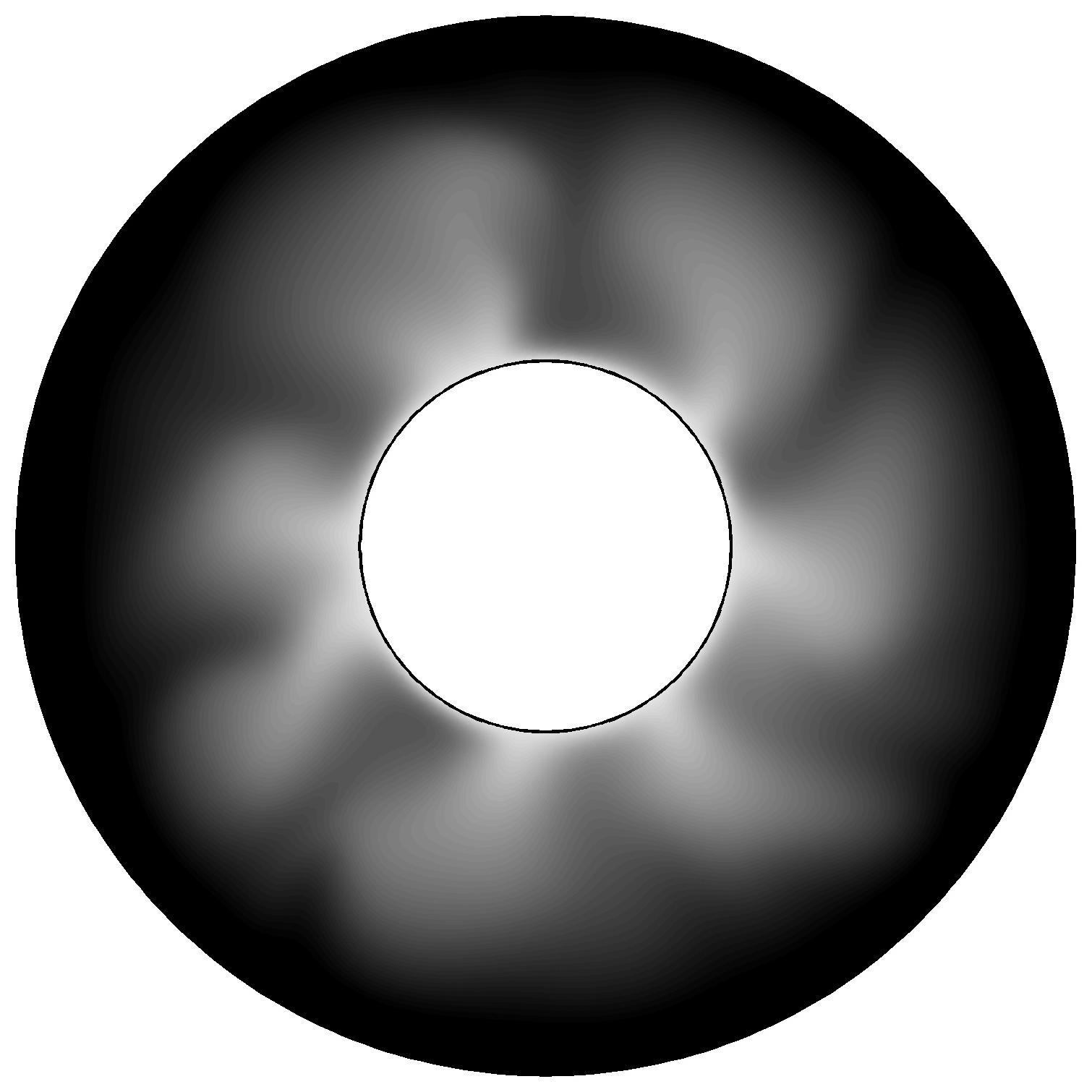}\\

   B&
   \includegraphics[width=1.5in]{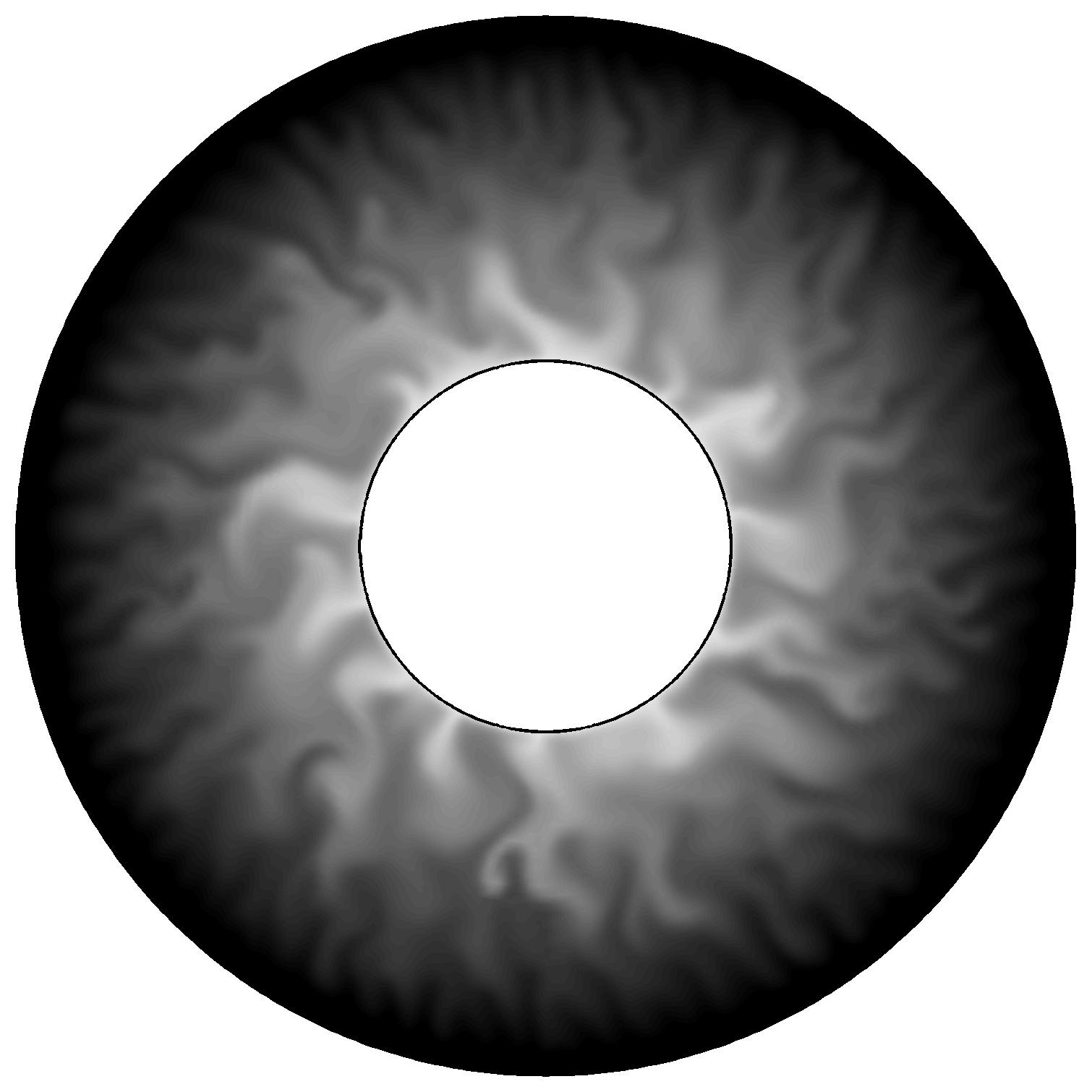}&
   \includegraphics[width=1.5in]{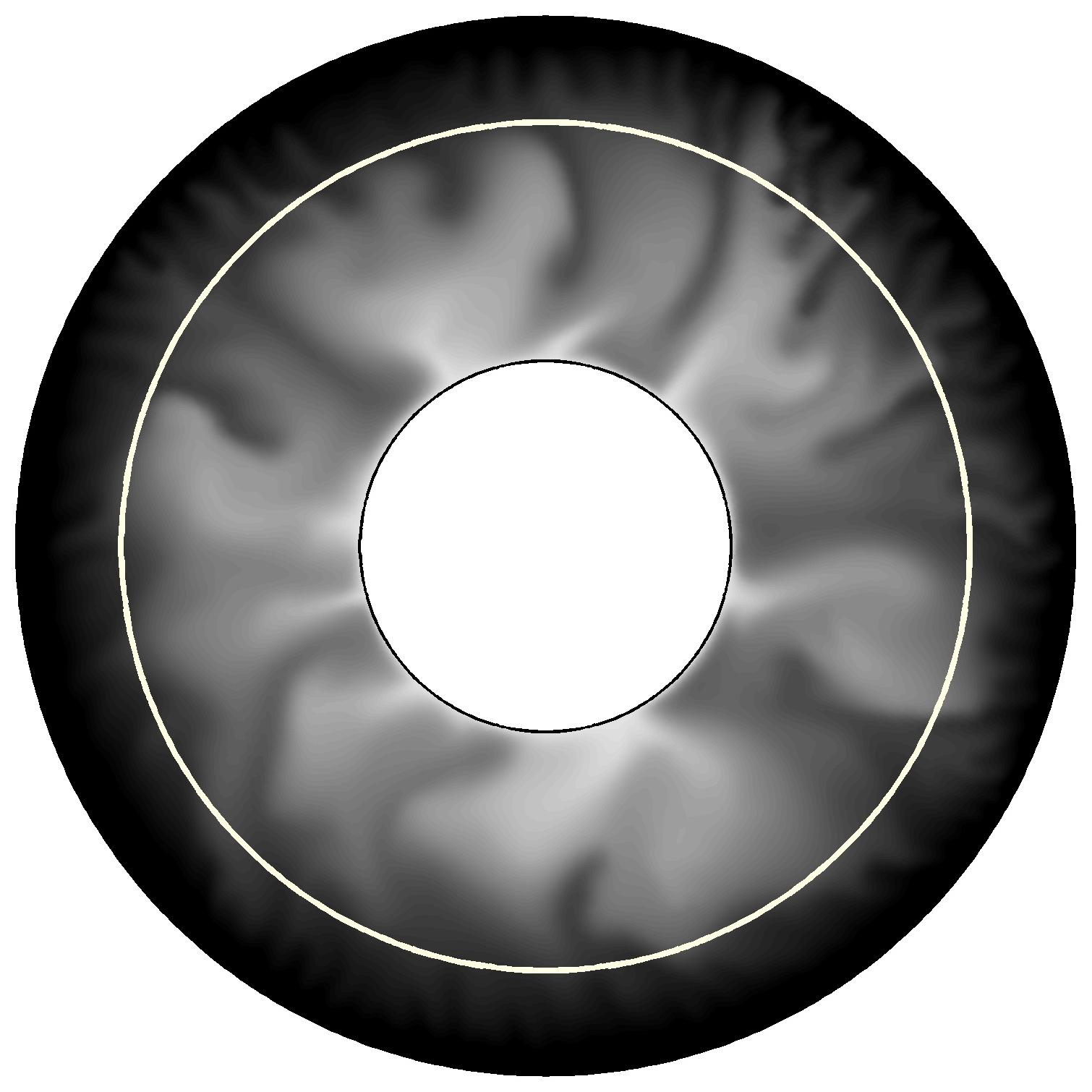}&
   \includegraphics[width=1.5in]{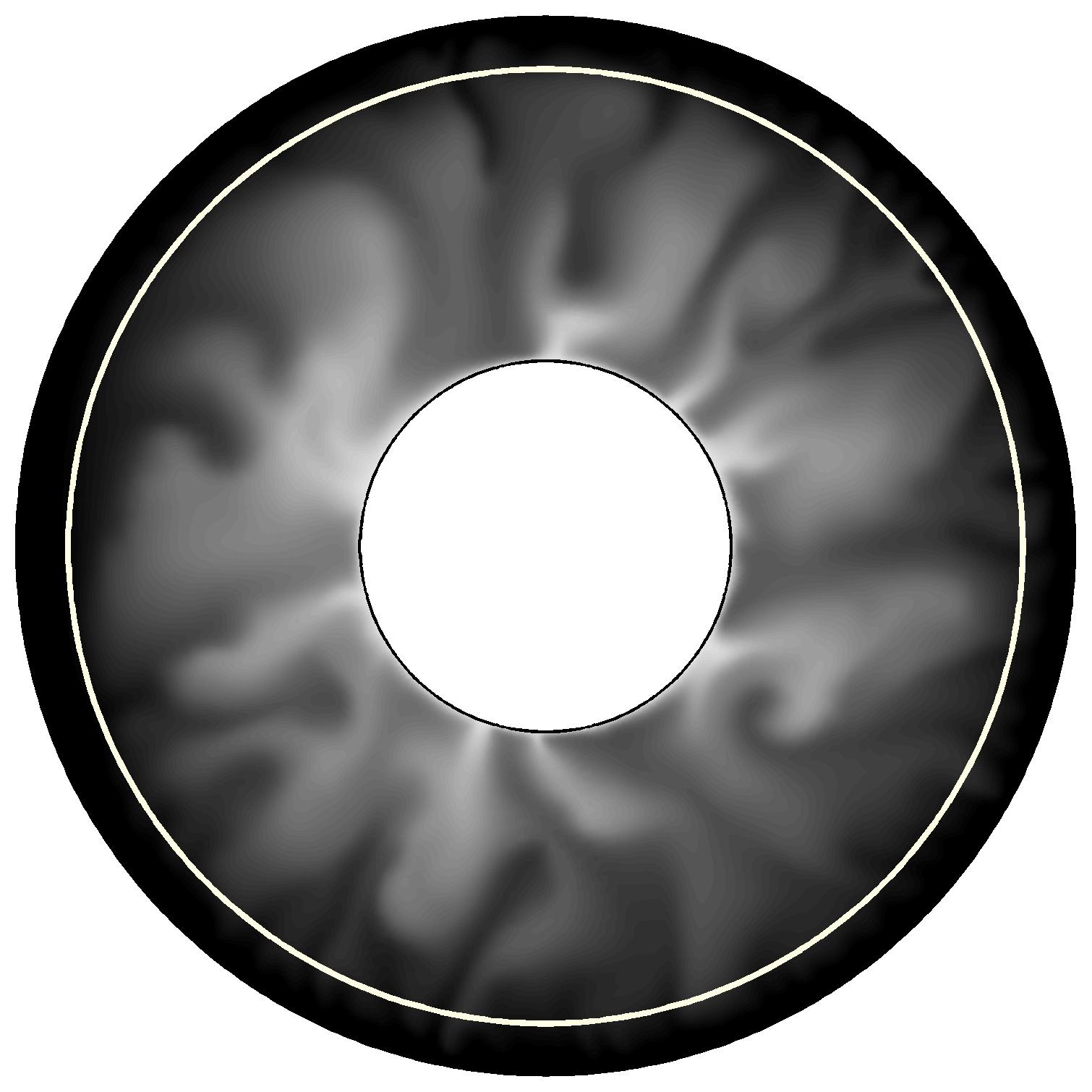}&
   \includegraphics[width=1.5in]{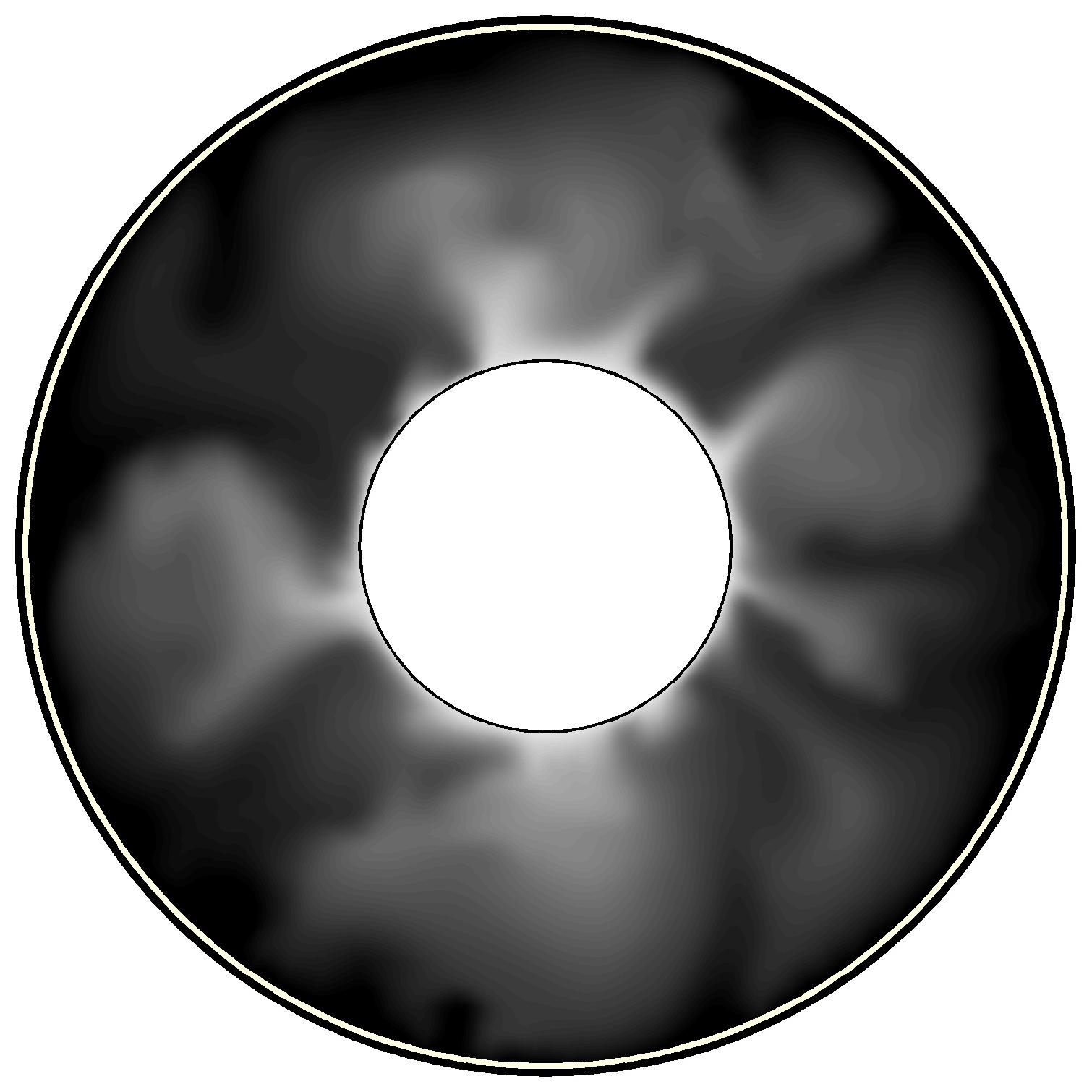}&
   \includegraphics[width=1.5in]{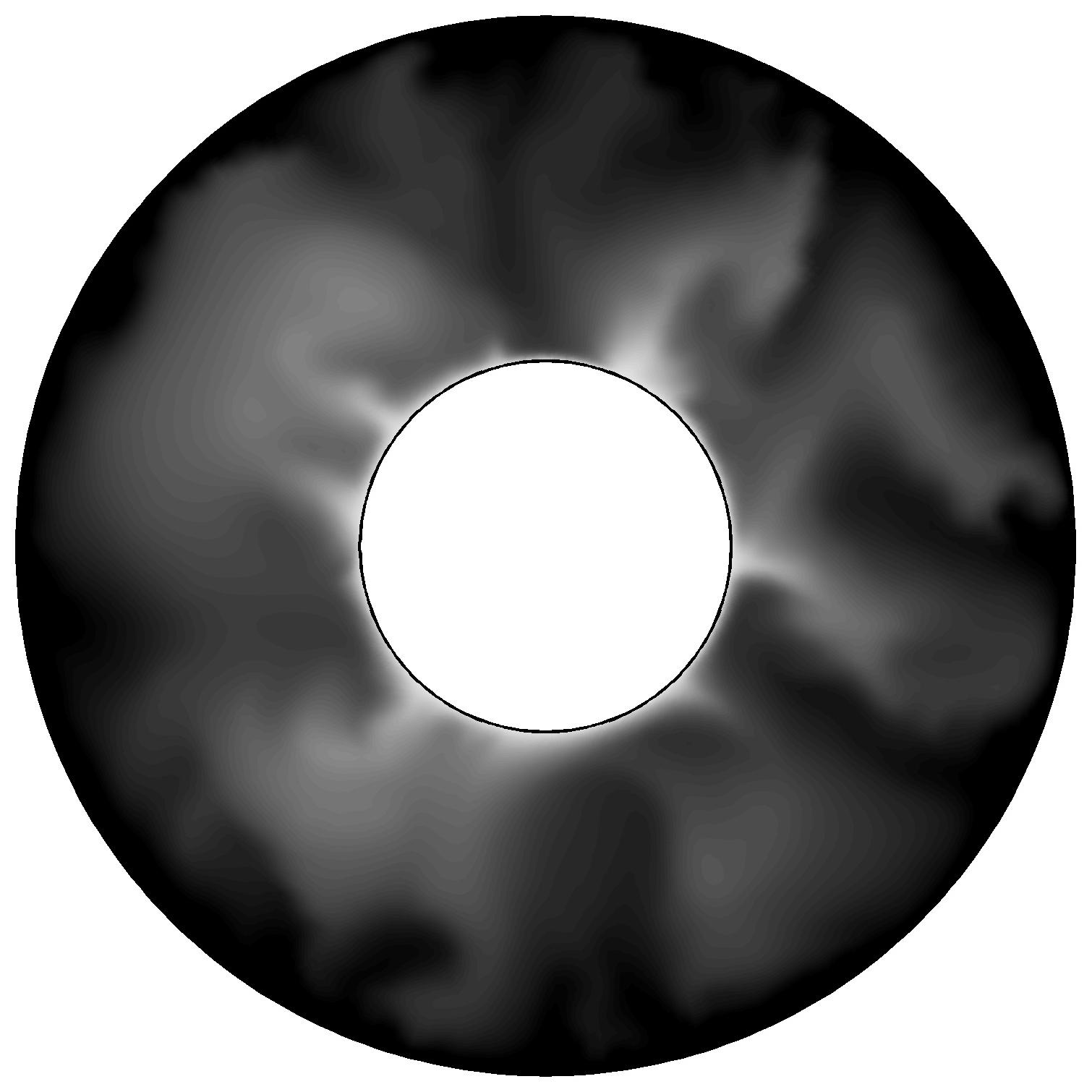}
   \\

   C&
   \includegraphics[width=1.5in]{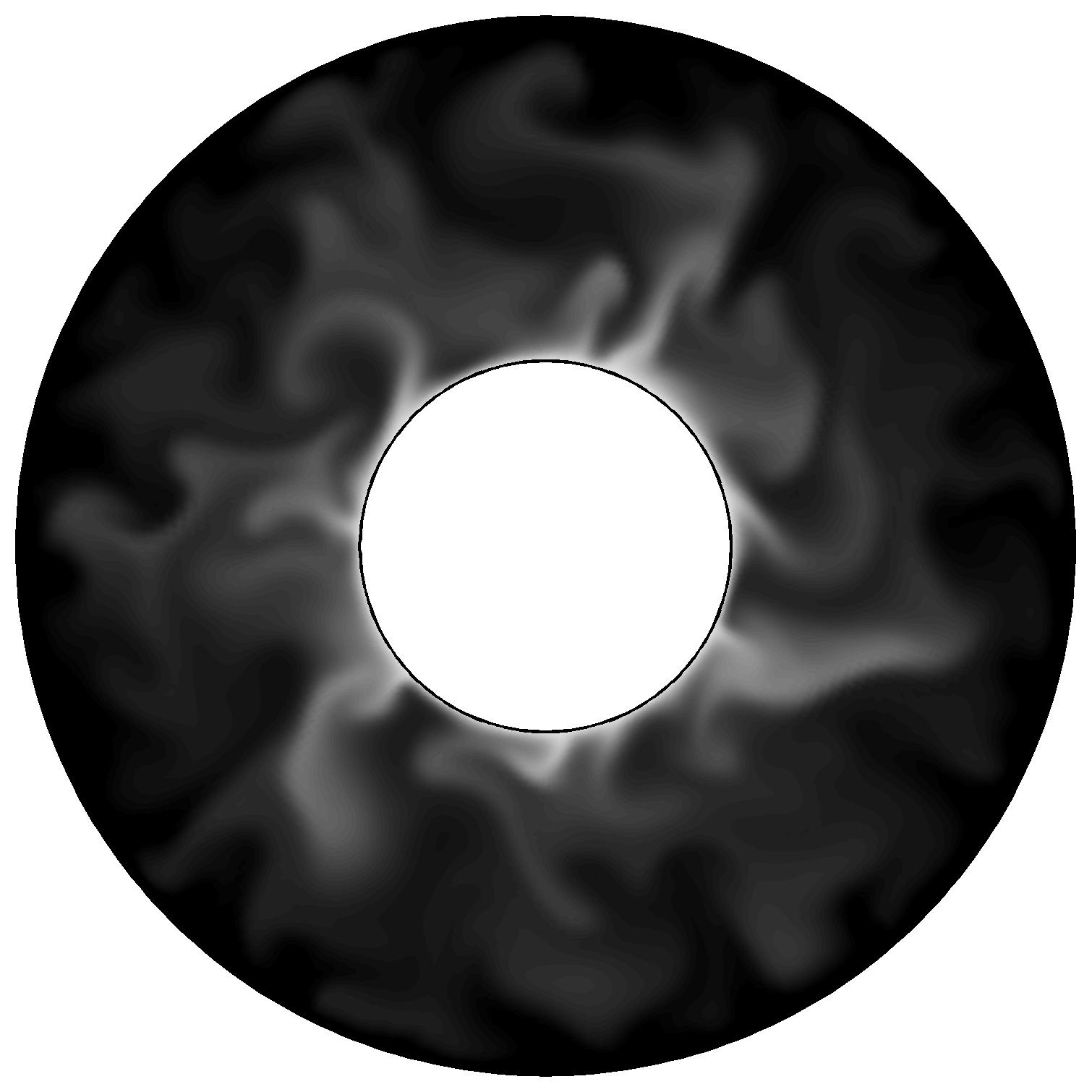}&
   \includegraphics[width=1.5in]{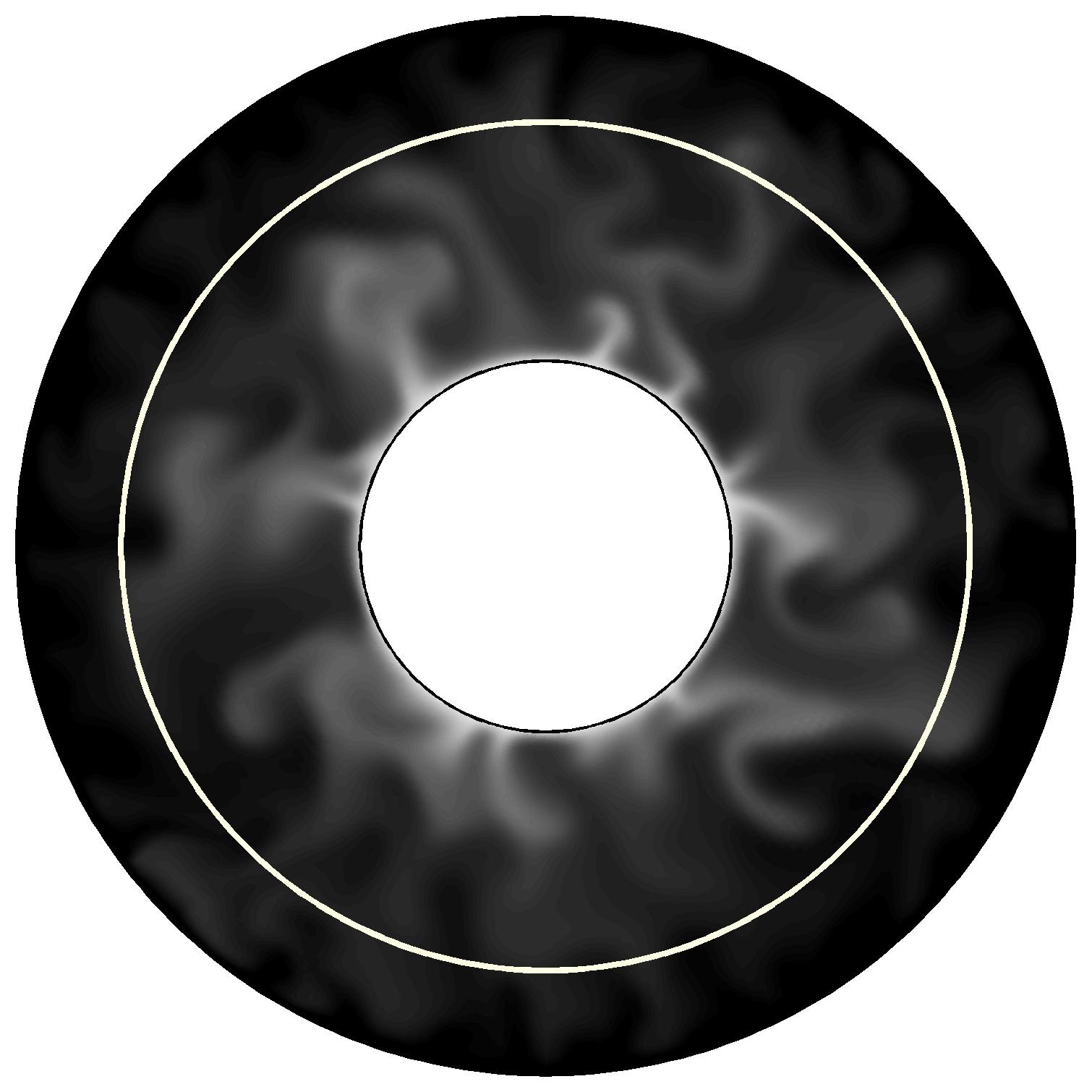}&
   \includegraphics[width=1.5in]{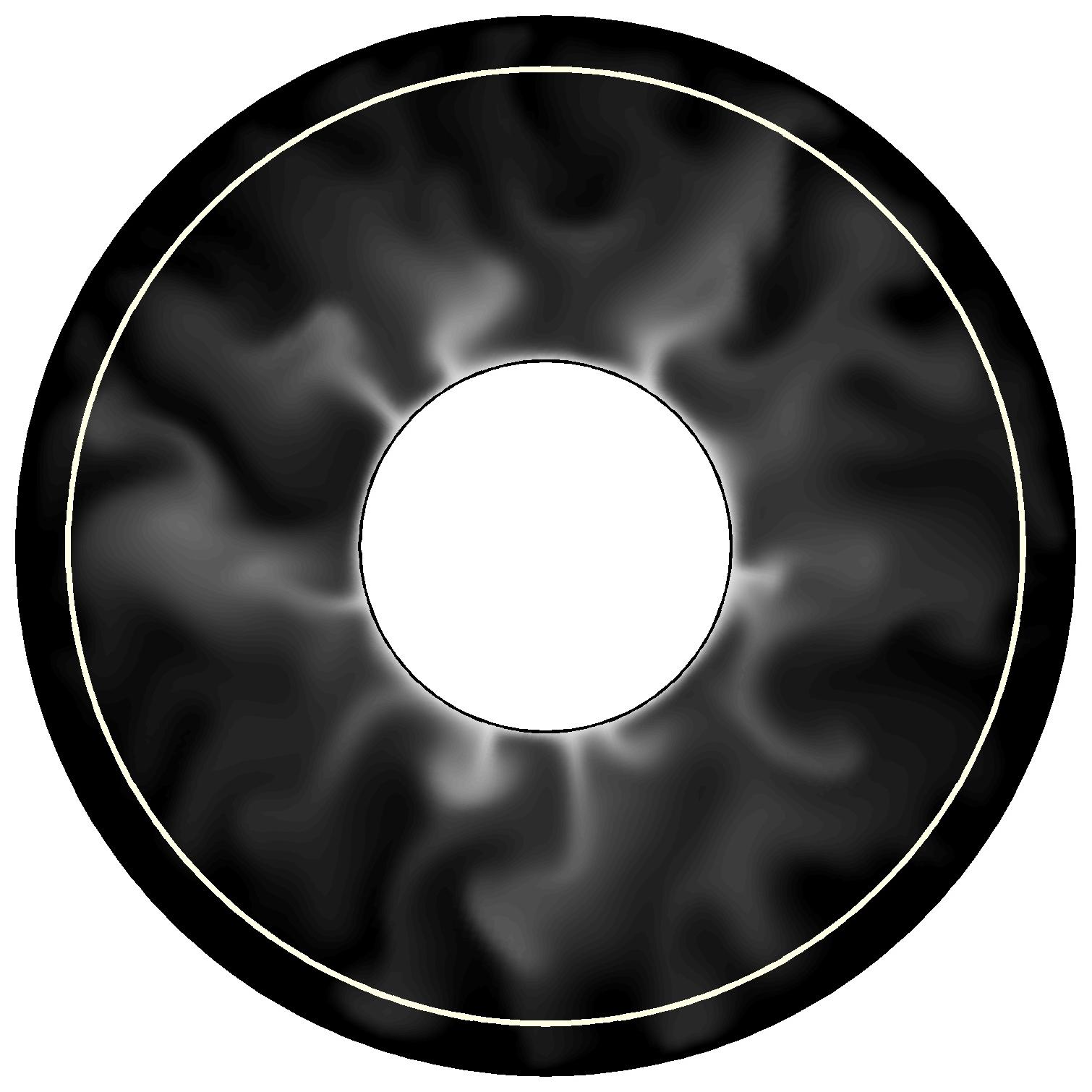}&
   \includegraphics[width=1.5in]{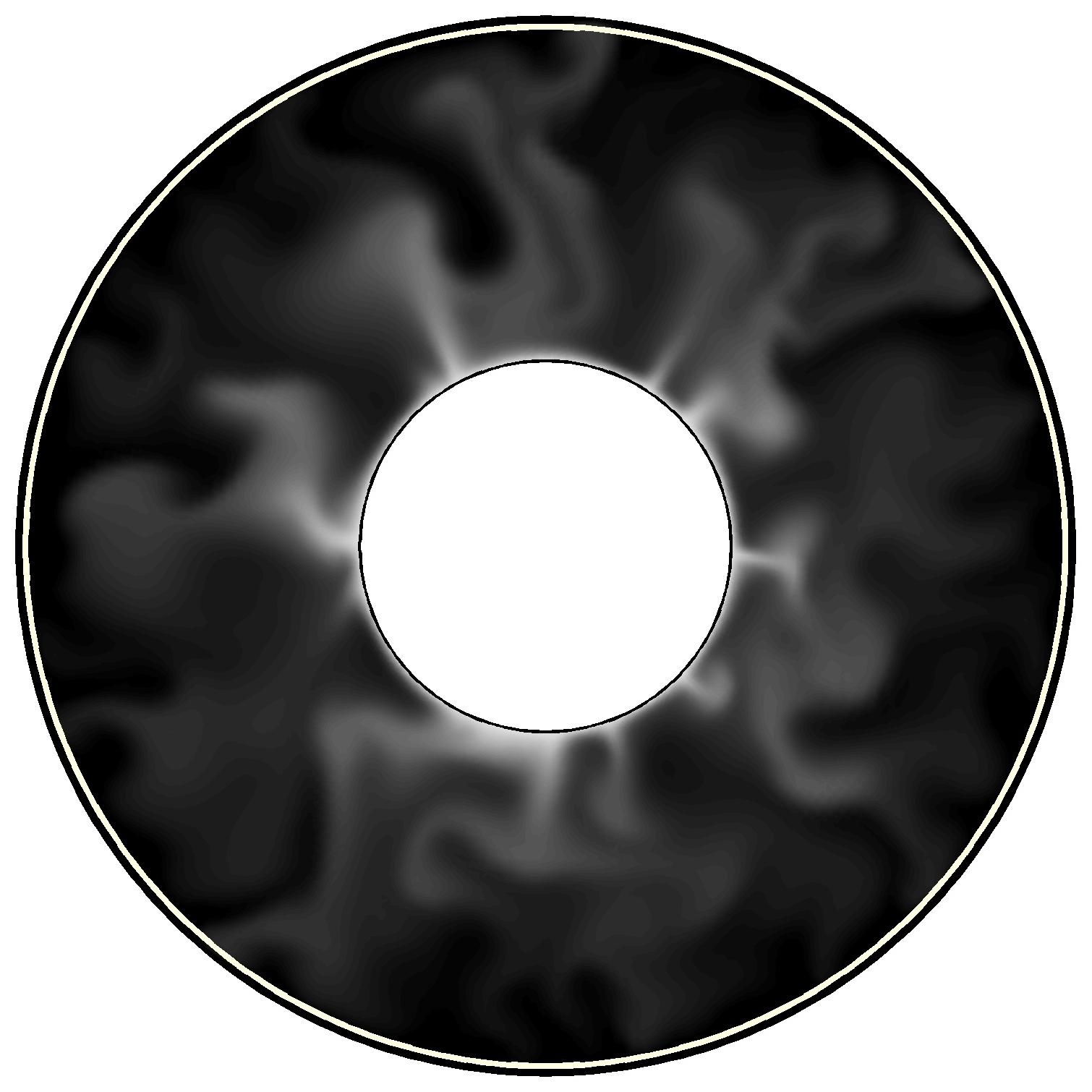}&
   \includegraphics[width=1.5in]{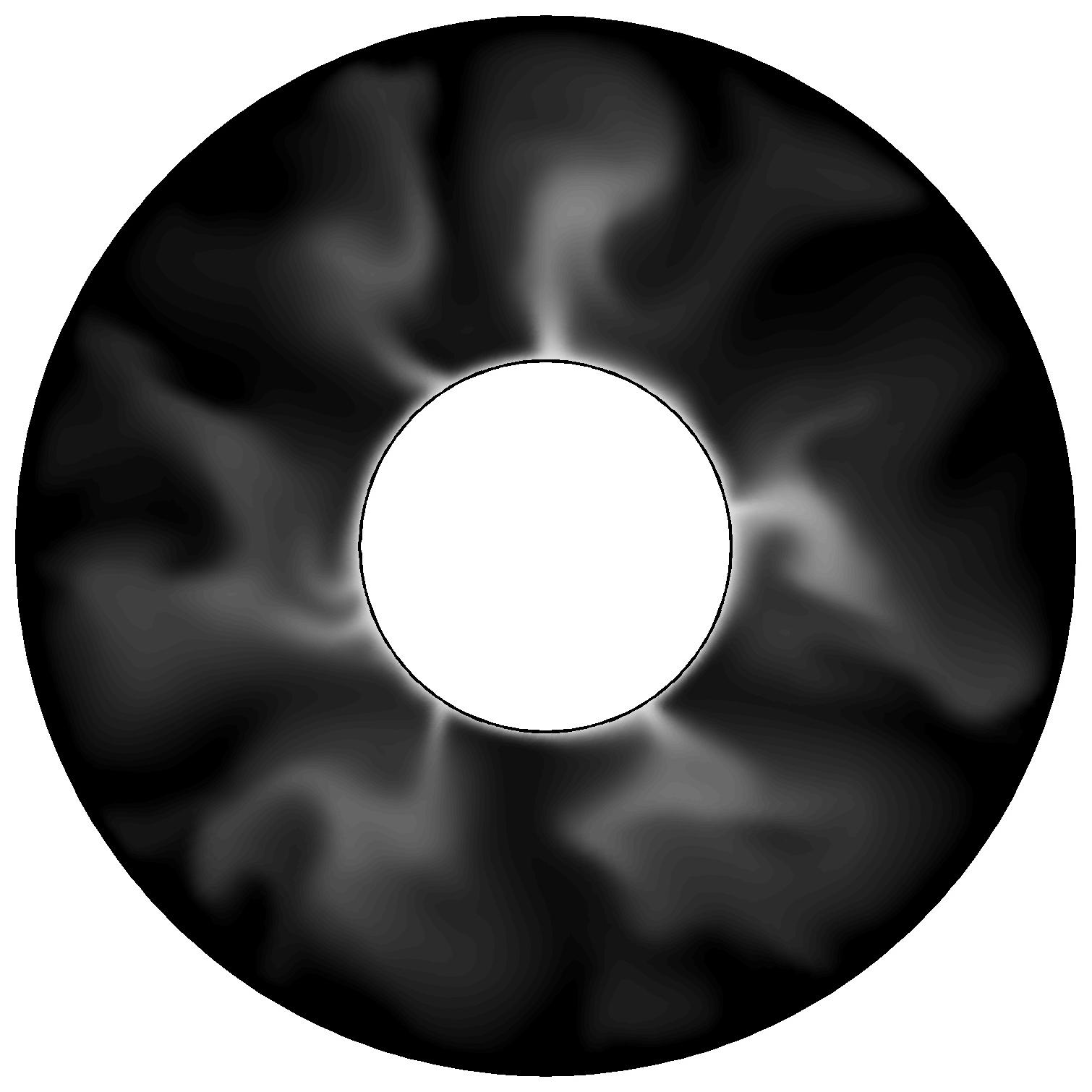}
   \\
   C'&
   \includegraphics[width=1.5in]{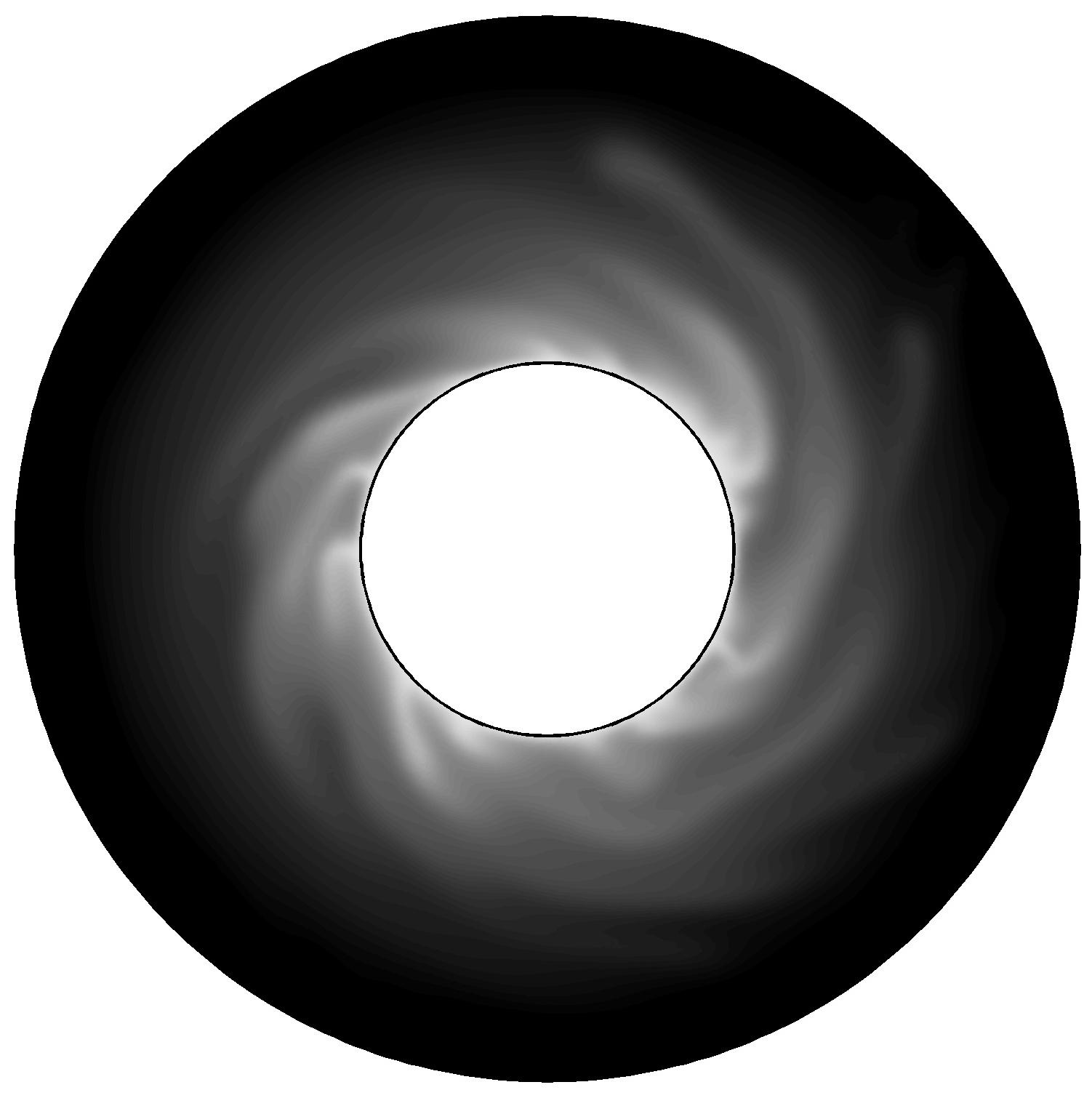}&
   \includegraphics[width=1.5in]{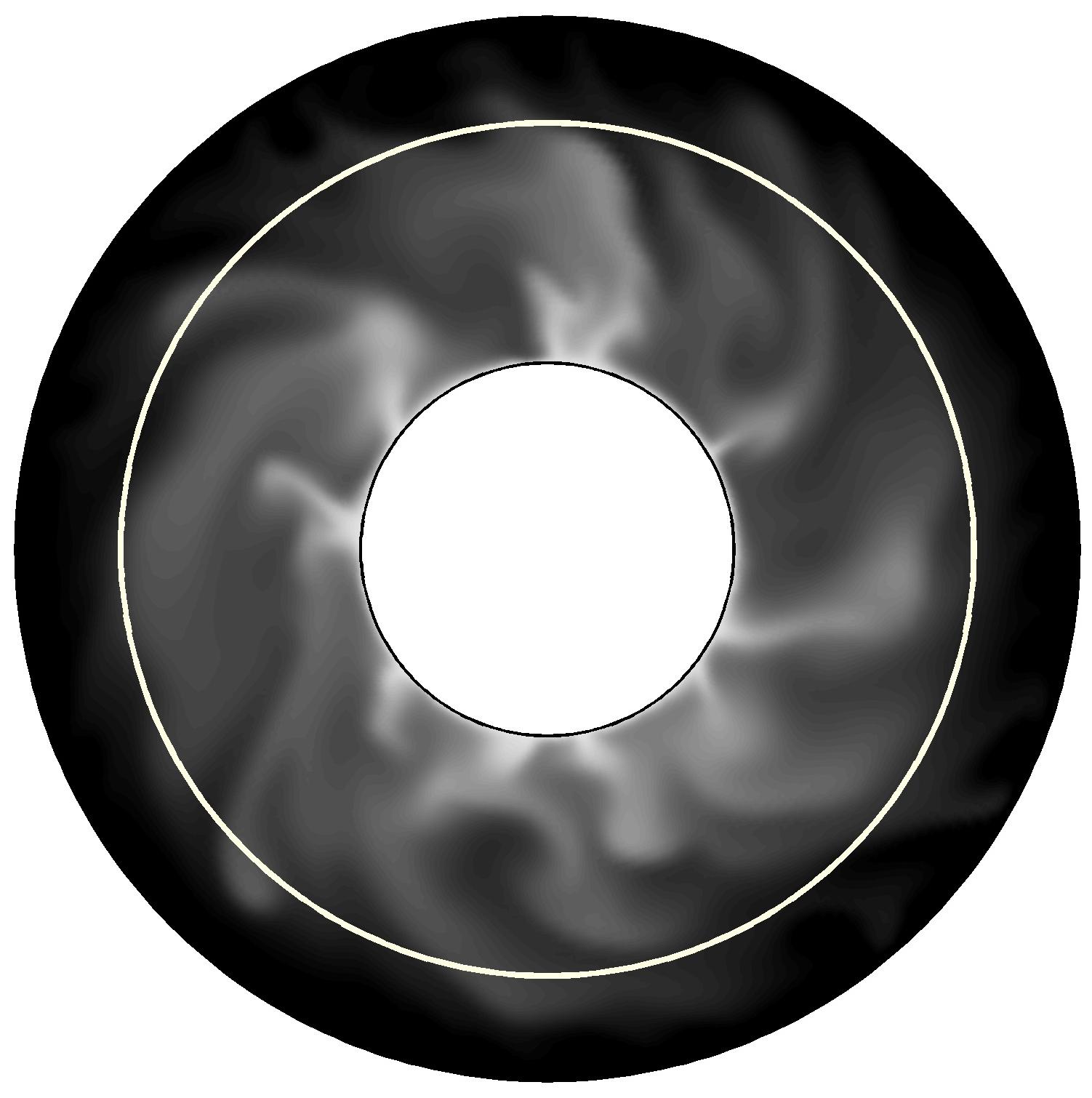}&
   \includegraphics[width=1.5in]{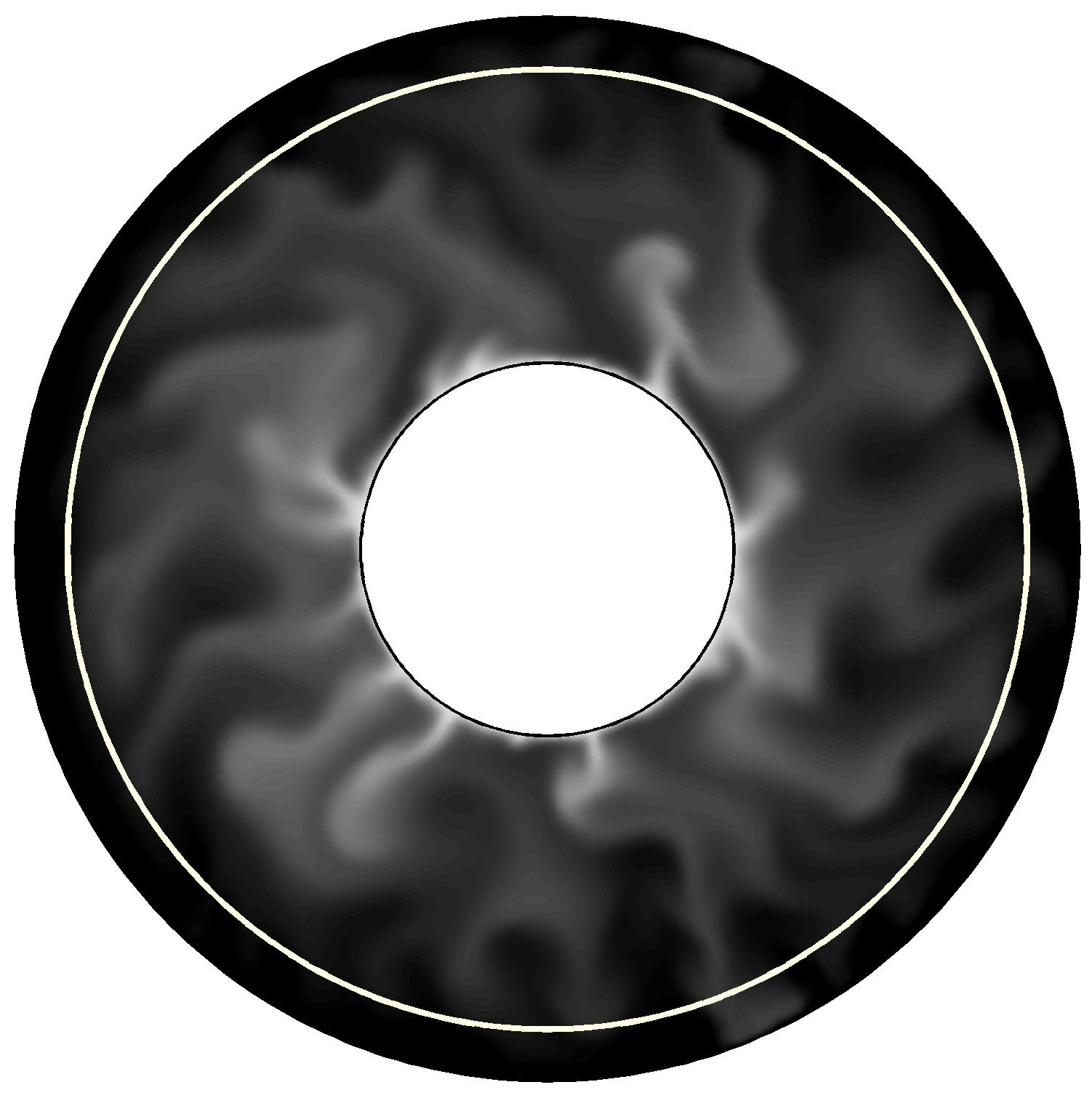}&
   \includegraphics[width=1.5in]{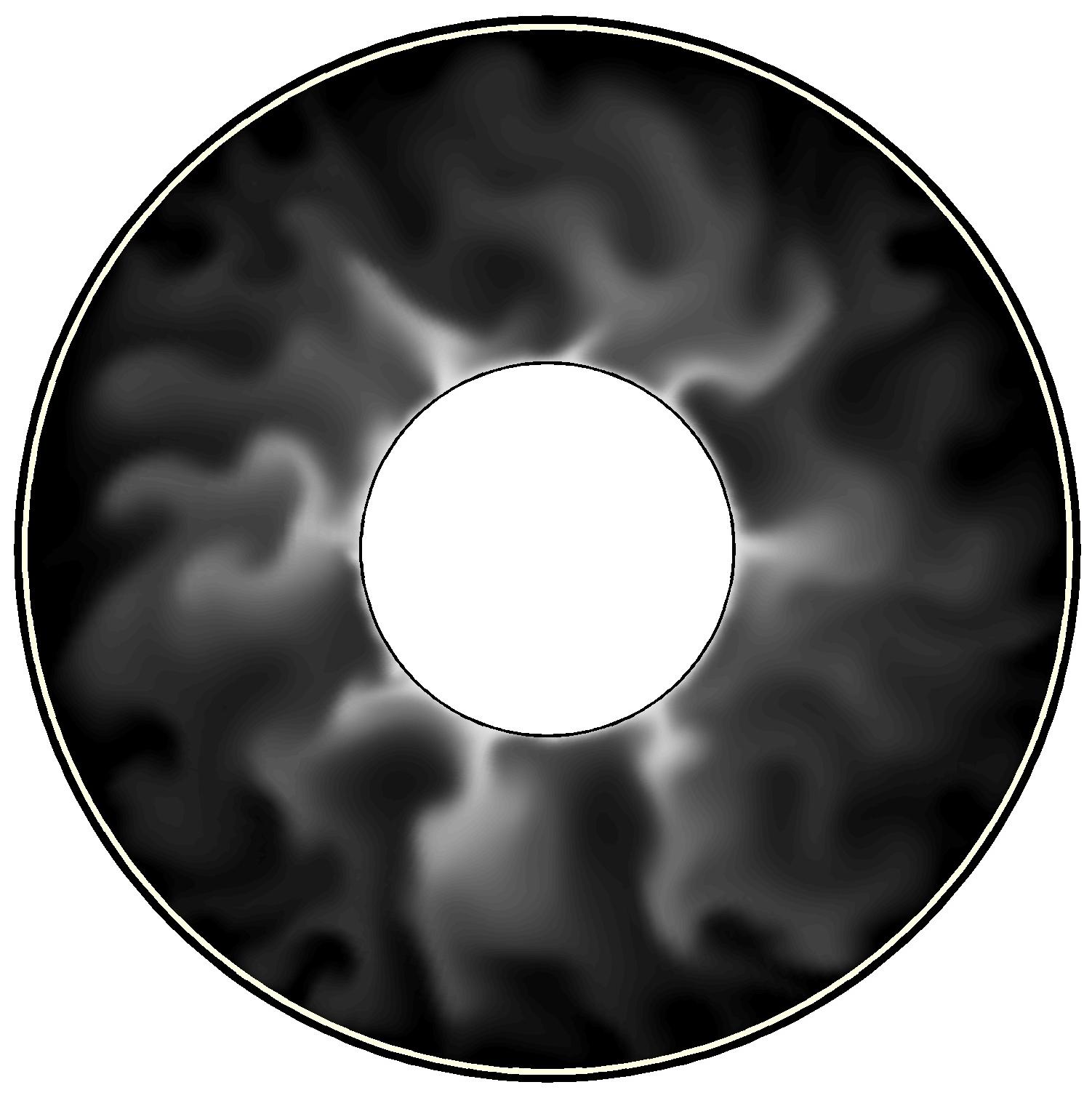}&
   \includegraphics[width=1.5in]{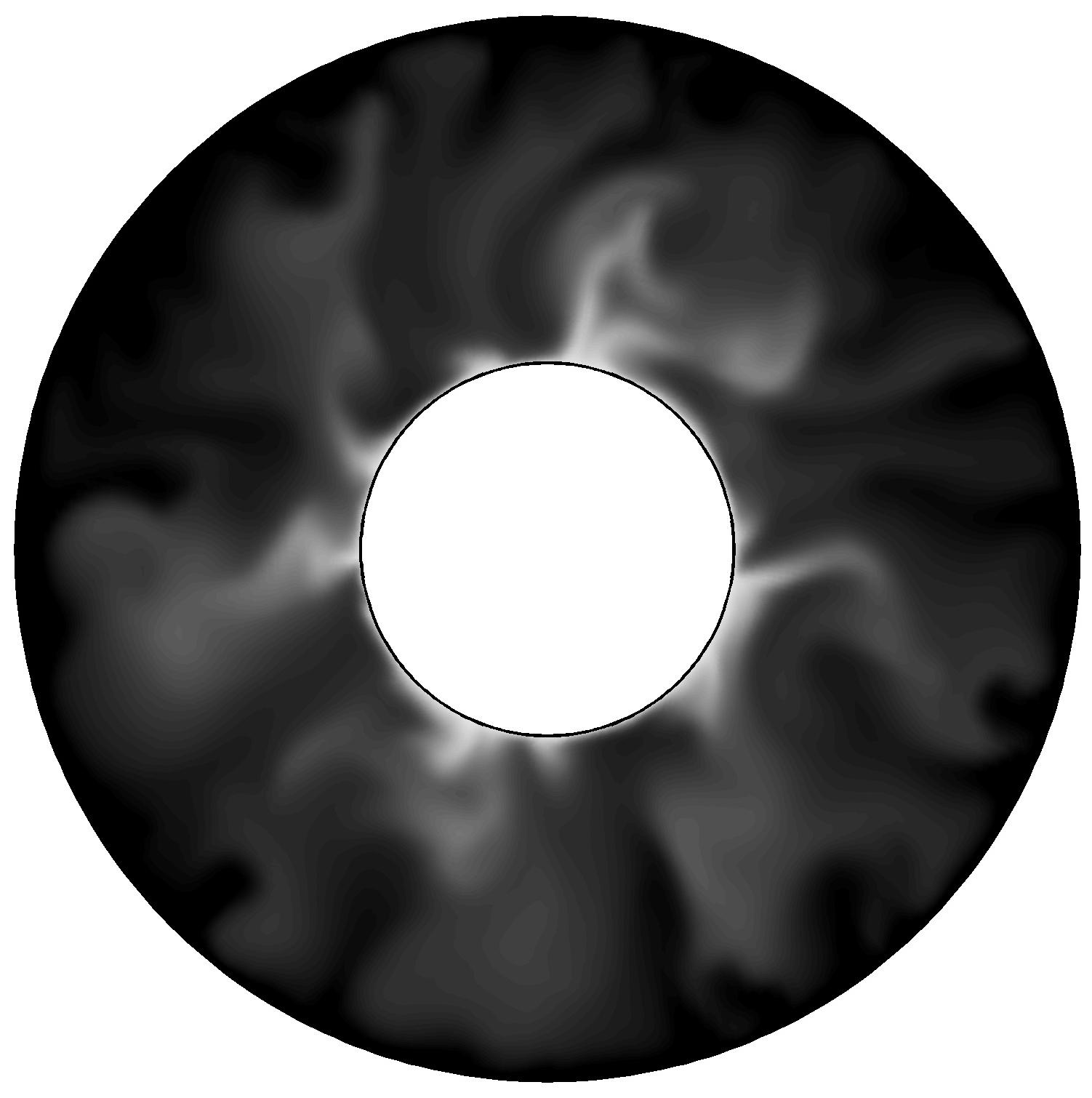}

   \end{tabular}
  \caption{Panels show randomly chosen snapshots of the
    equatorial temperature for the models  for  set A, B, C and C' 
    from top  to bottom rows. 
    From left to right: non-magnetic
    convective cases, variable conductivity models with $\chi_m=0.8$
    and $\chi_m=0.9$, $\chi_m=0.98$, and homogeneous non-zero electrical
    conductivity solutions. For the variable conductivity cases the
    circle of radius $r_m$ is shown with a white solid line.
    All panels show the normalized temperature from 0 (black) to 1 (white).}
   \label{fig:Temp_eq}
\end{sidewaysfigure}

In figure~\ref{fig:Temp_eq}  equatorial profiles of the temperature field are
shown.
A behaviour change is evident between the homogeneous non-magnetic and
magnetic solutions, with the variable conductivity solutions as
intermediate states.  For set A, columns one to three ($\chi_m=\emptyset,
0.8, 0.9$) exhibit a similar behaviour.  The presence of a mean zonal
flow bending the hot thermal plumes in the prograde direction is
visible, while for the fifth column ($\chi_m=\infty$) there is no
evidence for a strong axisymmetric zonal flow, and the plumes have a
dominant radial direction.  Column four ($\chi_m=0.98$) shows a weak
prograde tilt.
For set B, a transition is seen between $\chi_m=0.9$ and
$\chi_m=0.98$, although there is also a significant morphology
difference between the non-magnetic case and the variable conductivity
$\chi=0.8$ and $\chi_m=0.9$.
For the third row (set C) 
there is a difference between the 
finite $\chi_m$ runs and $\chi_m=\infty$ in 
the characteristic
azimuthal wave number of the plumes.  In $\chi_m=\infty$ of set C
there are six distinguishable hot plumes while for all others the wave
number is as high as nine.

\begin{sidewaysfigure}
   \centering
   \begin{tabular}{m{0.1in}|m{1.5in}|m{1.5in}|m{1.5in}|m{1.5in}|m{1.5in}}
    &\multicolumn{1}{c|}{$\chi_m=\emptyset$}  
    &\multicolumn{1}{c|}{ $\chi_m=0.8$} 
    &\multicolumn{1}{c|}{ $\chi_m=0.9$ }
    &\multicolumn{1}{c|}{ $\chi_m=0.98$} 
    &\multicolumn{1}{c}{$\chi_m=\infty$} \\

     A&
   \includegraphics[width=1.5in]{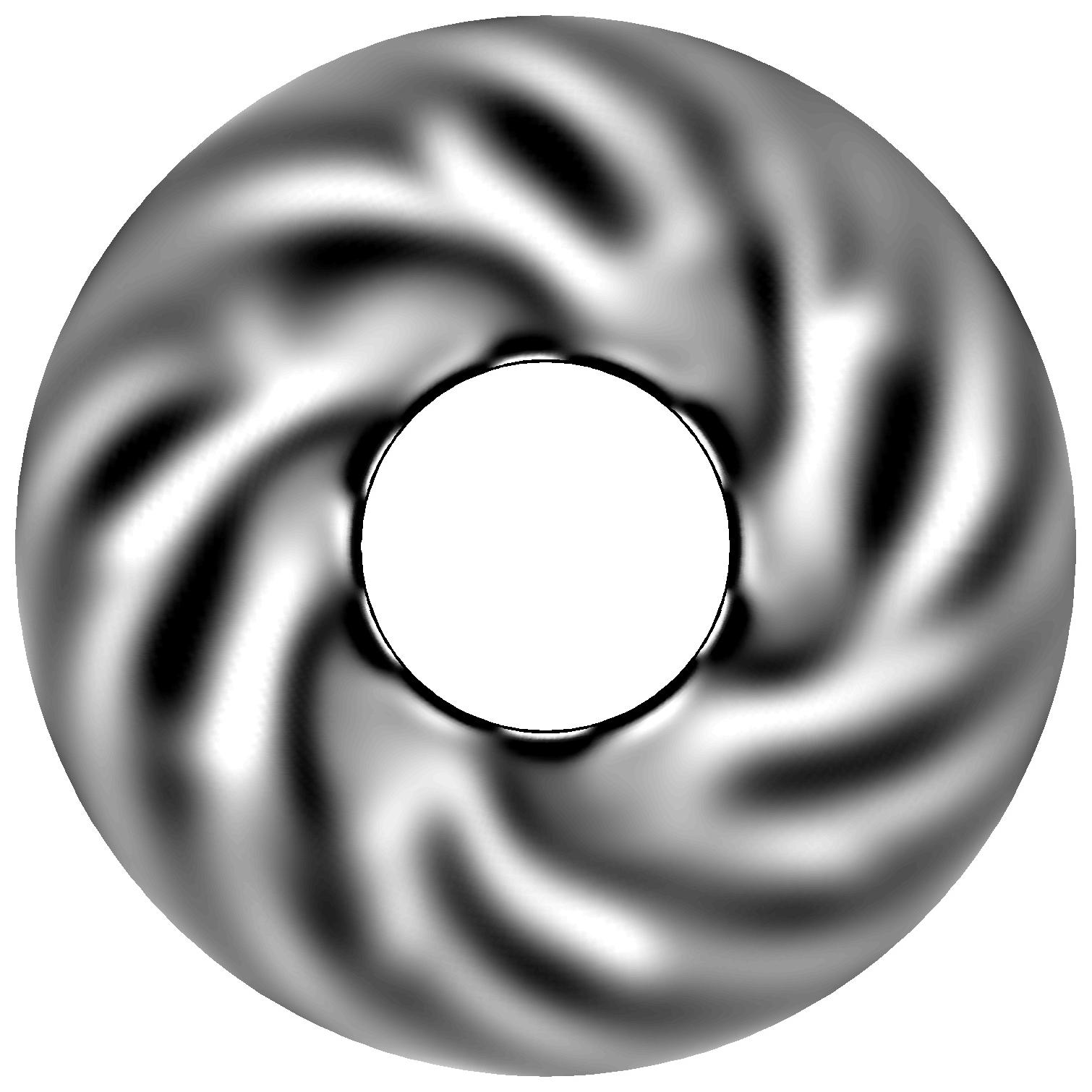}&
   \includegraphics[width=1.5in]{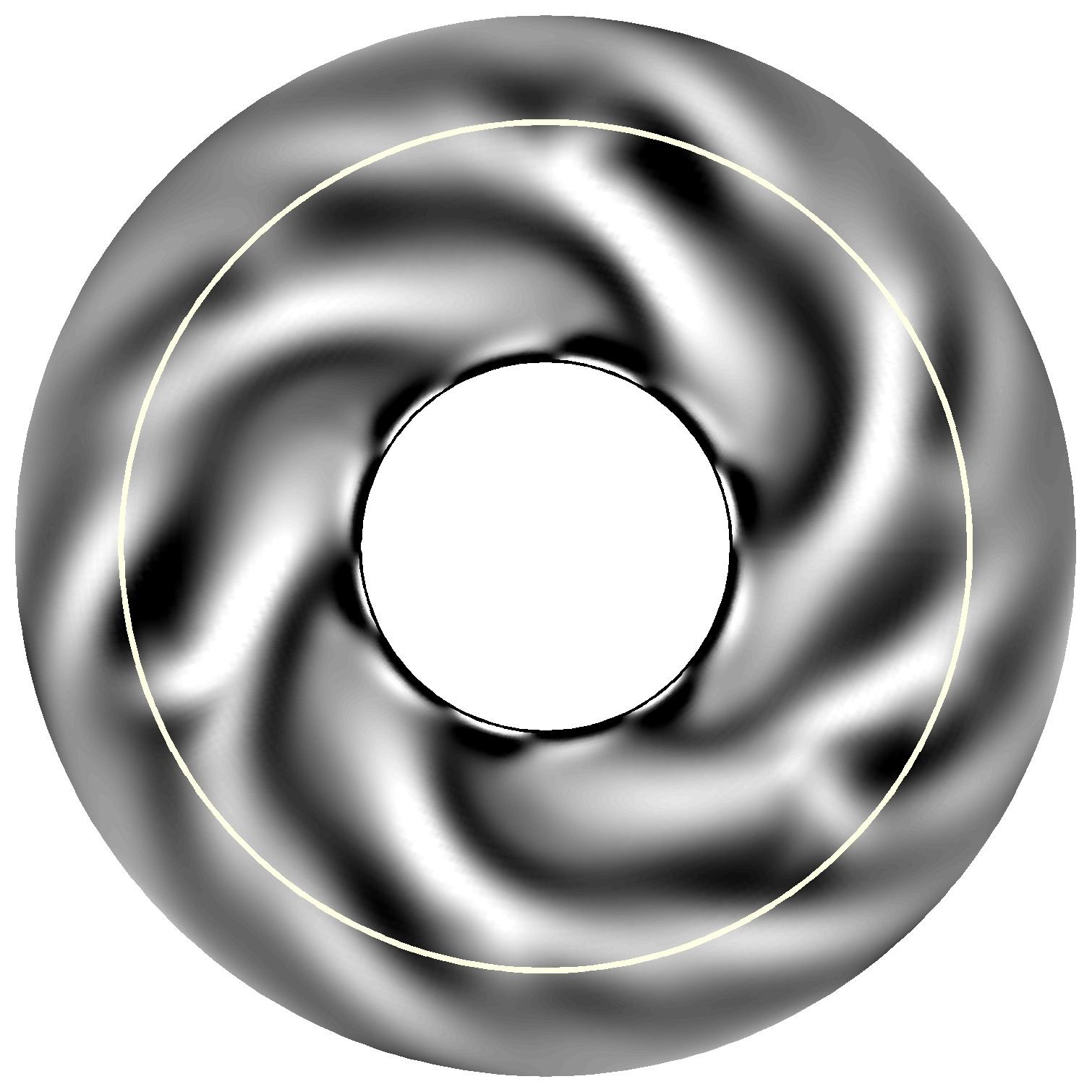}&
   \includegraphics[width=1.5in]{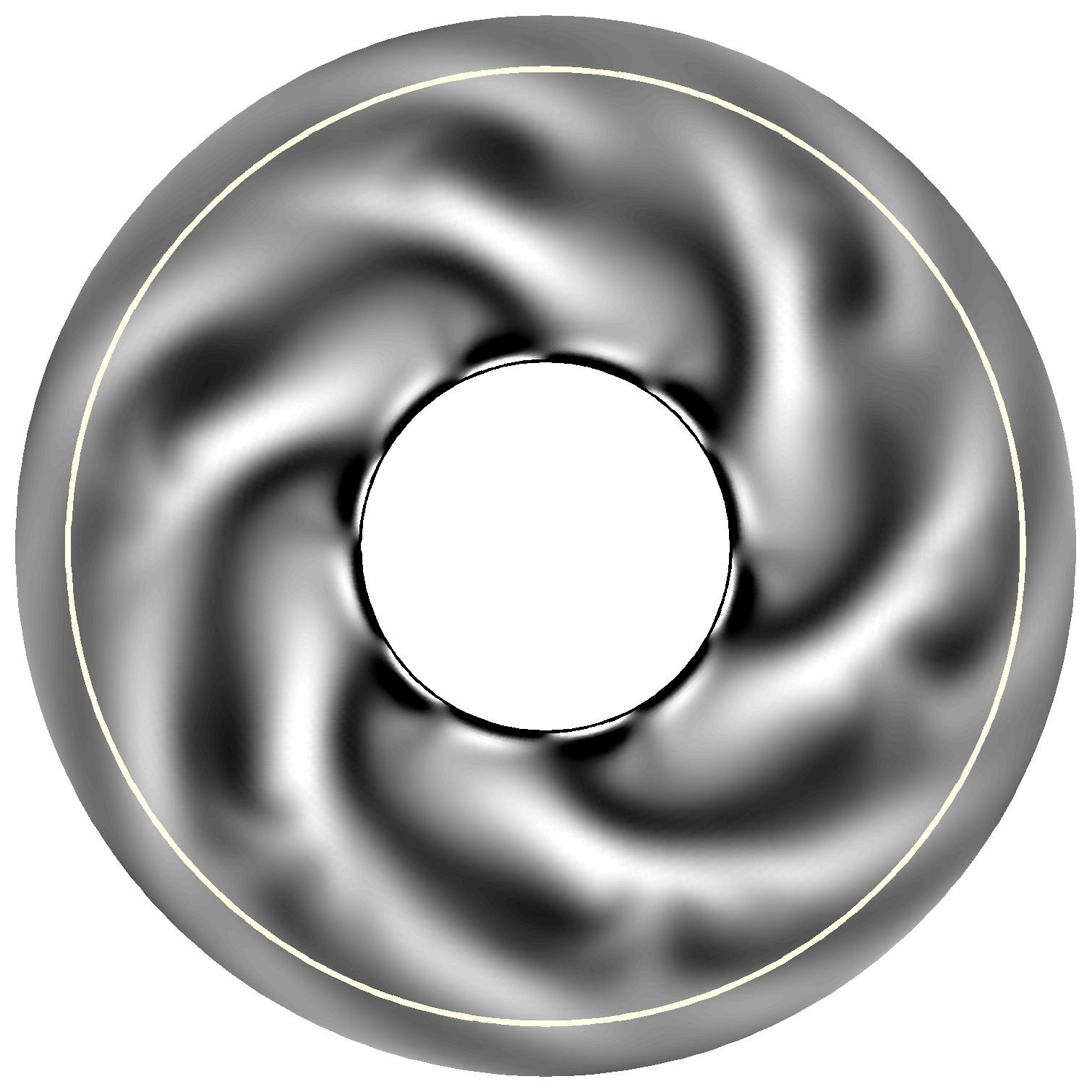}&
   \includegraphics[width=1.5in]{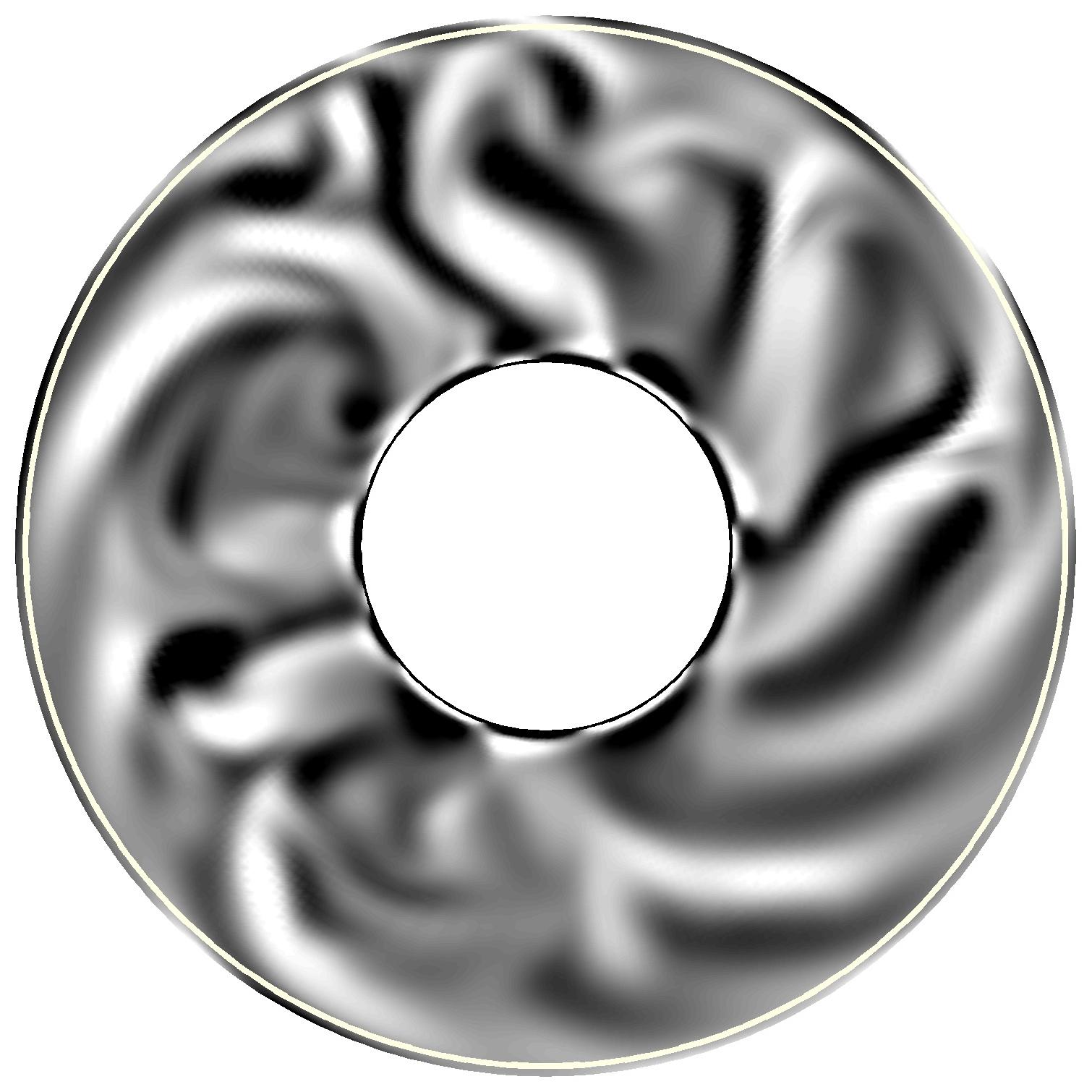}&
   \includegraphics[width=1.5in]{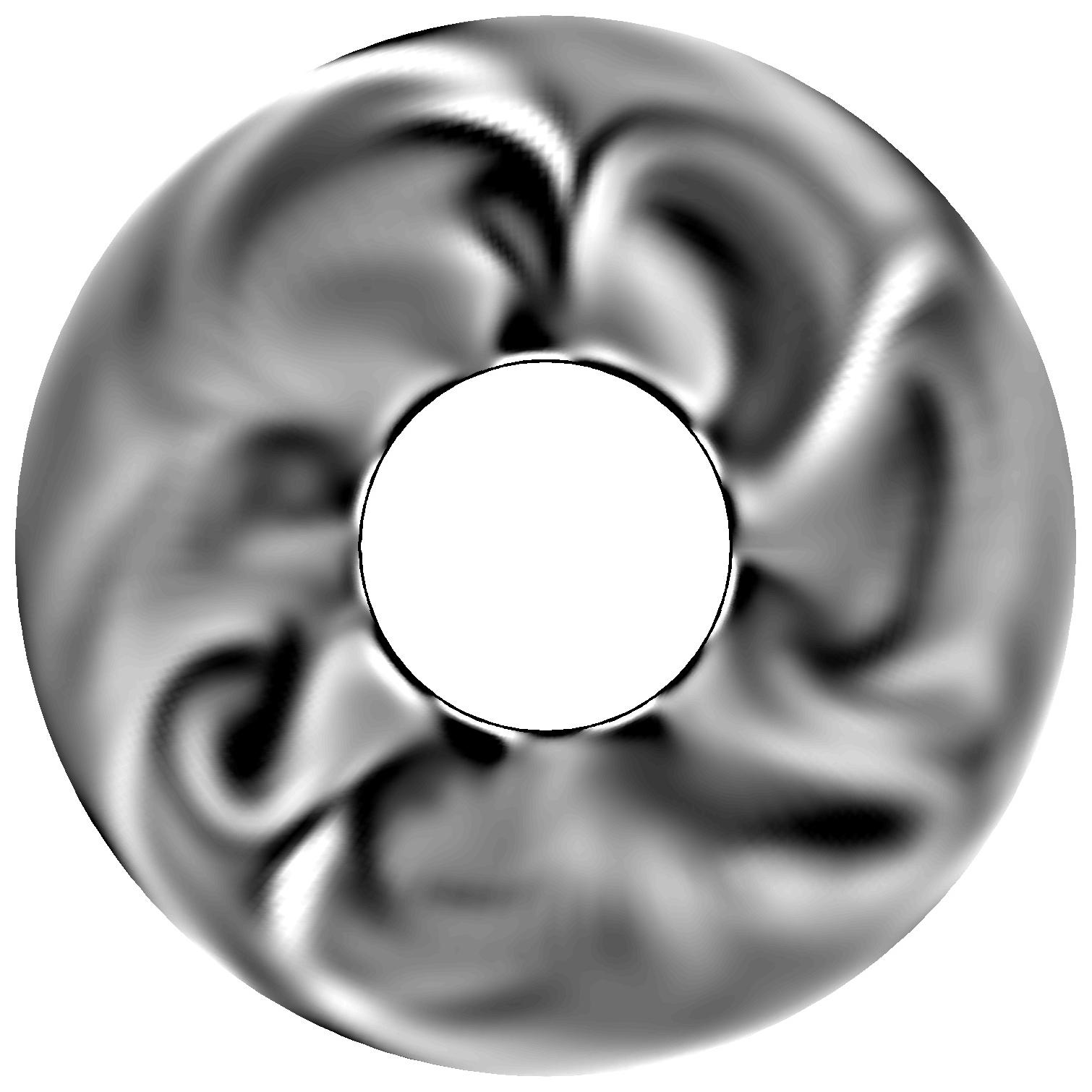}\\

   B&
   \includegraphics[width=1.5in]{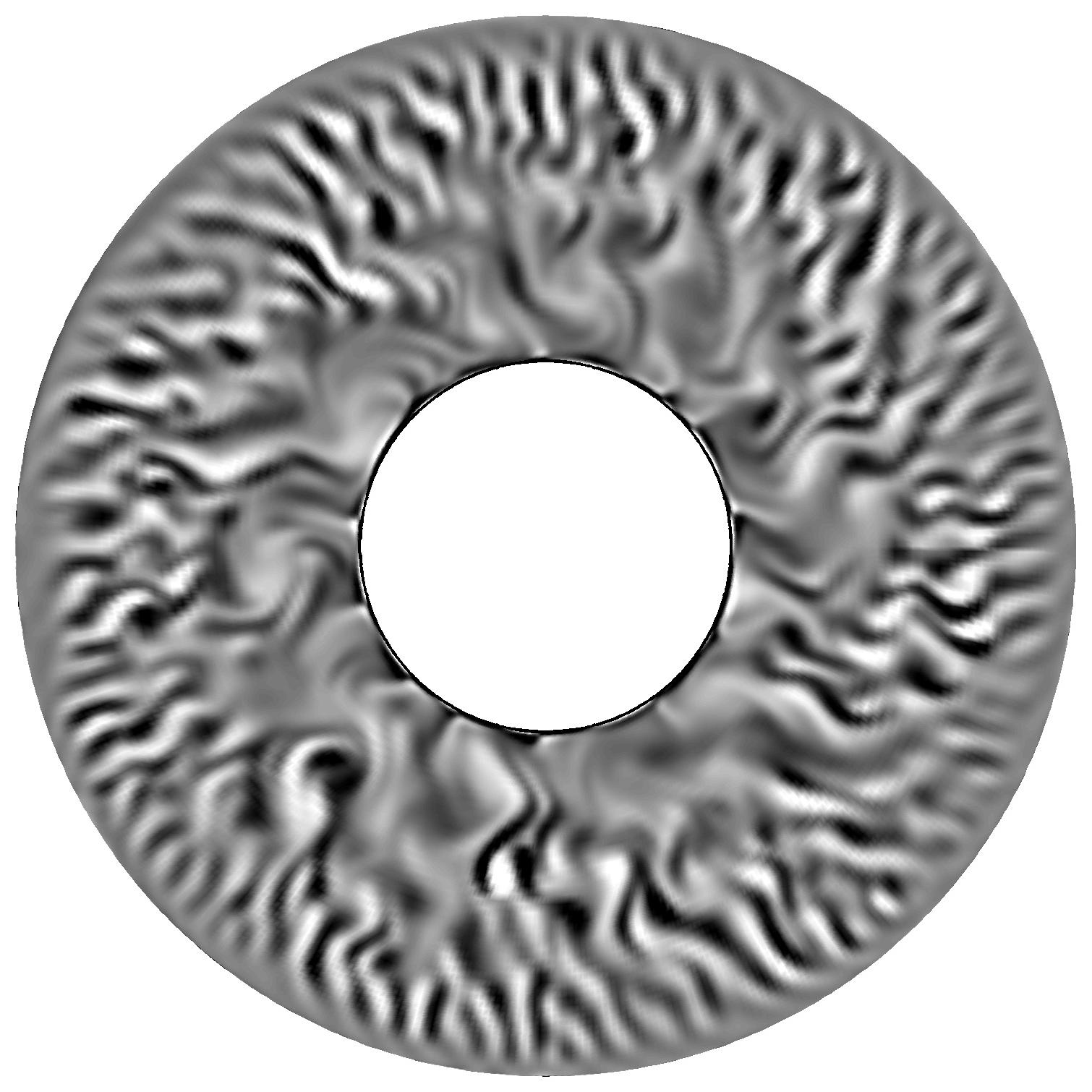}&
   \includegraphics[width=1.5in]{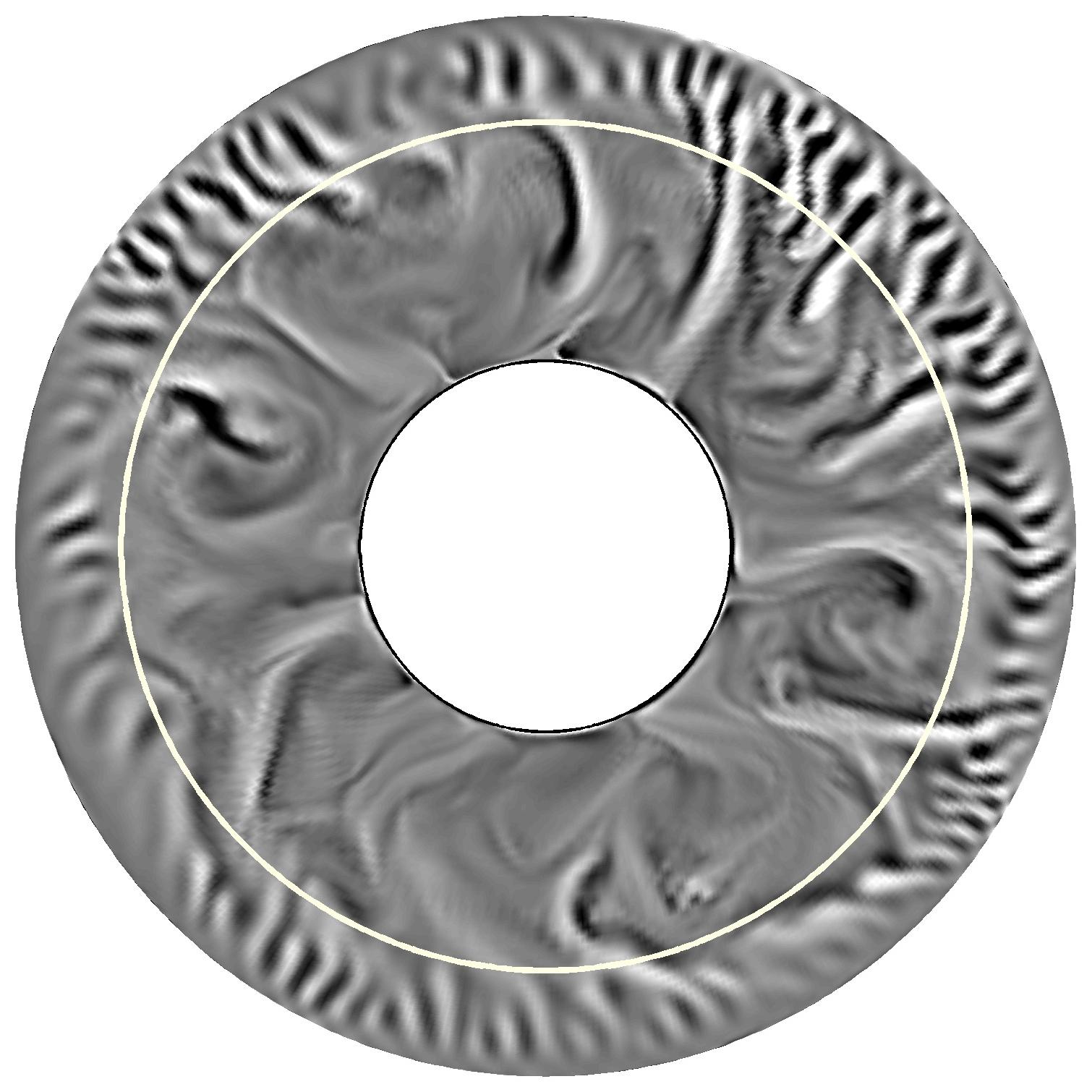}&
   \includegraphics[width=1.5in]{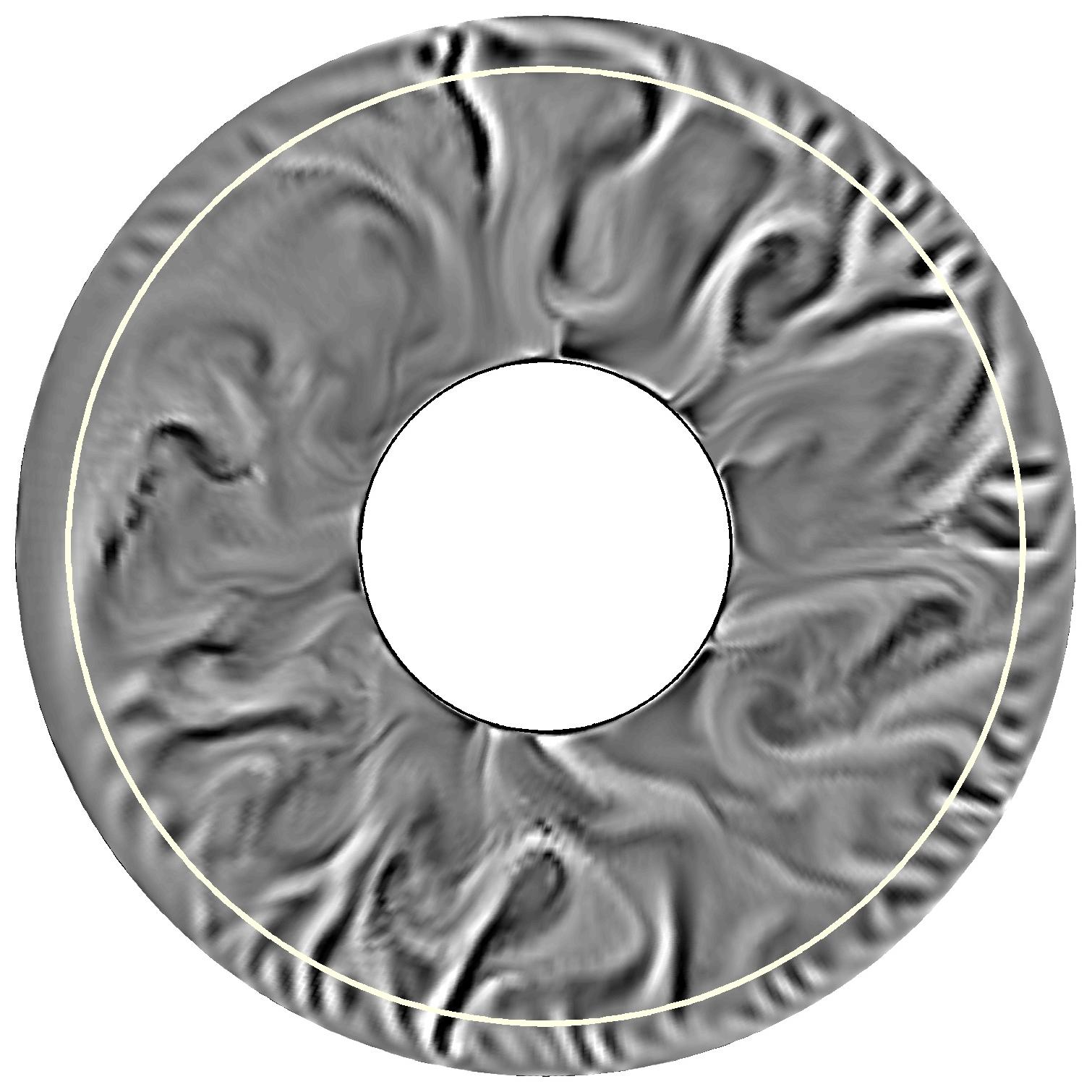}&
   \includegraphics[width=1.5in]{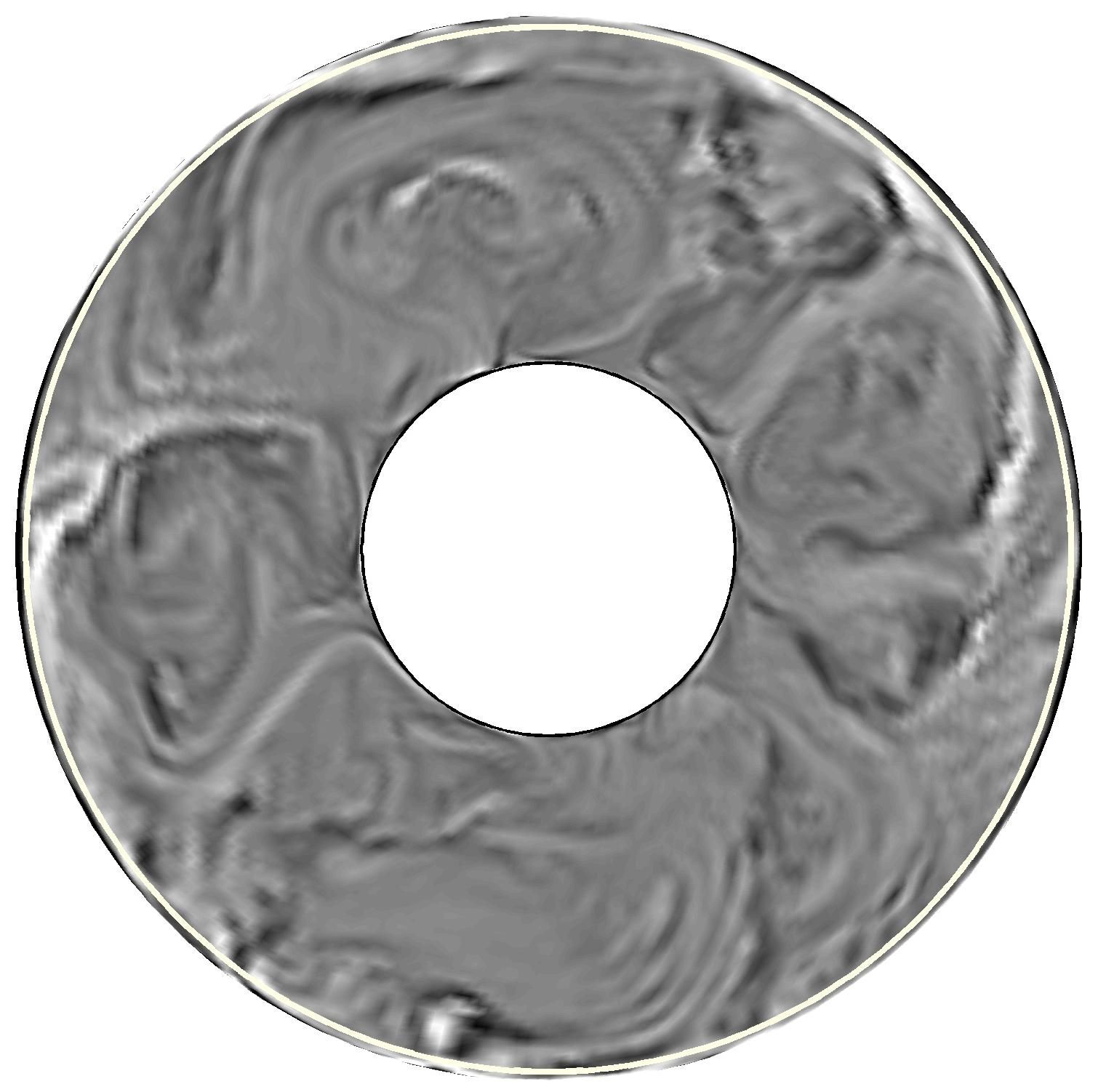}&
   \includegraphics[width=1.5in]{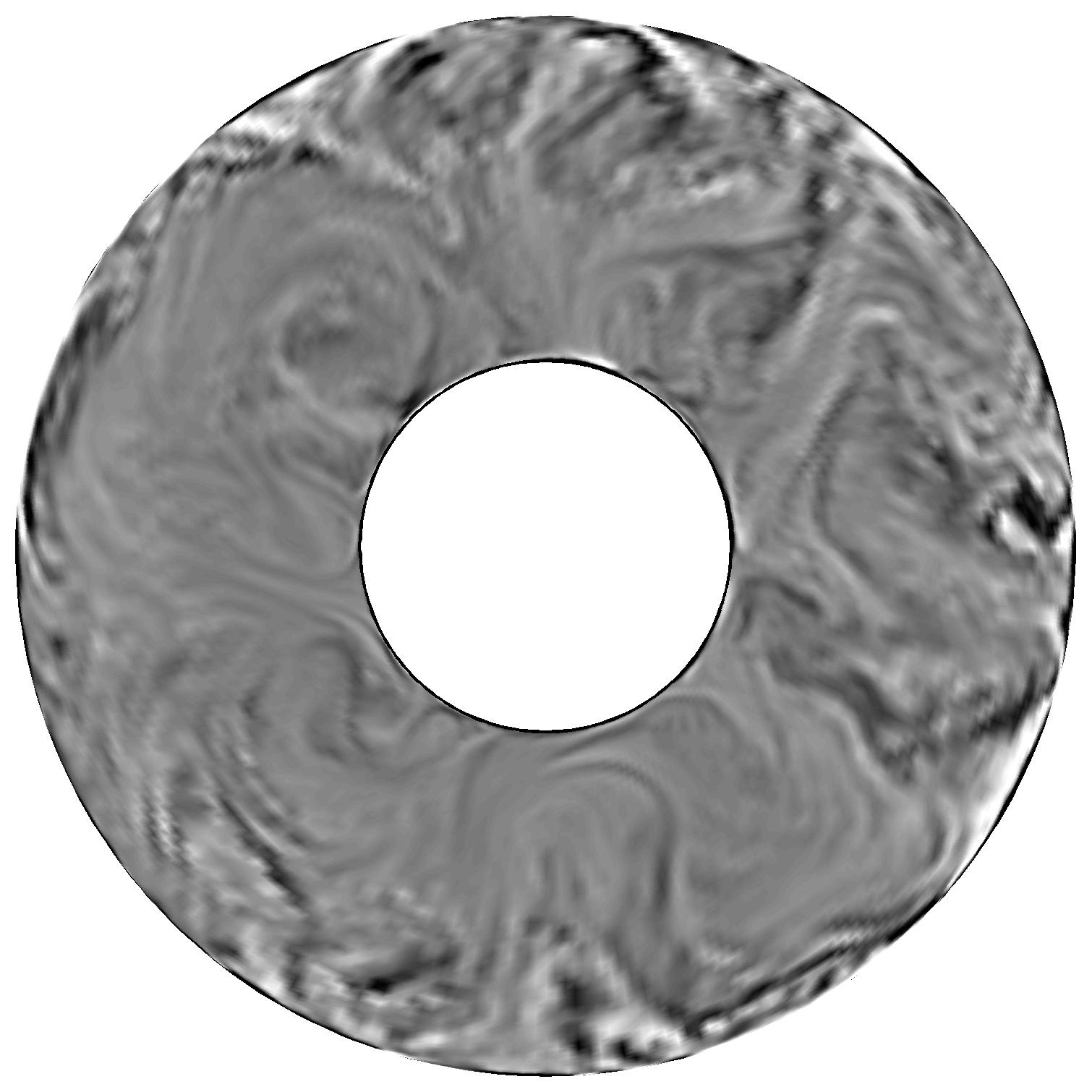}
   \\

   C&
   \includegraphics[width=1.5in]{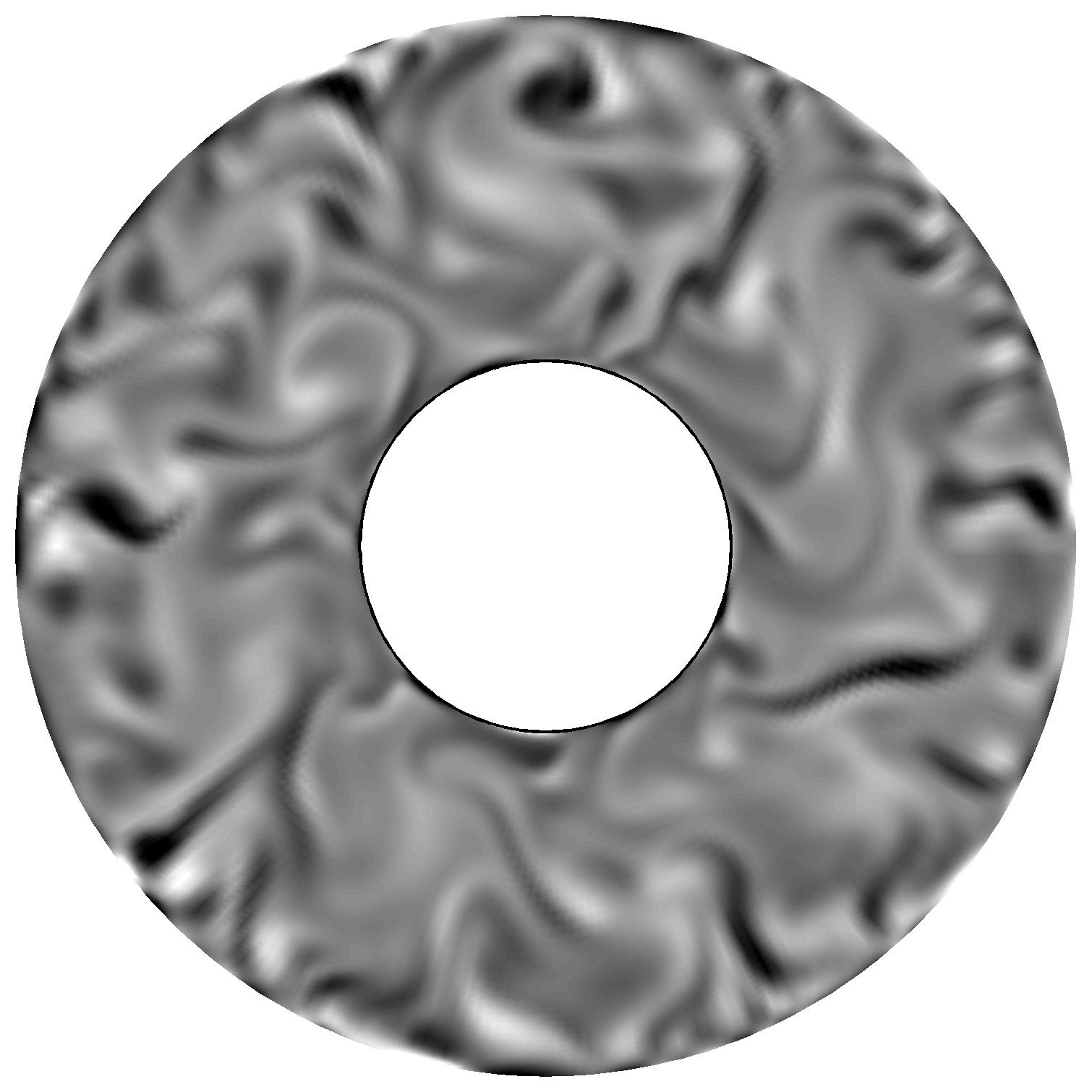}&
   \includegraphics[width=1.5in]{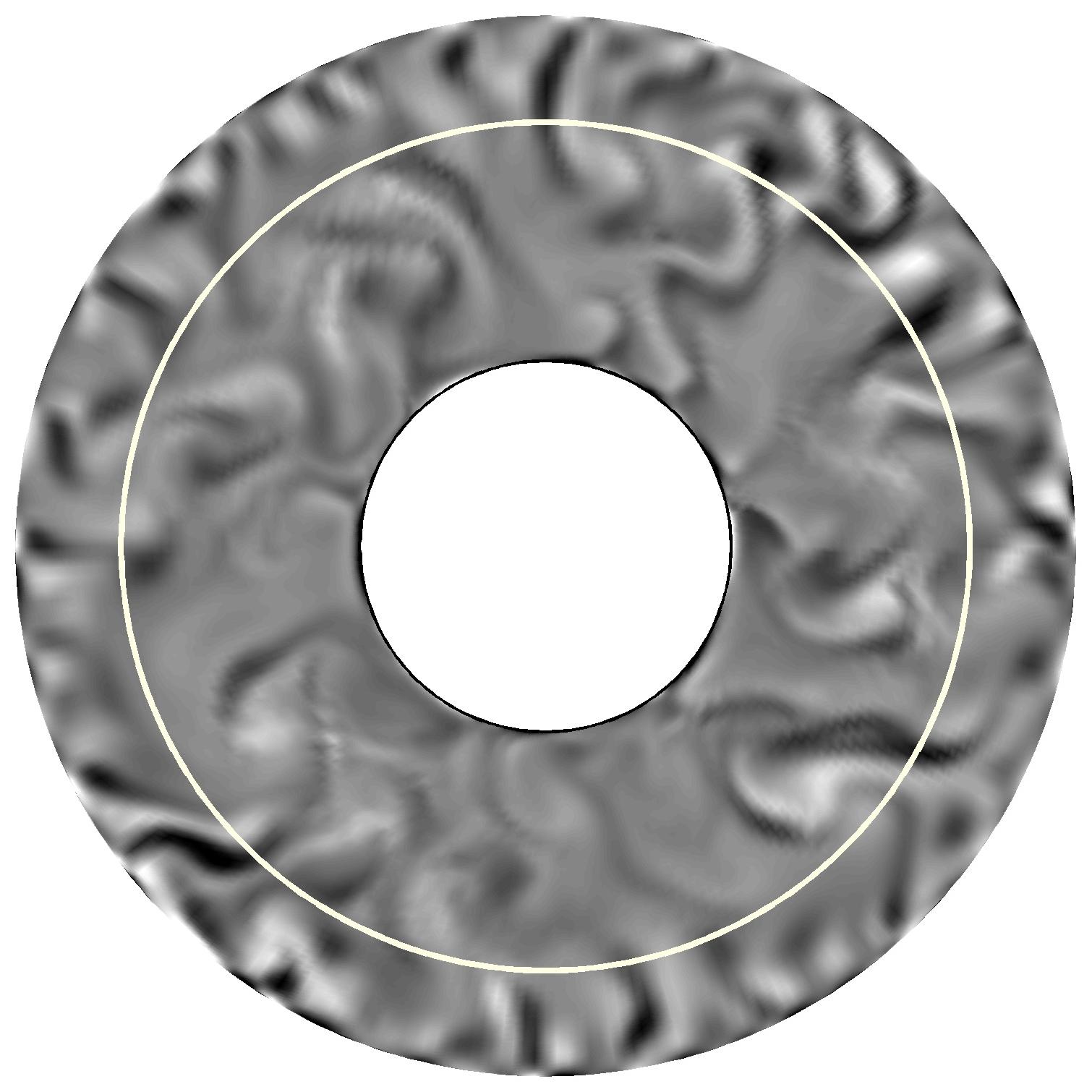}&
   \includegraphics[width=1.5in]{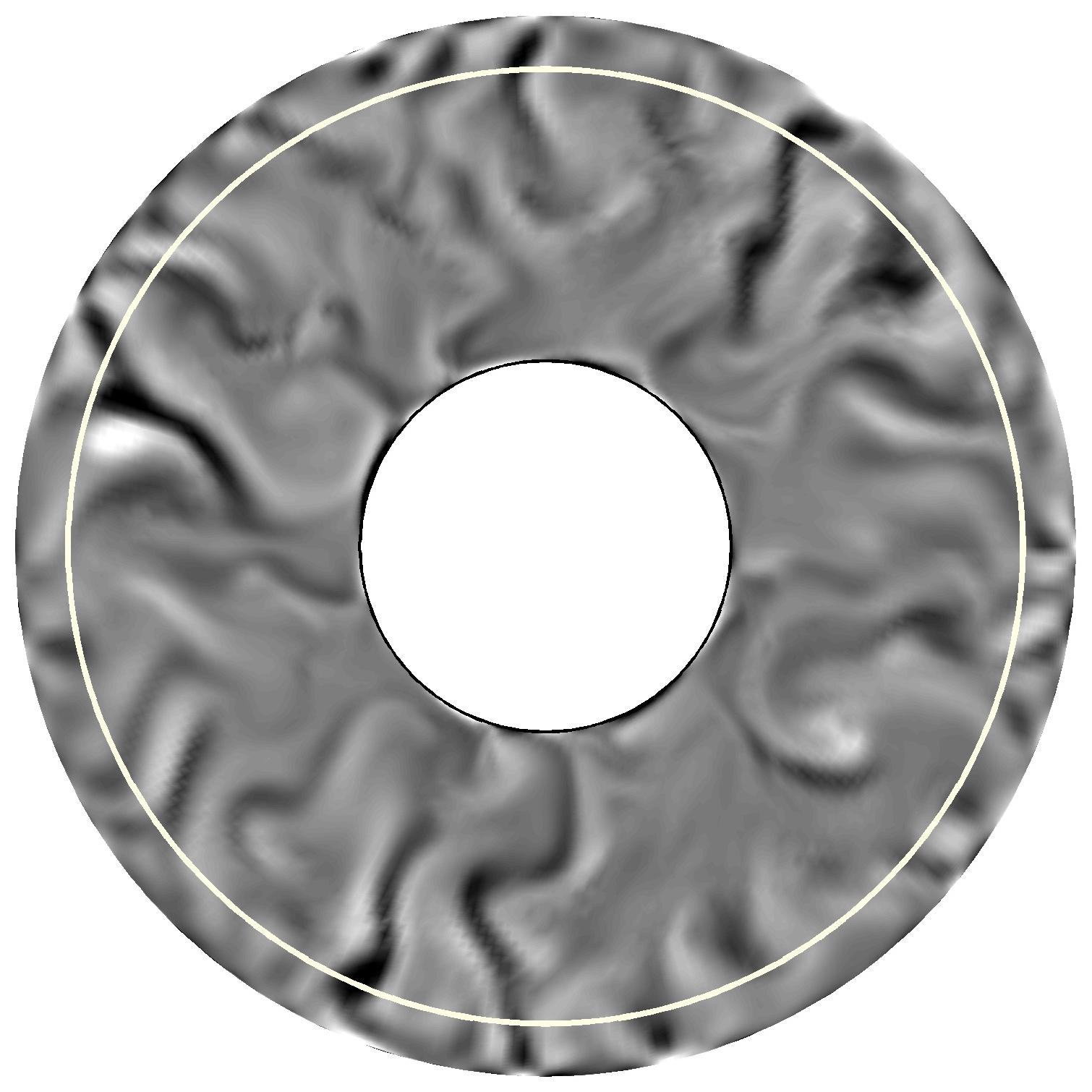}&
   \includegraphics[width=1.5in]{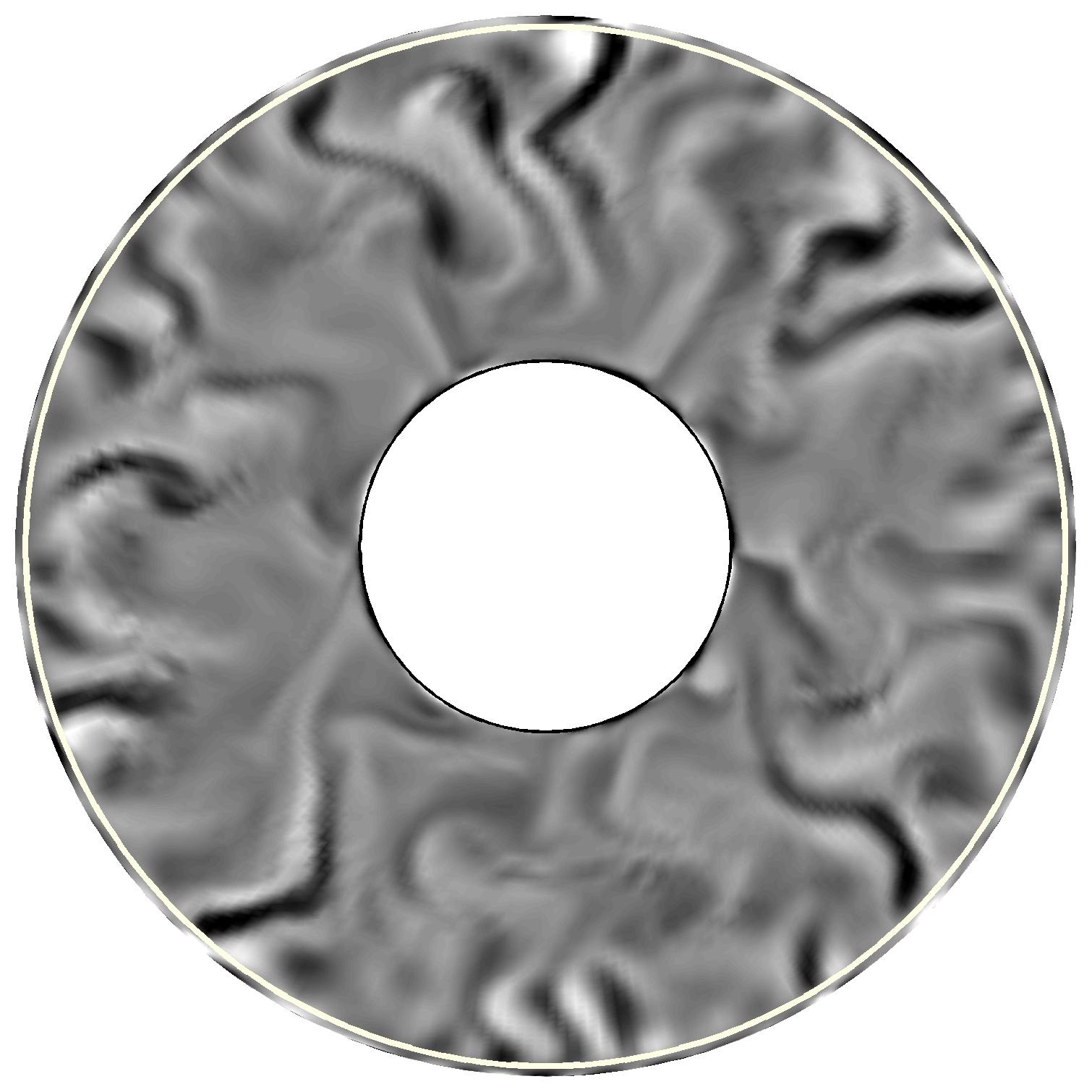}&
   \includegraphics[width=1.5in]{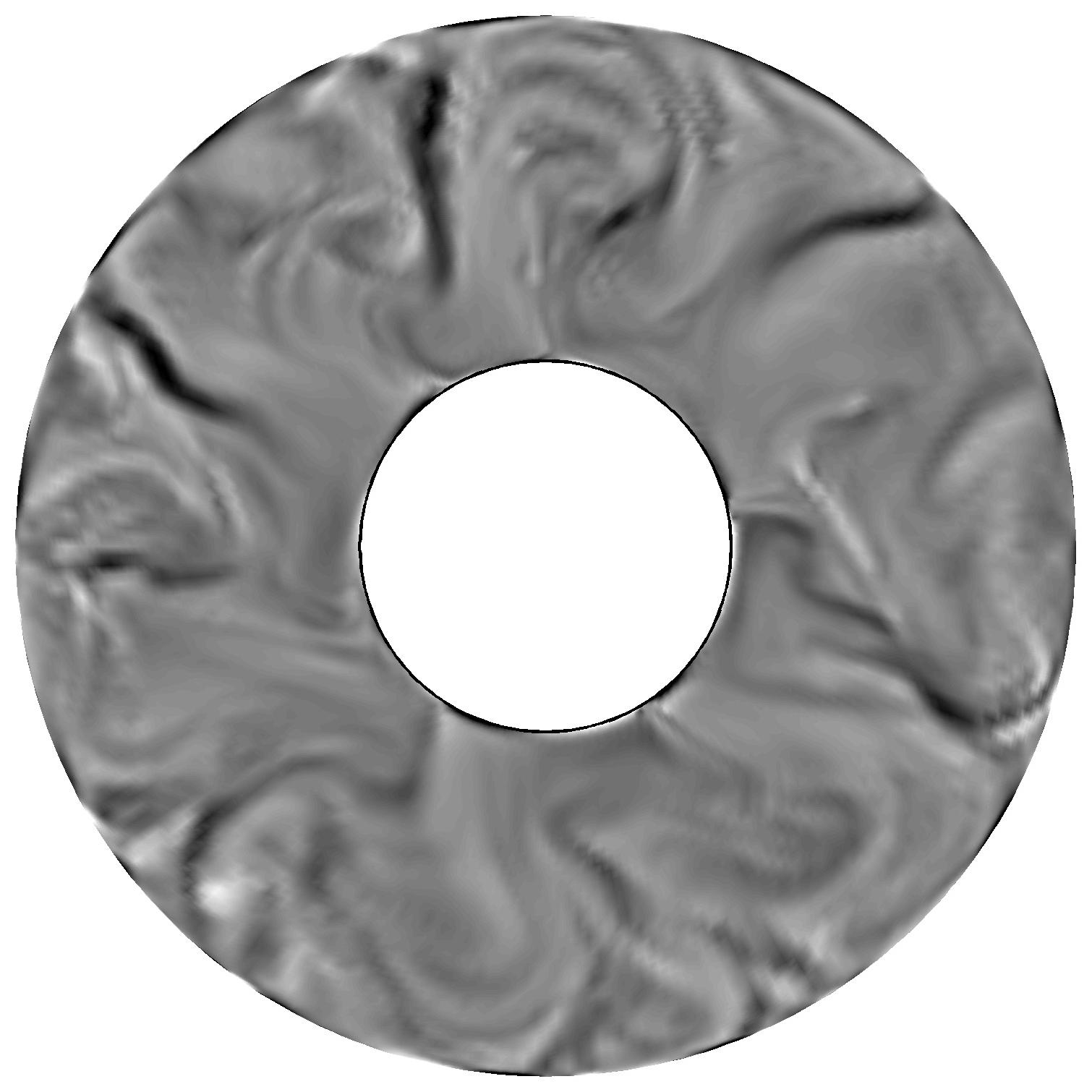}
      \\
   C'&
   \includegraphics[width=1.5in]{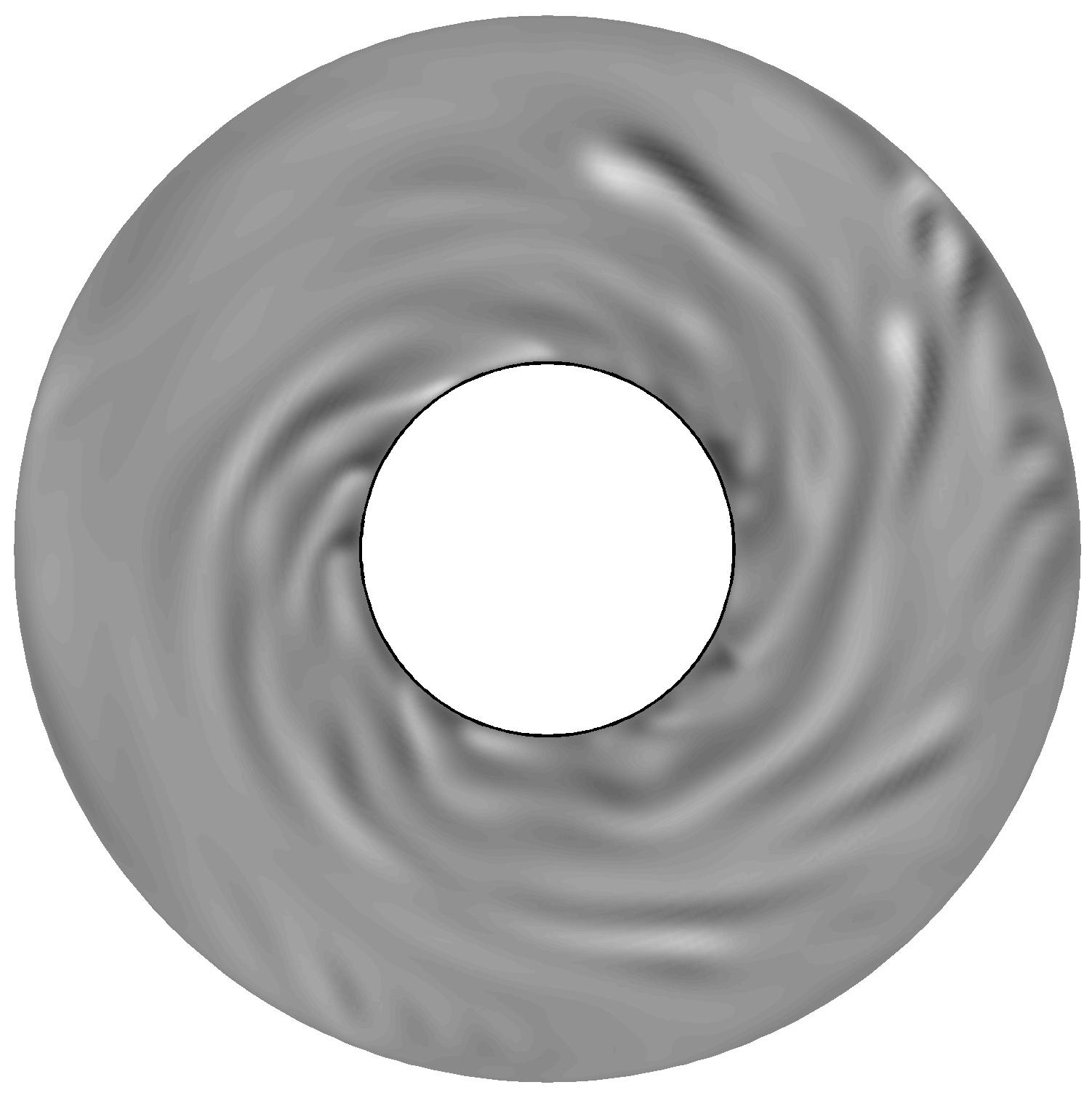}&
   \includegraphics[width=1.5in]{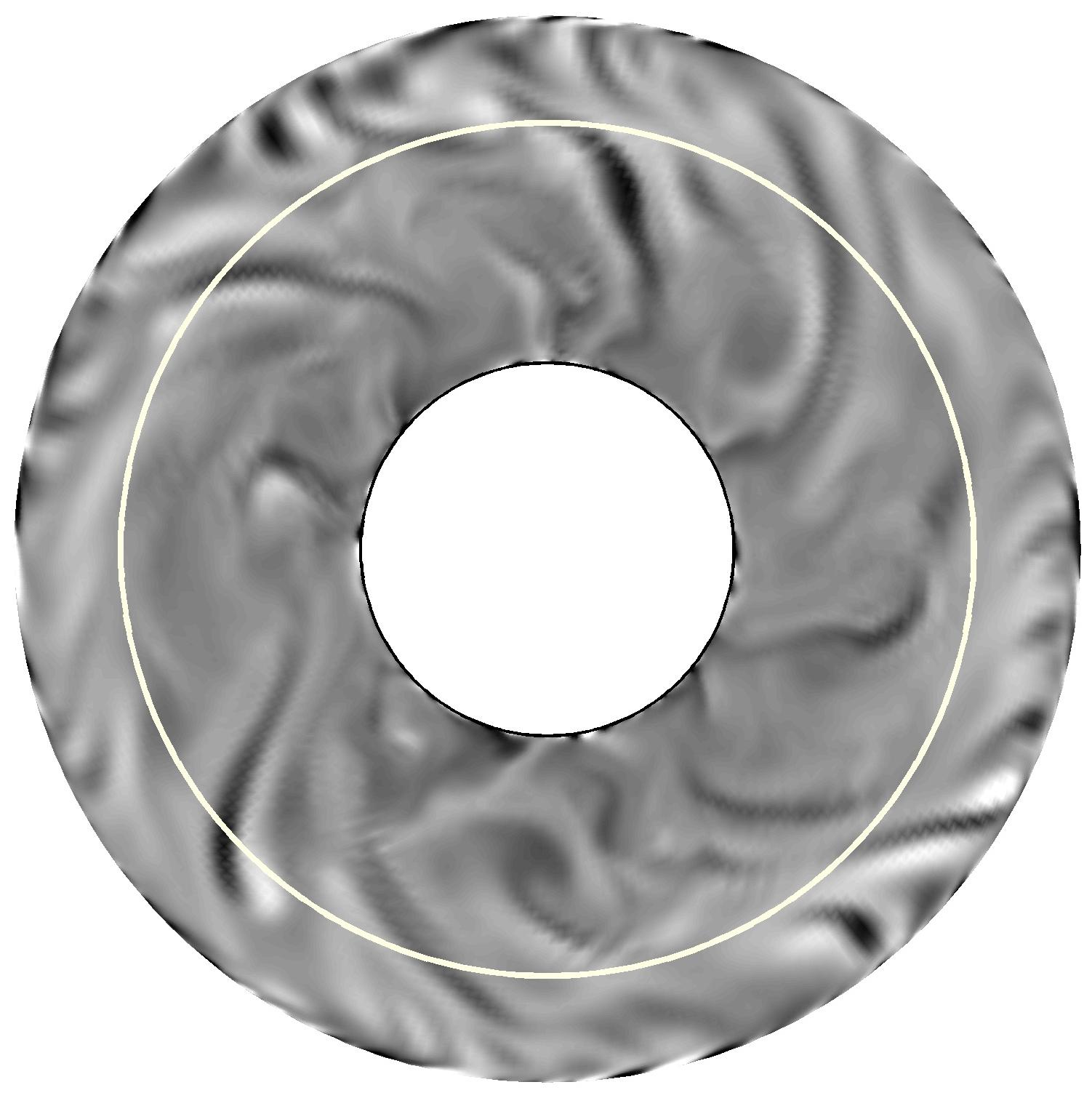}&
   \includegraphics[width=1.5in]{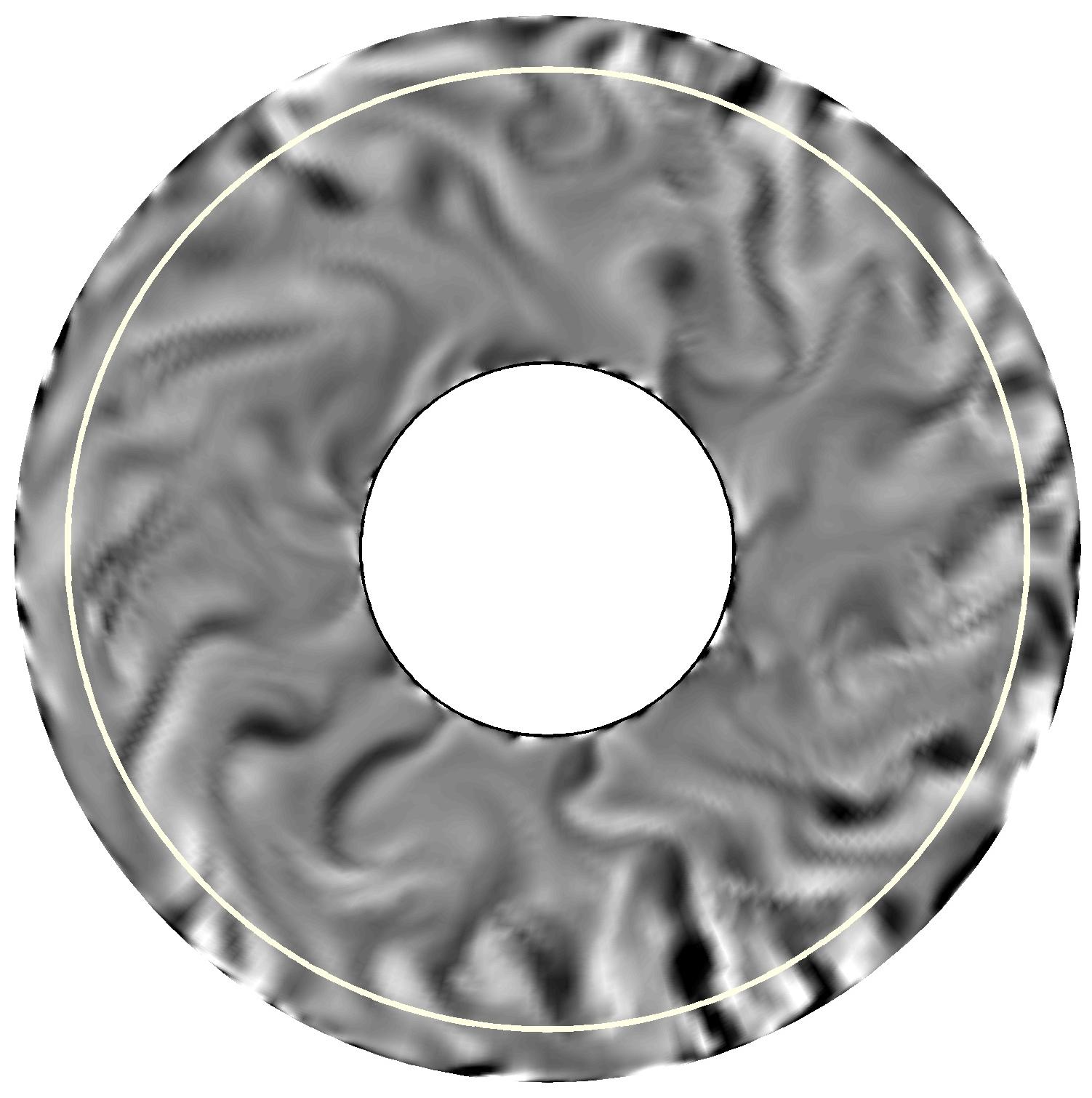}&
   \includegraphics[width=1.5in]{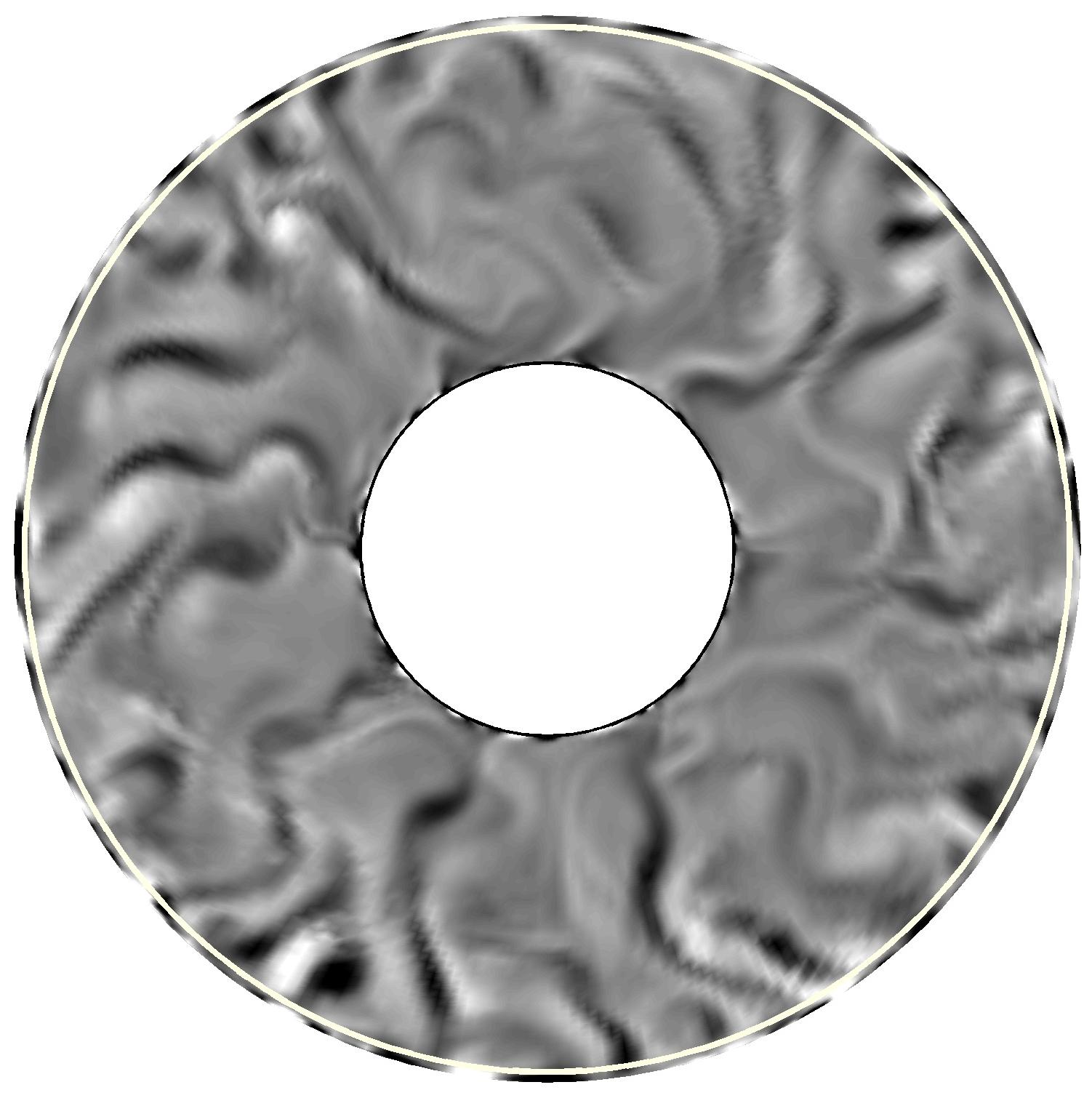}&
   \includegraphics[width=1.5in]{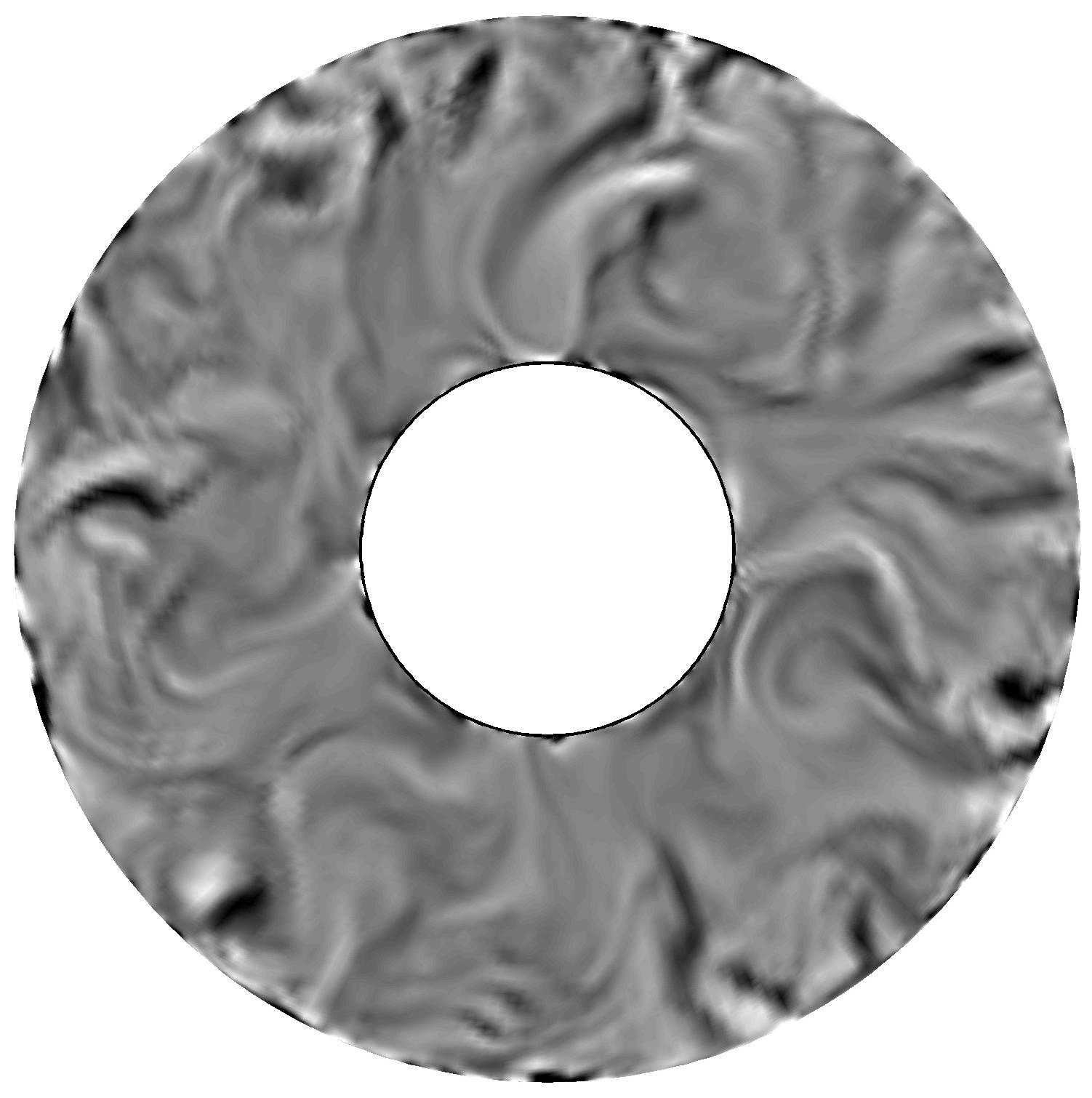}
   
   \end{tabular}
  \caption{ For the same snapshots as in figure~\ref{fig:Temp_eq},
    equatorial cuts of the axial vorticity,
    $\omega_z=\mathbf{z}\cdot(\nabla\times\mathbf{u})$.
    The colors represent the same values for
    each row but it is scaled between different sets.  The circle with
    $r=r_m$ is drawn in white for each panel of the non-homogeneous
    electrical conductivity.  }
   \label{fig:Vort_eq}

\end{sidewaysfigure}

\begin{sidewaysfigure}
   \centering
   \begin{tabular}{m{0.1in}|m{1.5in}|m{1.5in}|m{1.5in}|m{1.5in}|m{1.5in}}
    &\multicolumn{1}{c|}{ $\chi_m=0.8$} 
    &\multicolumn{1}{c|}{ $\chi_m=0.9$ }
    &\multicolumn{1}{c|}{ $\chi_m=0.98$} 
    &\multicolumn{1}{c}{$\chi_m=\infty$} \\

     A&
   \includegraphics[width=1.5in]{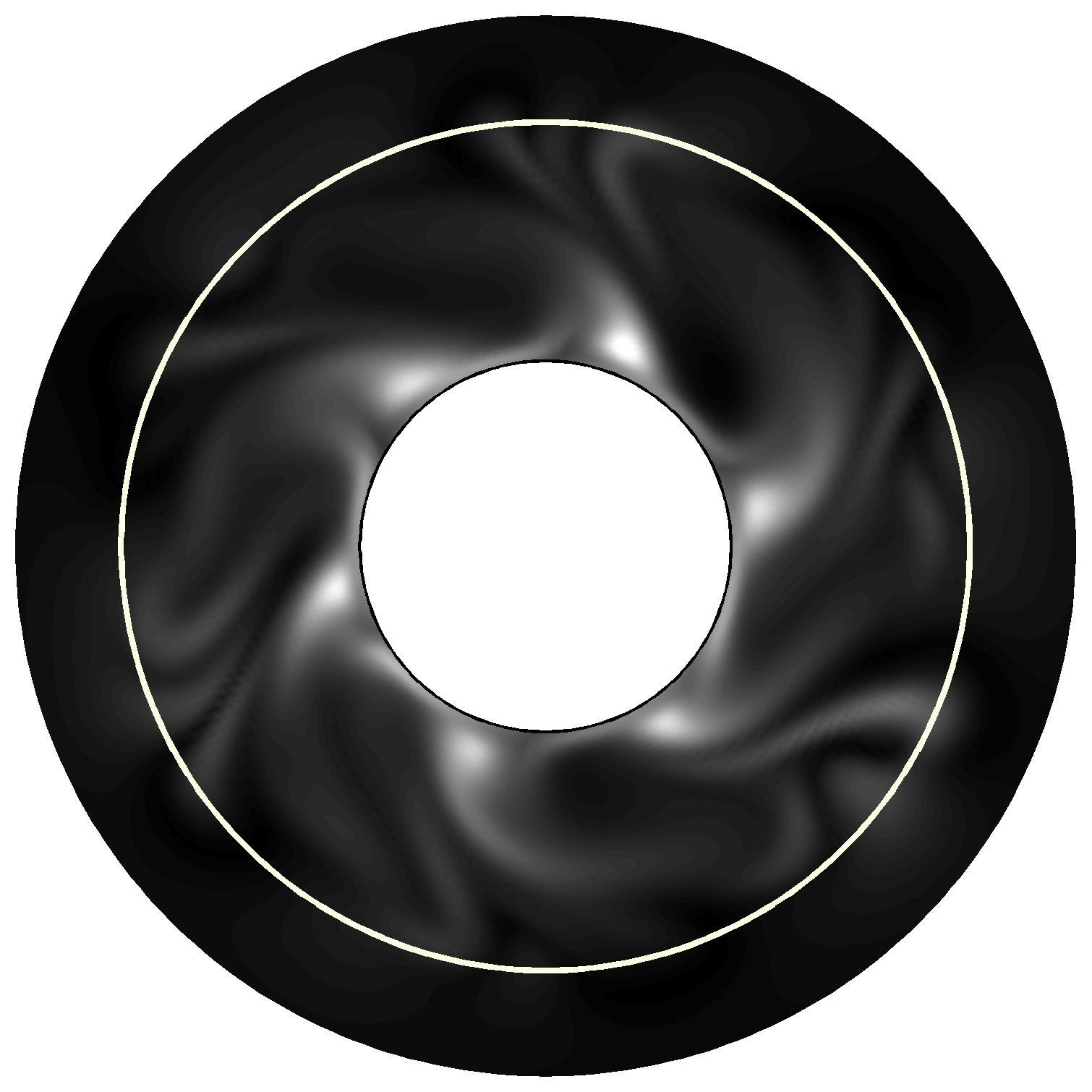}&
   \includegraphics[width=1.5in]{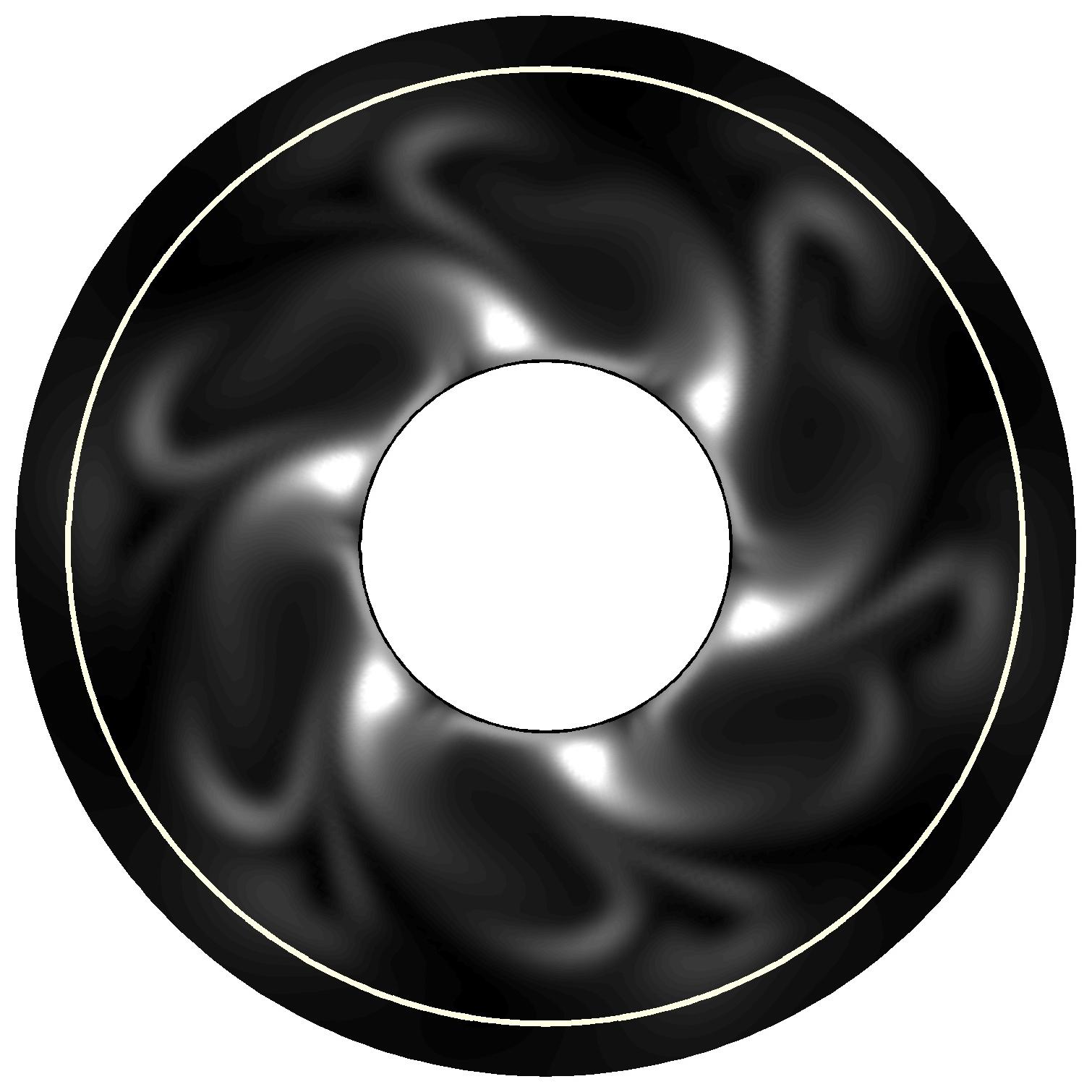}&
   \includegraphics[width=1.5in]{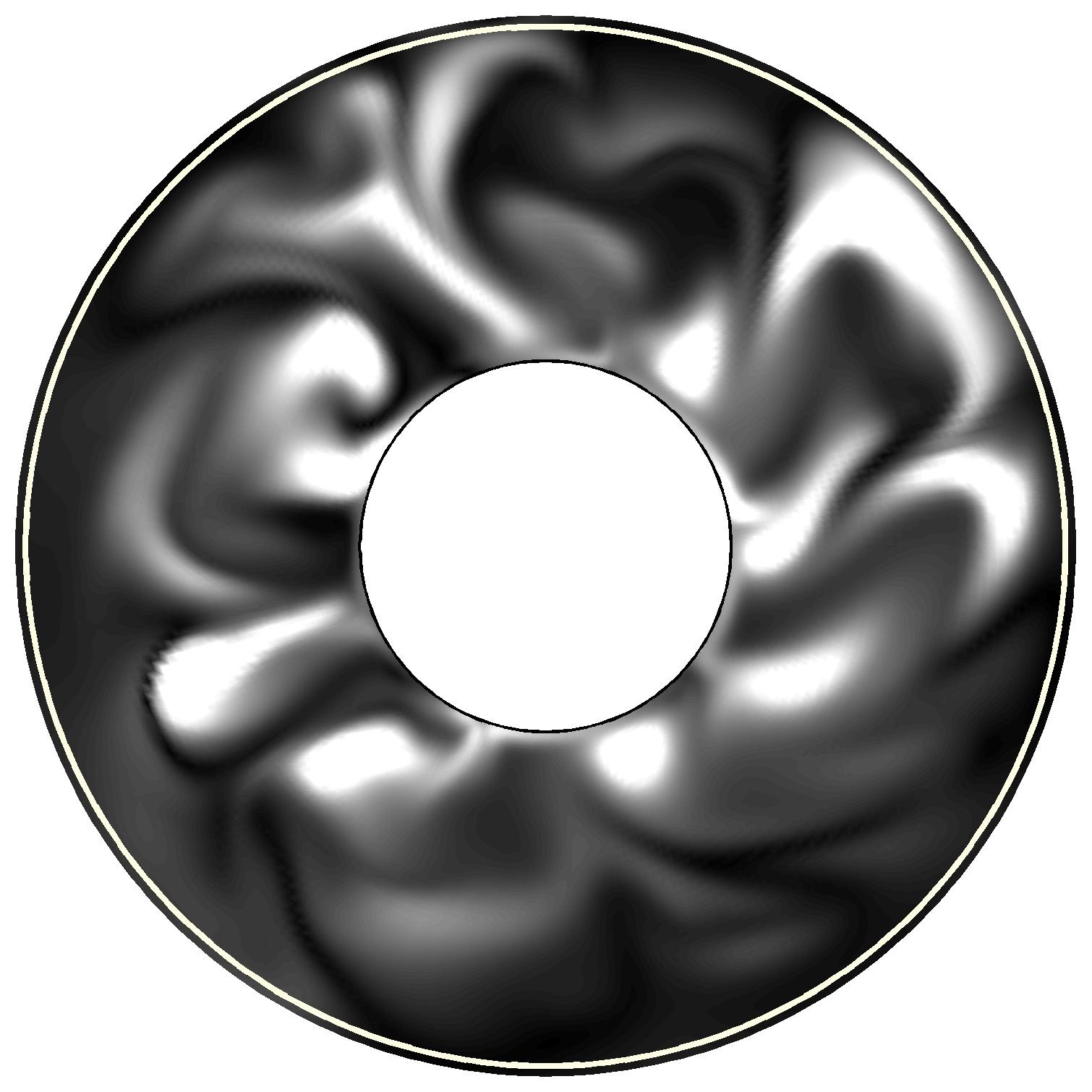}&
   \includegraphics[width=1.5in]{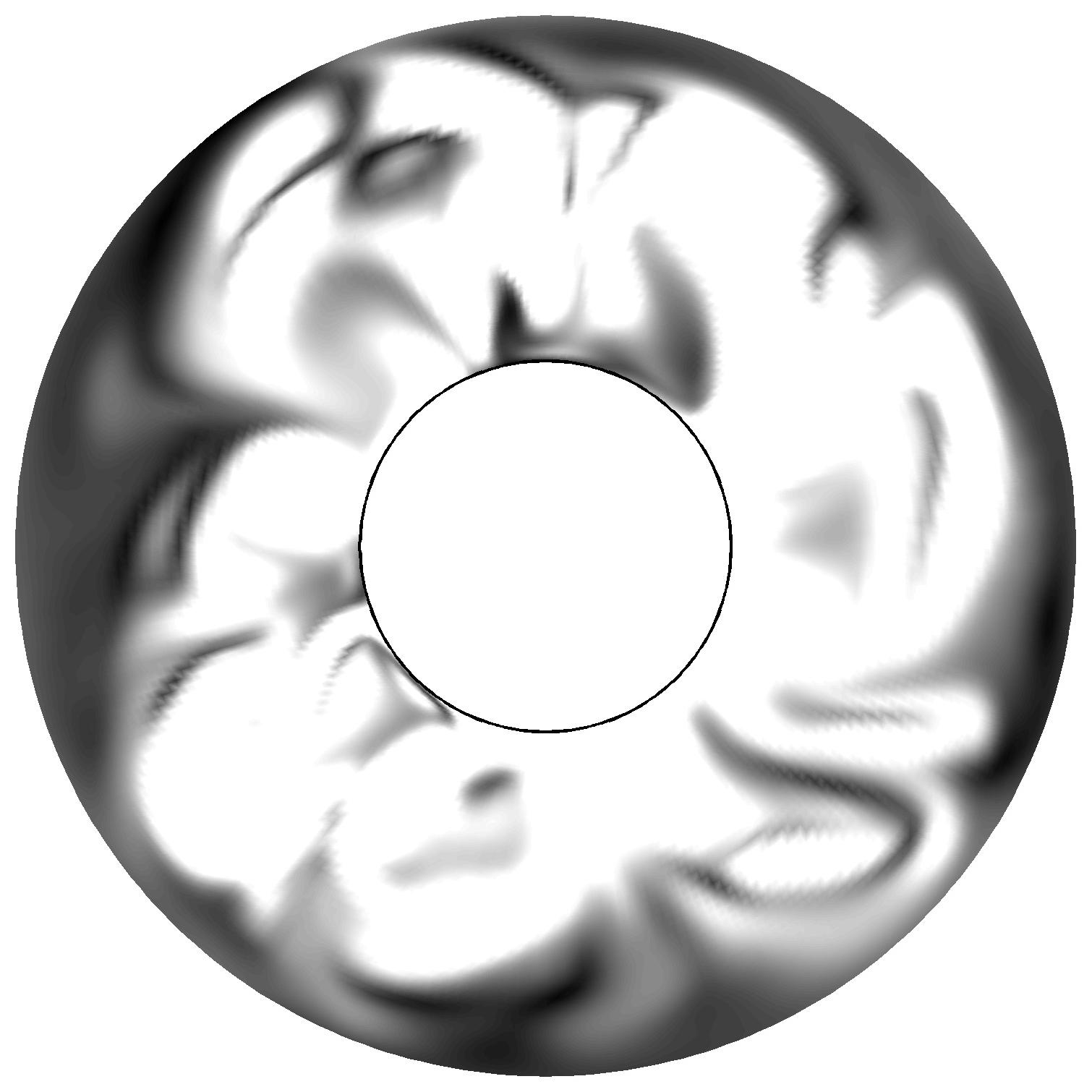}\\

   B&
   \includegraphics[width=1.5in]{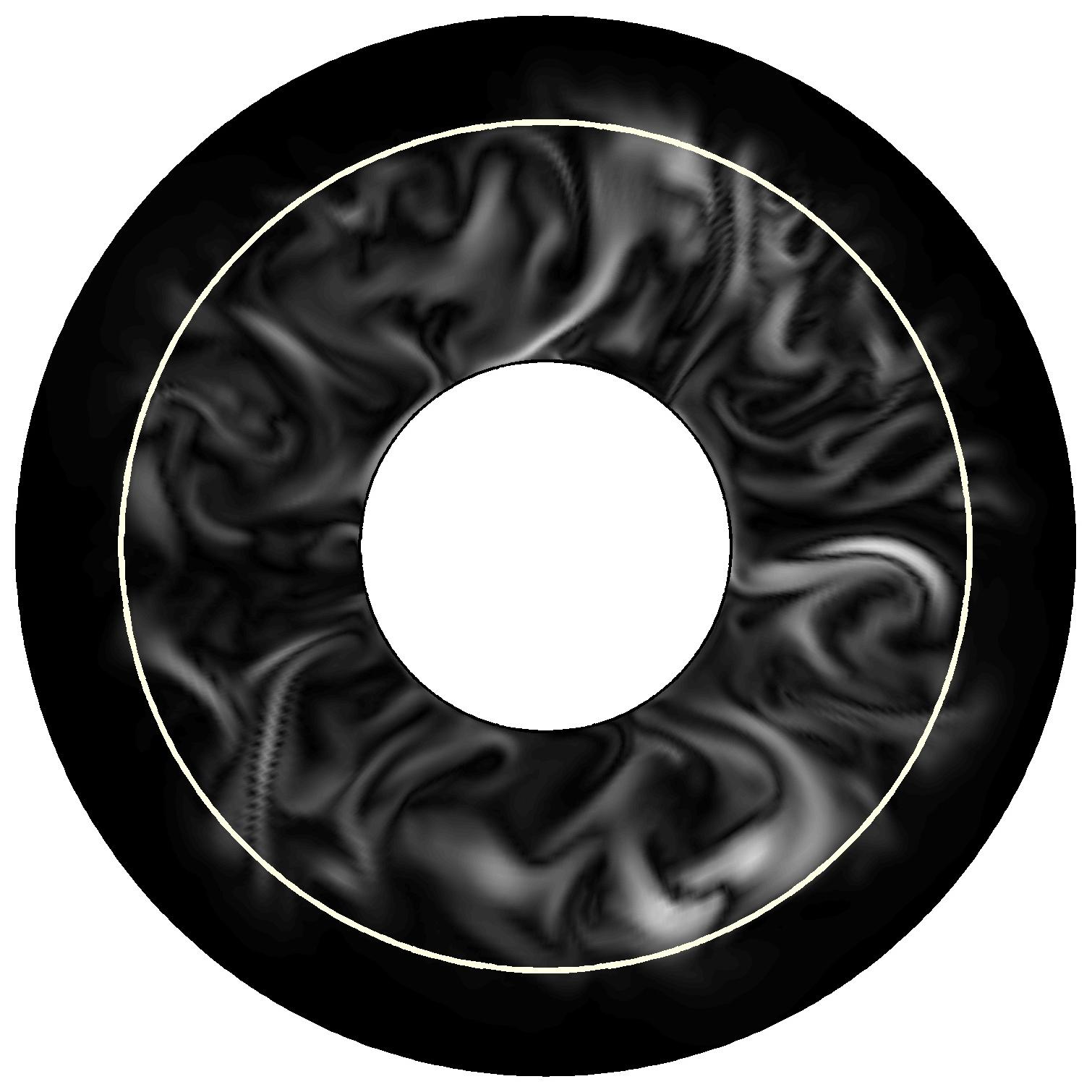}&
   \includegraphics[width=1.5in]{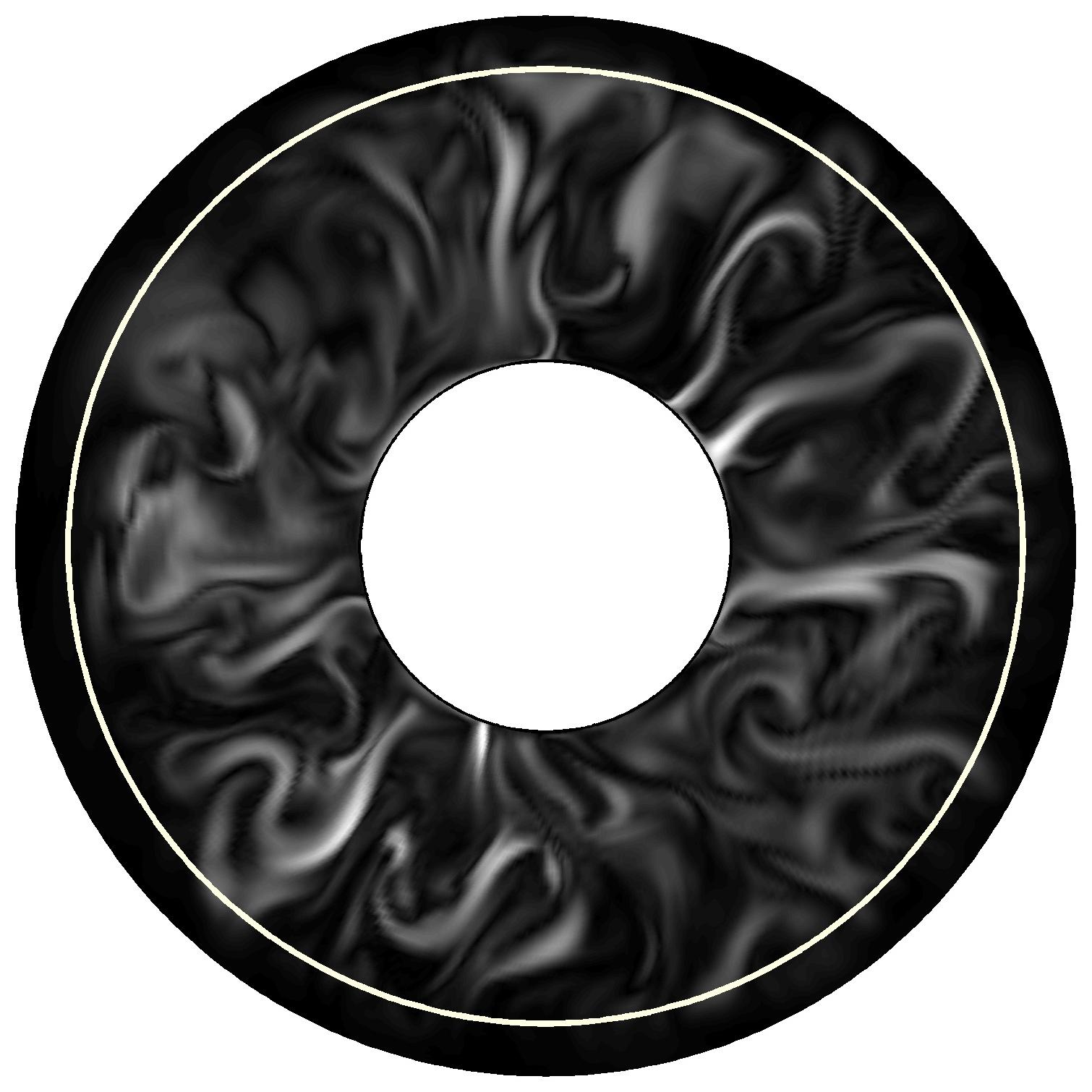}&
   \includegraphics[width=1.5in]{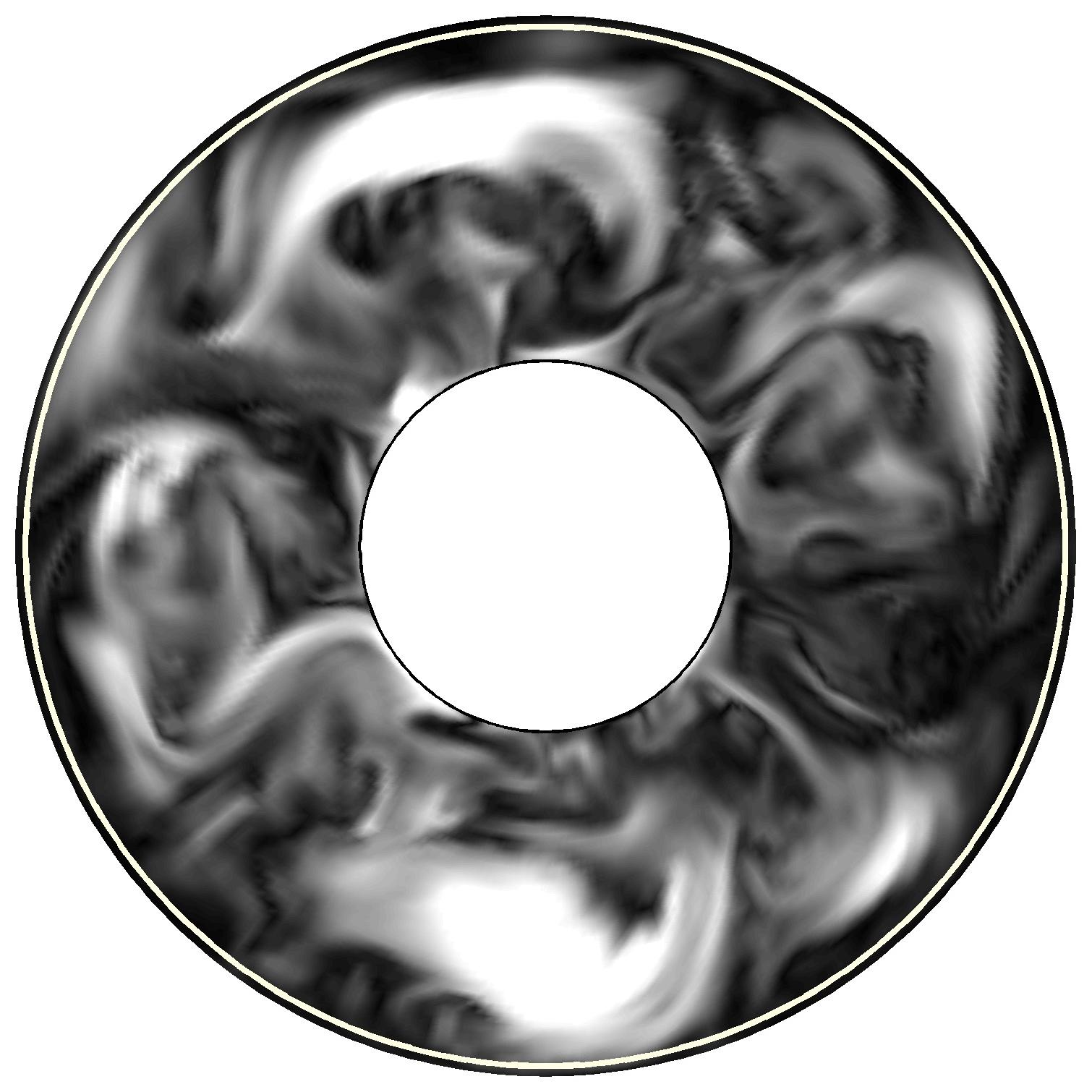}&
   \includegraphics[width=1.5in]{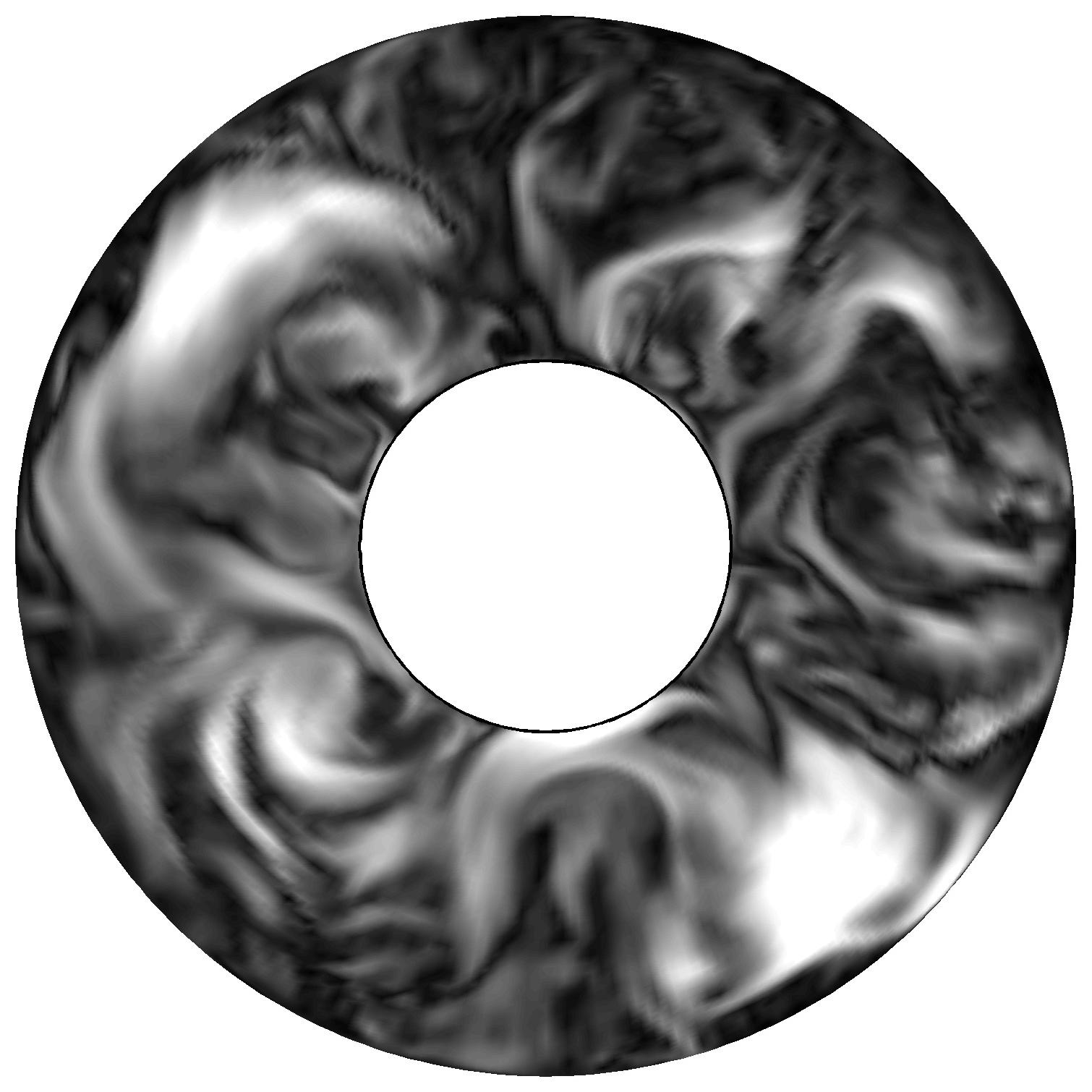}
   \\

   C&
   \includegraphics[width=1.5in]{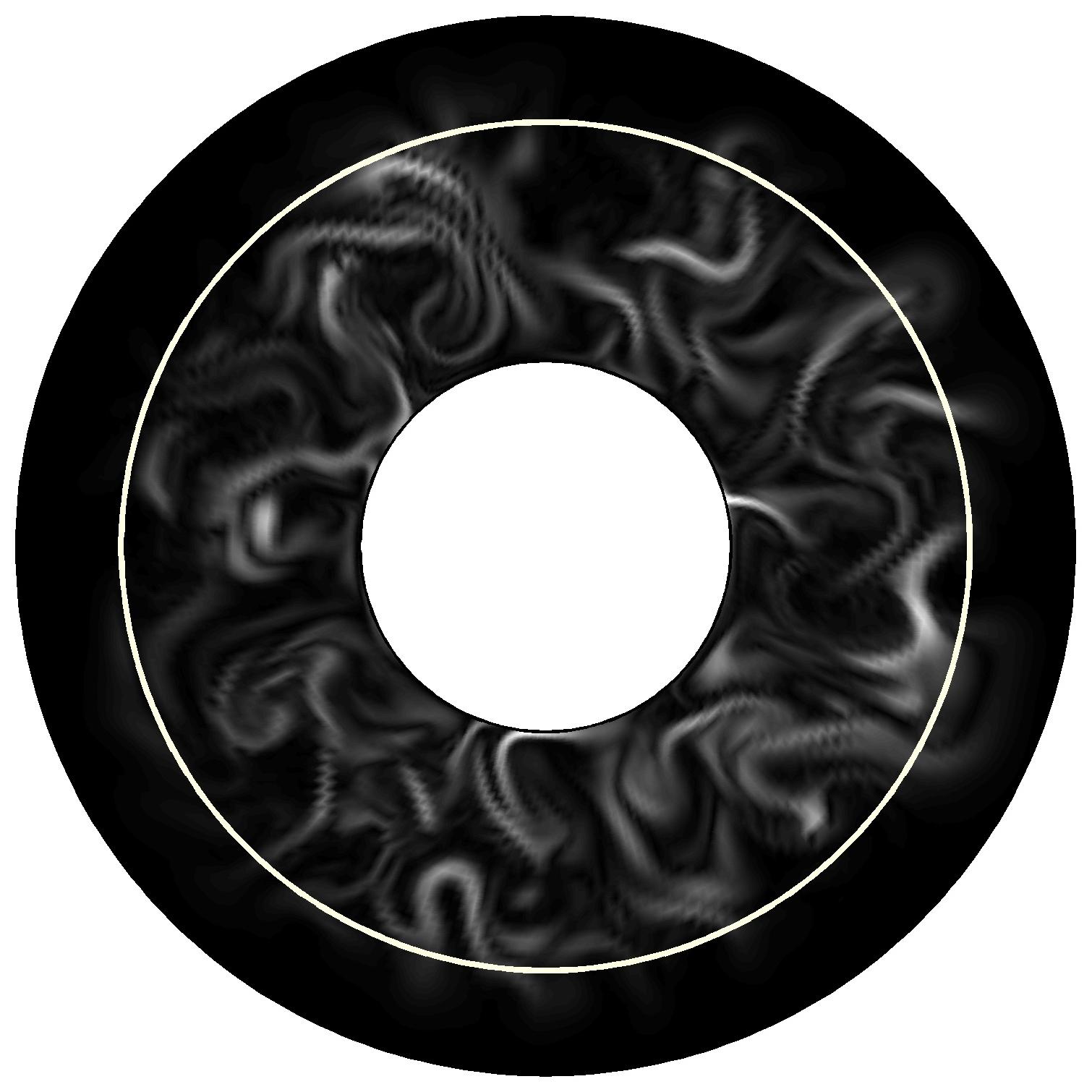}&
   \includegraphics[width=1.5in]{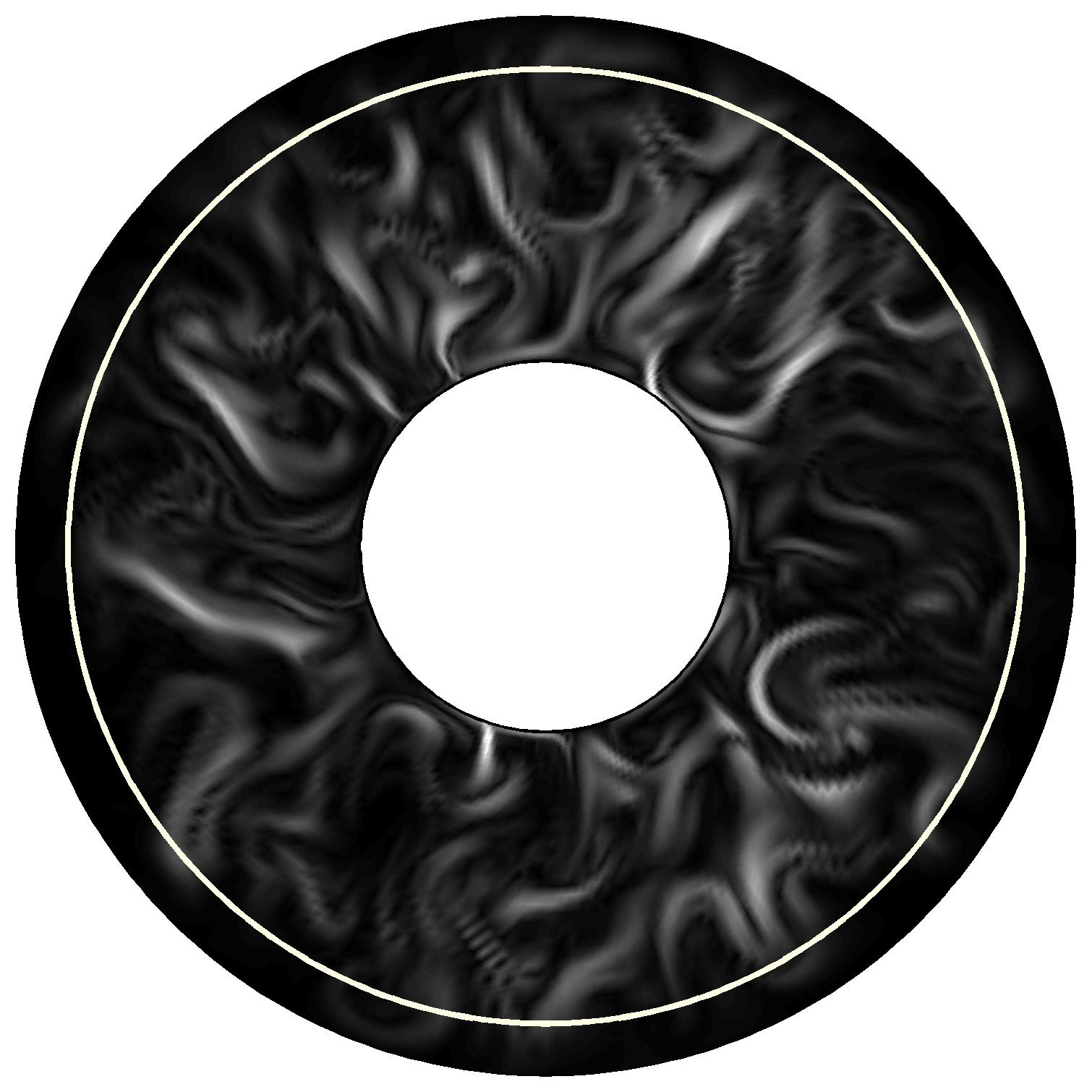}&
   \includegraphics[width=1.5in]{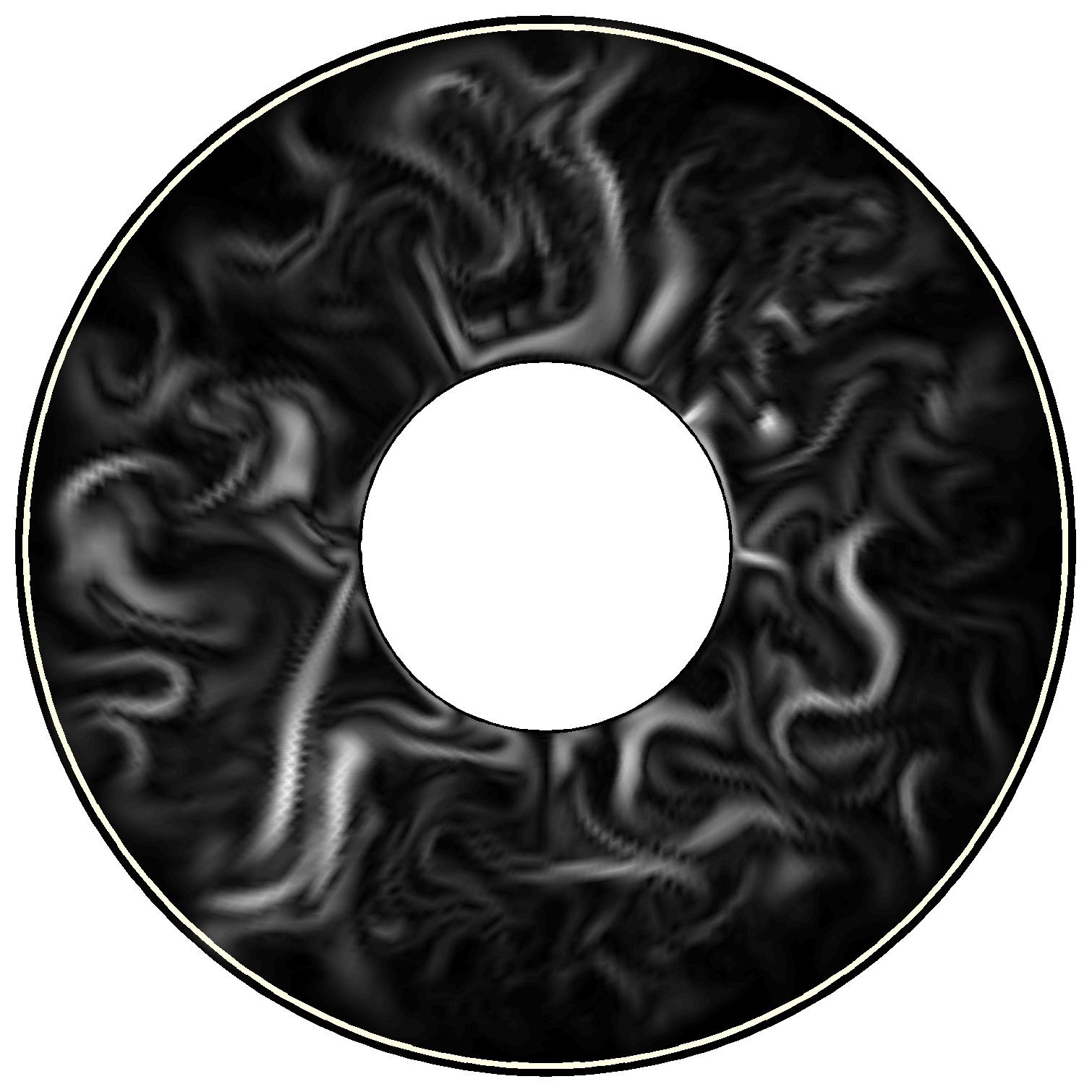}&
   \includegraphics[width=1.5in]{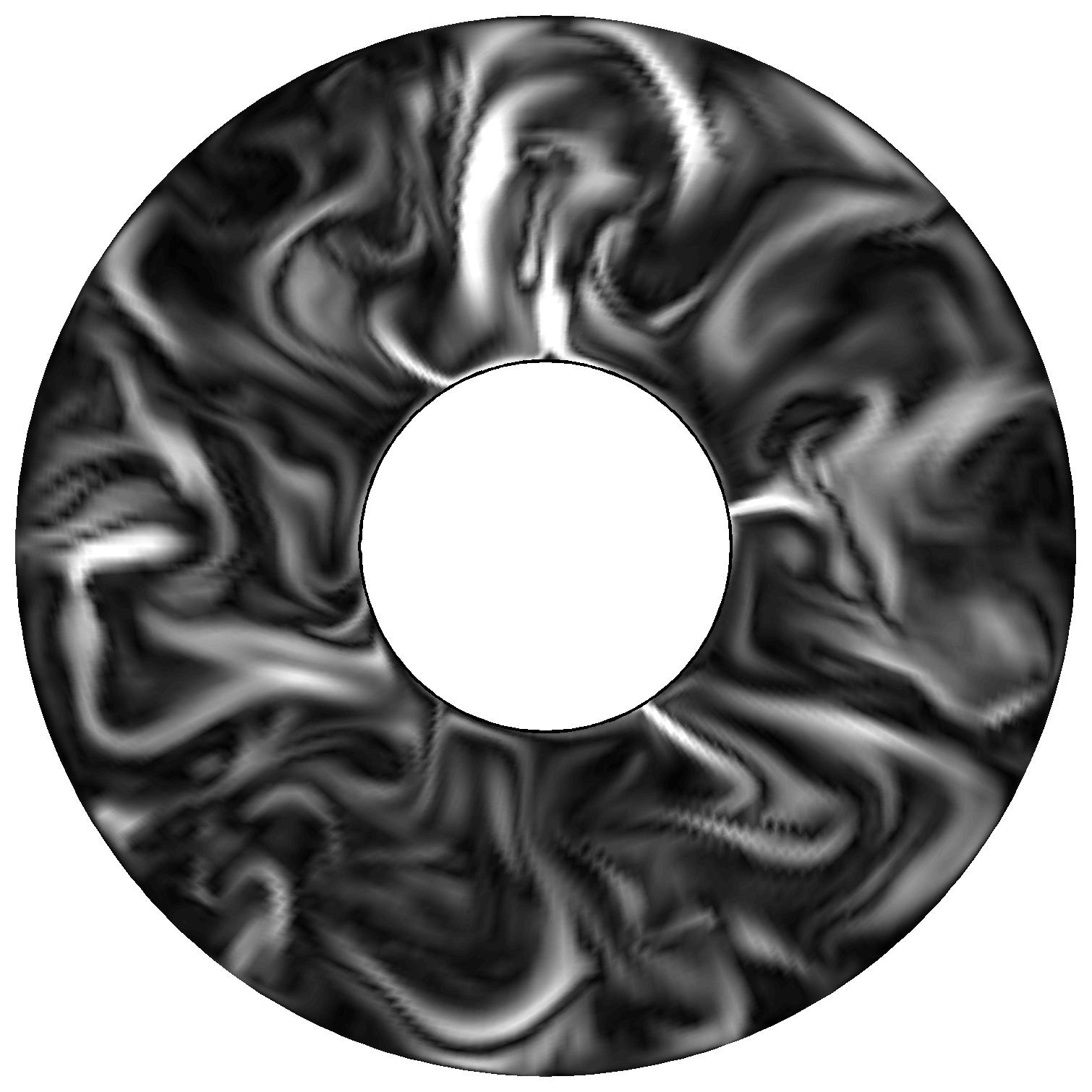}
      \\
   C'&
   \includegraphics[width=1.5in]{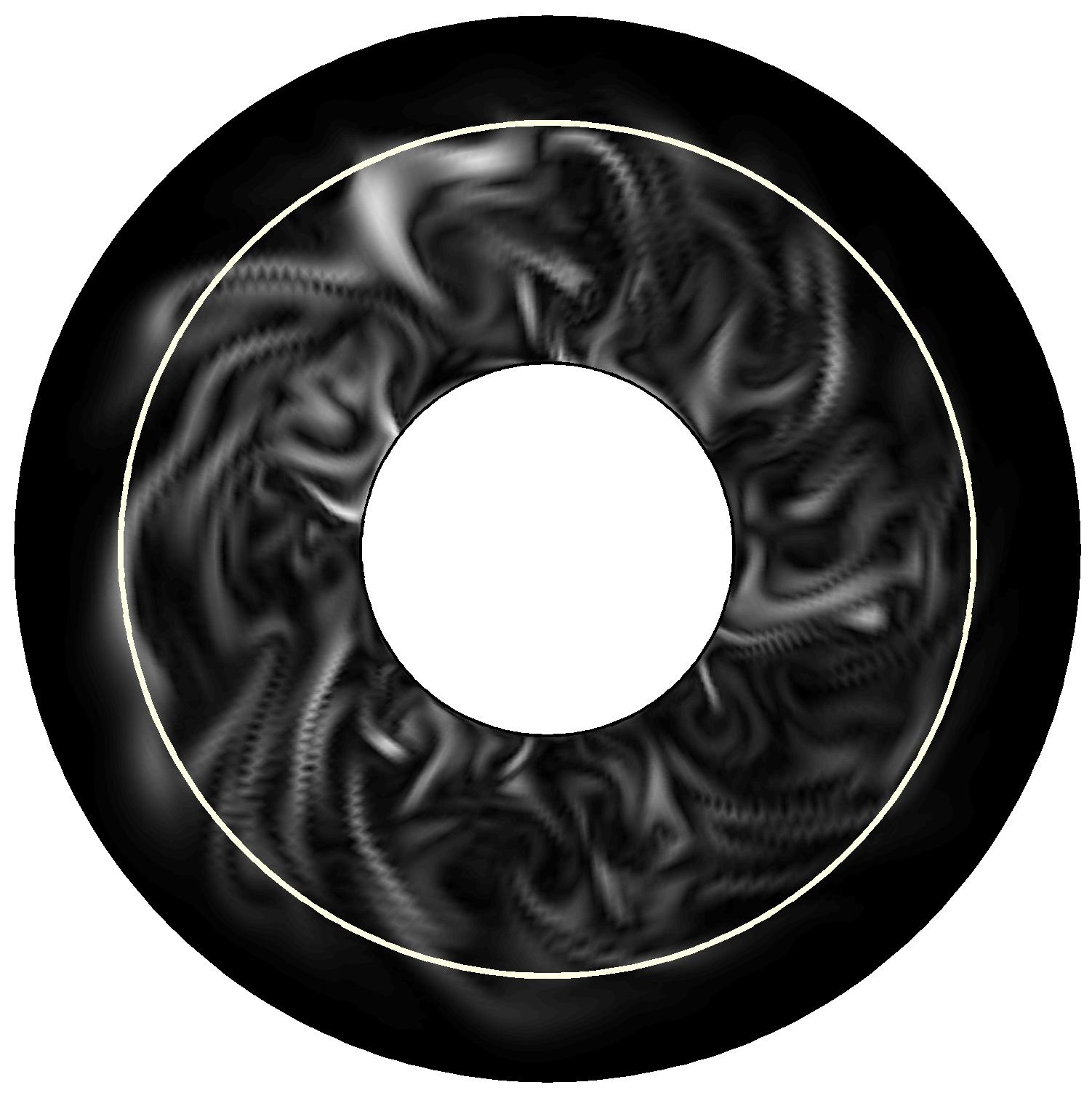}&
   \includegraphics[width=1.5in]{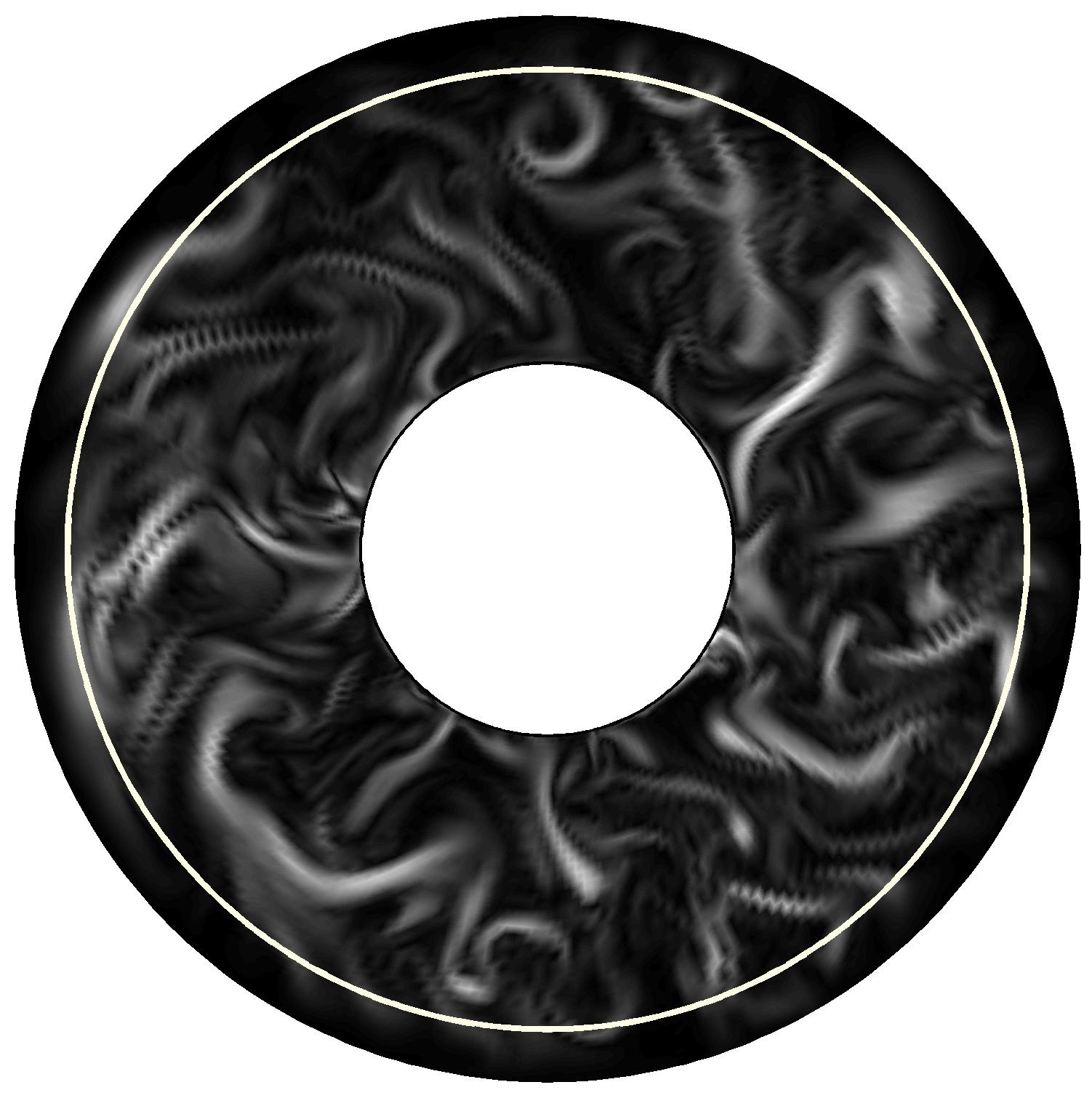}&
   \includegraphics[width=1.5in]{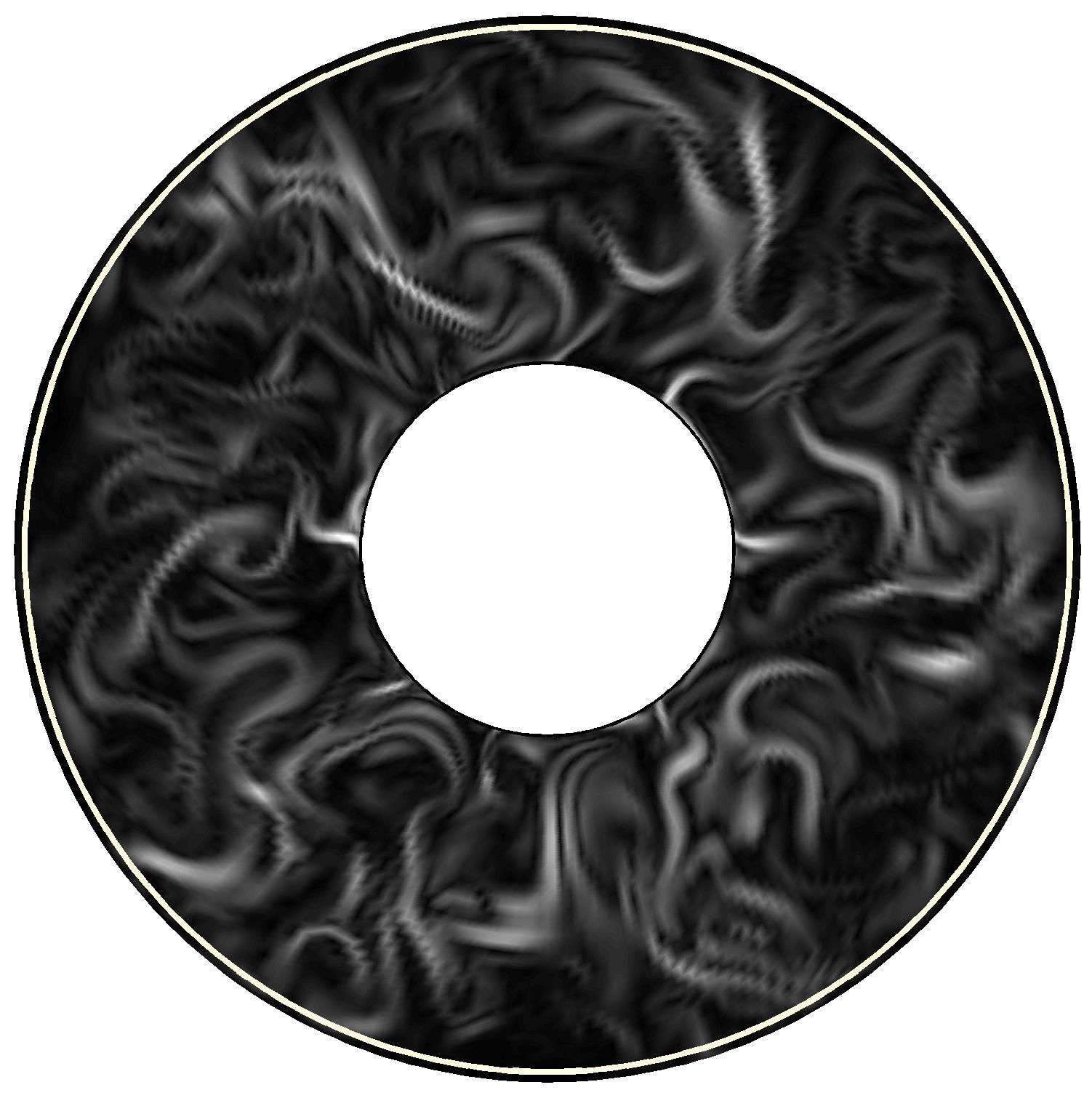}&
   \includegraphics[width=1.5in]{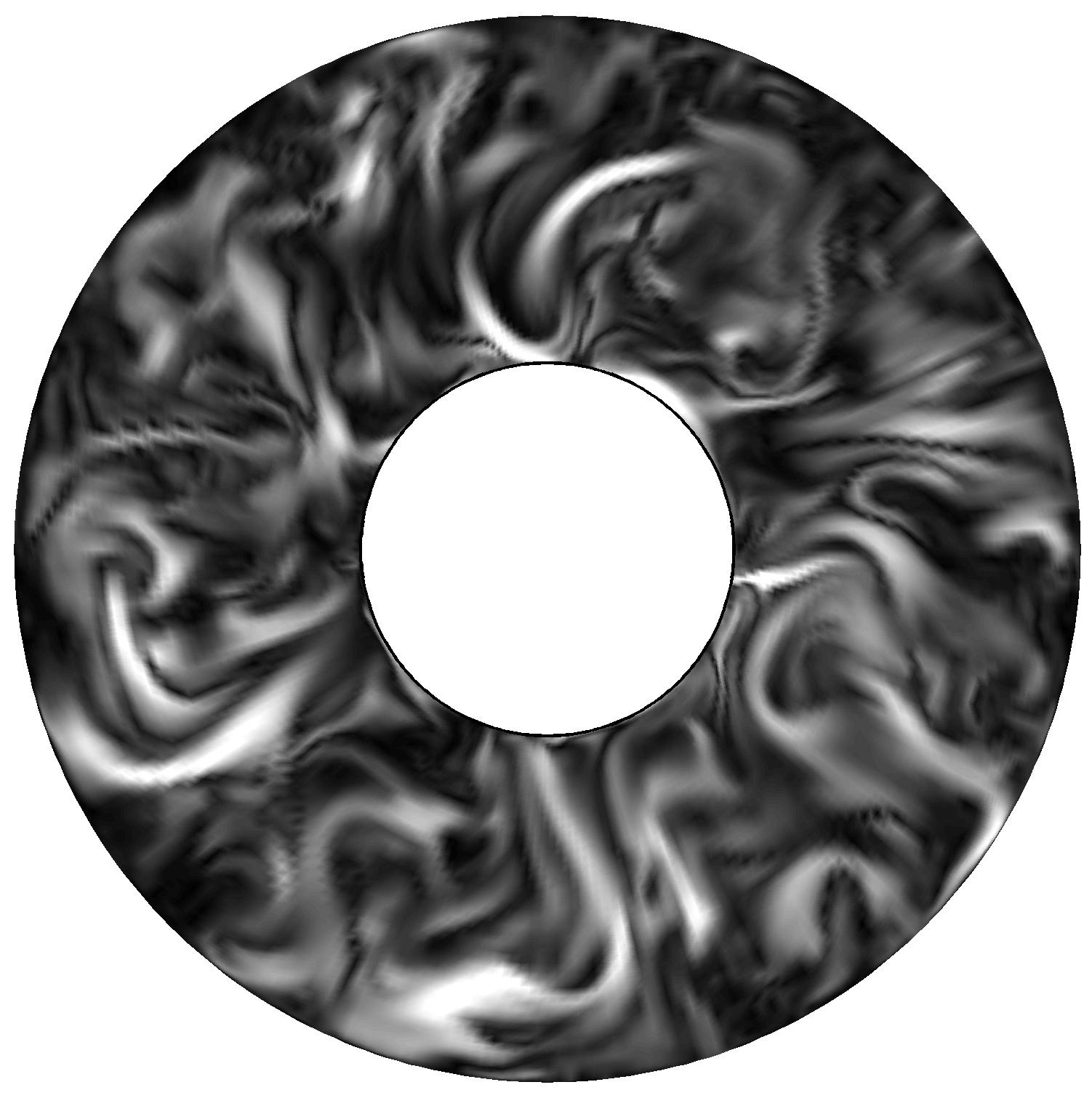}

   \end{tabular}
  \caption{ For the same time steps chosen for
    figures~\ref{fig:Temp_eq} and \ref{fig:Vort_eq}, snapshots of
    equatorial cuts of the magnetic field magnitude, $|\mathbf{B}|$
    for the models in sets A, B, C and C' from top to bottom rows.  From
    left to right, variable conductivity models with: $\chi_m=0.8$ and
    $\chi_m=0.9$, $\chi_m=0.98$, and homogeneous non-zero electrical
    conductivity solutions. For the variable conductivity cases the
    circle of radius $ r_m$ is shown with a white solid line.
    The color scale is the same for each row but changes between
    sets, between 0 (black) and 1 (white) for set A, and 0 (black) and
    15 (white) for sets B, C and C', in units of the Elsasser number. }
   \label{fig:sqrBsq_eq}
\end{sidewaysfigure}

In figure~\ref{fig:Vort_eq},
snapshots (at the same instants as for figure~\ref{fig:Temp_eq}) of
equatorial profiles of the axial vorticity are shown. 
Corresponding equatorial profiles of
$|\mathbf{B}|$ in figure~\ref{fig:sqrBsq_eq}  are included for all sets.

In set A there is a clear correlation between the cyclonic vortices
(in black) with the magnetic field magnitude maxima (in white).
Similarly to the temperature, the wave number of the vorticity
profiles of the variable conductivity cases is equal to that of the
non-magnetic case and allows for smaller scale features.  For set A,
as well as seen in the temperature profiles, the presence of zonal
flow bends the vortices in a prograde direction.  Runs with
$\chi_m=0.8$ and $\chi_m=0.9$ mimic this prograde tendency and the
magnetic field is regularly organized in azimuth with a dominant wave
number of six. Runs $\chi_m=\infty$ and $\chi_m=0.98$, do not show a
clear prograde tilt with increasing radius, and have a higher dominant
wave number than runs with lower values of $\chi_m$.
The correlation between cyclonic vortices and magnetic field maxima
can be inferred for sets B, C and C', but it is not as evident as for set
A.  The length scale of the magnetic field decreases with decreasing
$\chi_m$.

\begin{sidewaysfigure}
   \centering
   \begin{tabular}{m{0.1in}|m{2in}|m{2in}|m{2in}|m{2in}|m{2in}}
    &\multicolumn{1}{c|}{ $\chi_m=0.8$} 
    &\multicolumn{1}{c|}{ $\chi_m=0.9$ }
    &\multicolumn{1}{c|}{ $\chi_m=0.98$} 
    &\multicolumn{1}{c}{$\chi_m=\infty$} \\

   A&
   \includegraphics[width=2in]{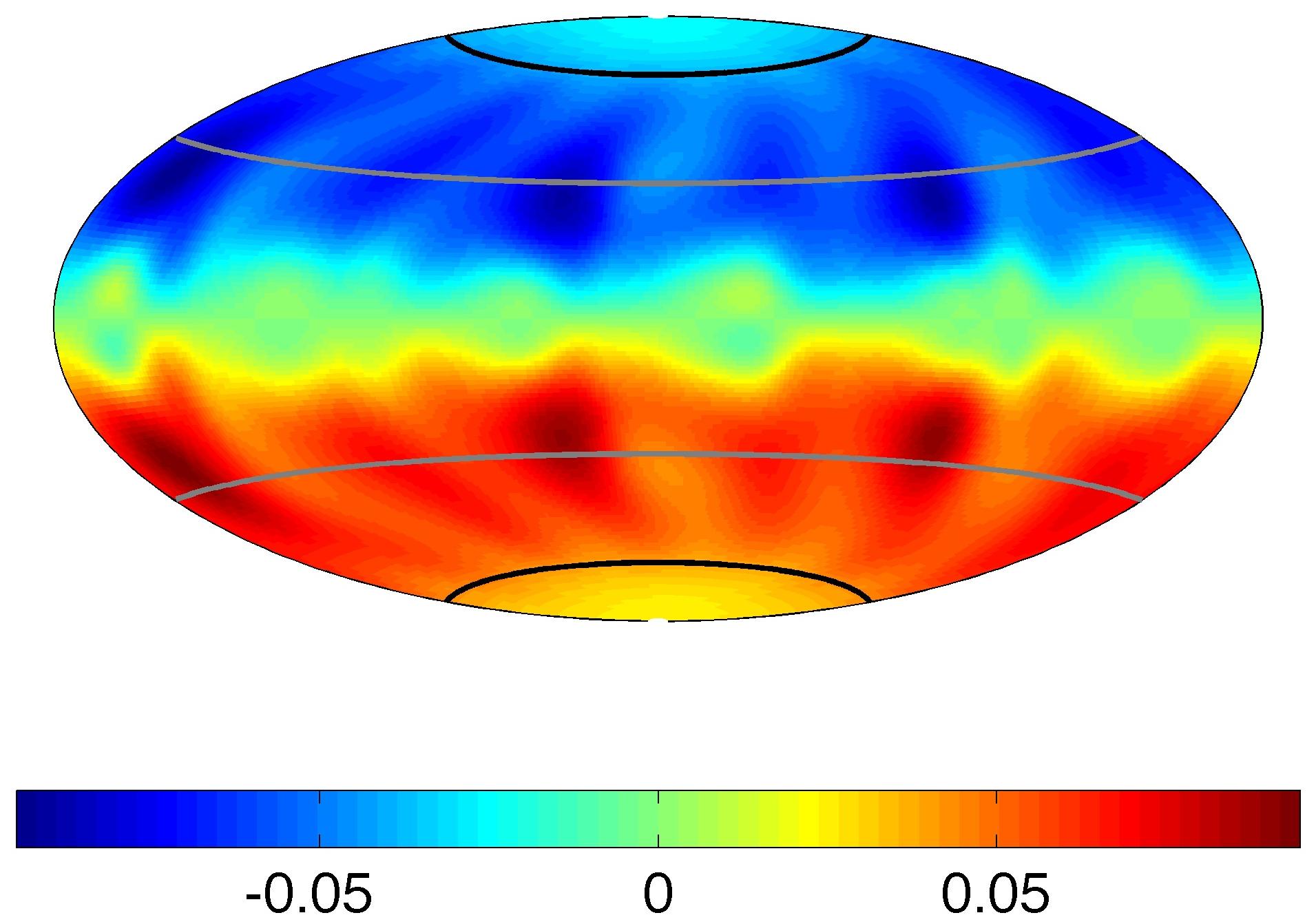}&
   \includegraphics[width=2in]{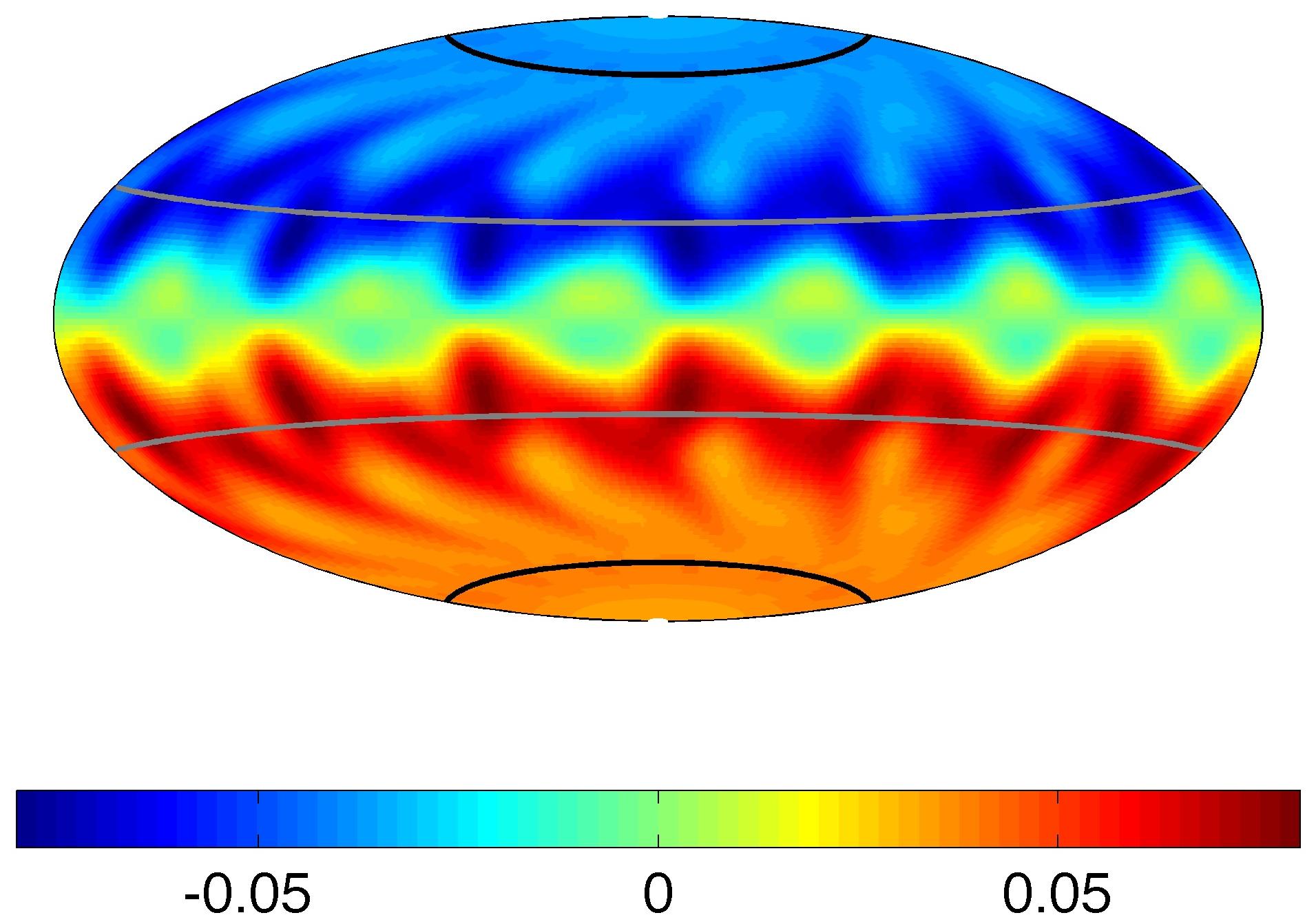}&
   \includegraphics[width=2in]{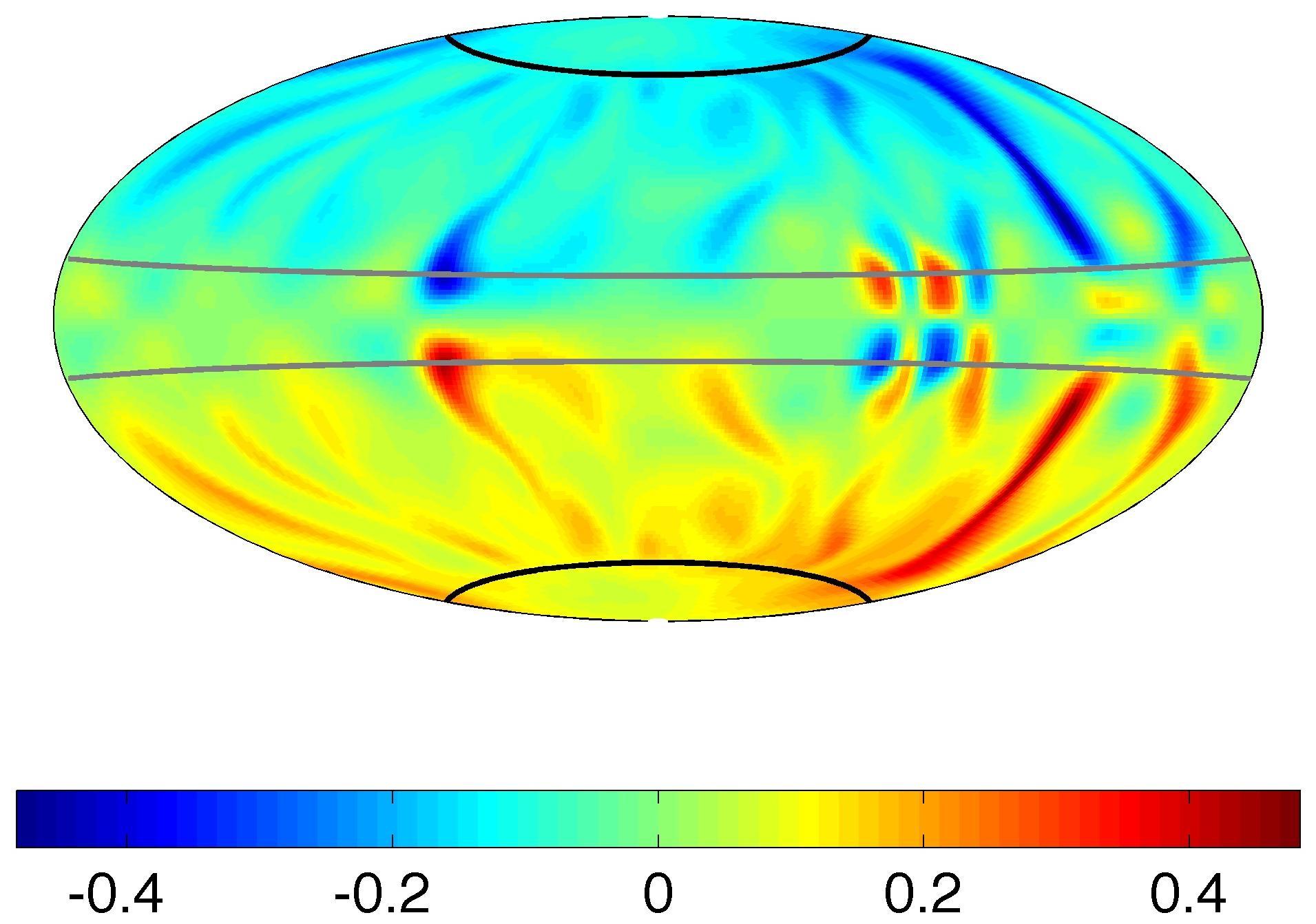}&
   \includegraphics[width=2in]{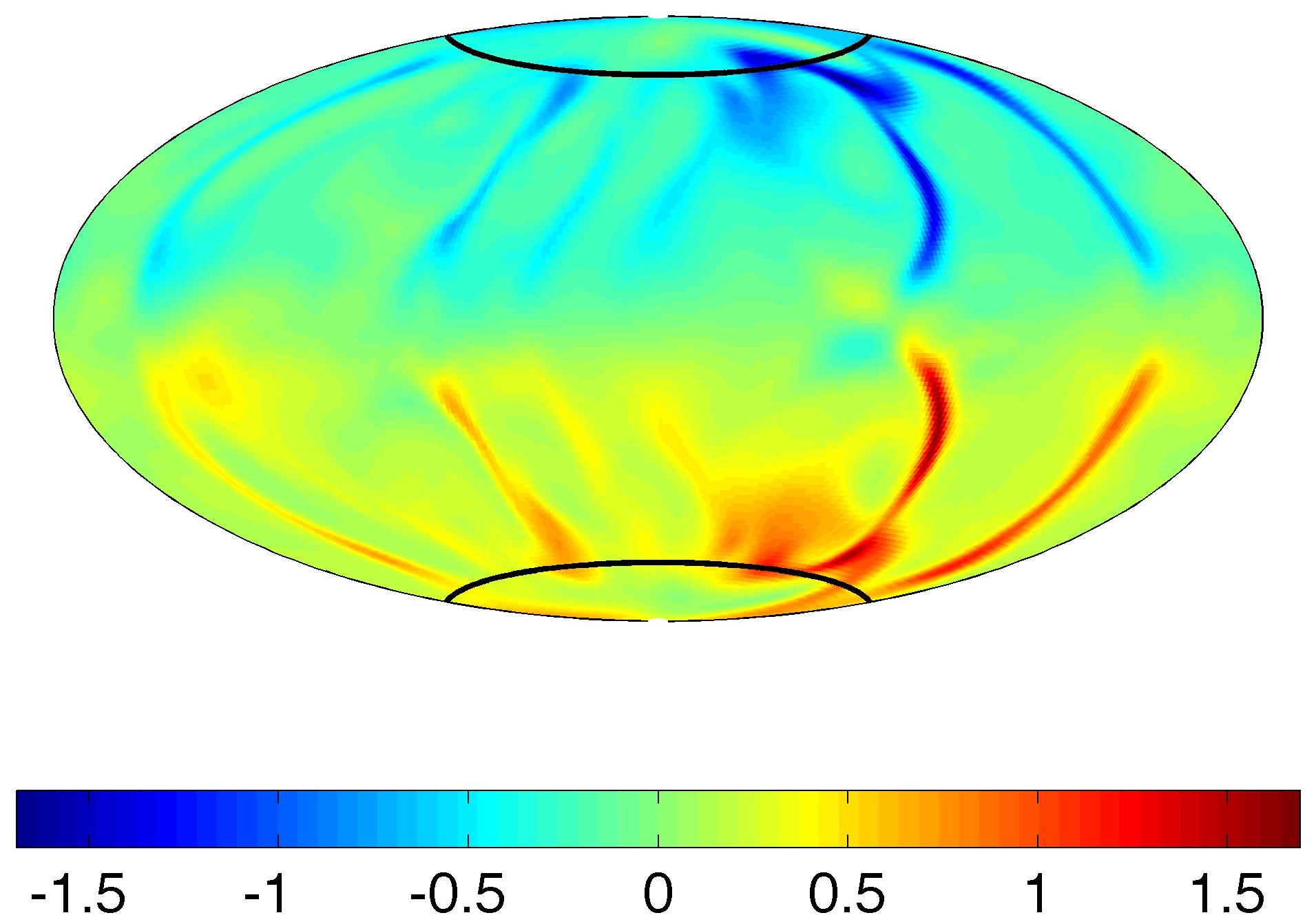}\\

   B&
   \includegraphics[width=2in]{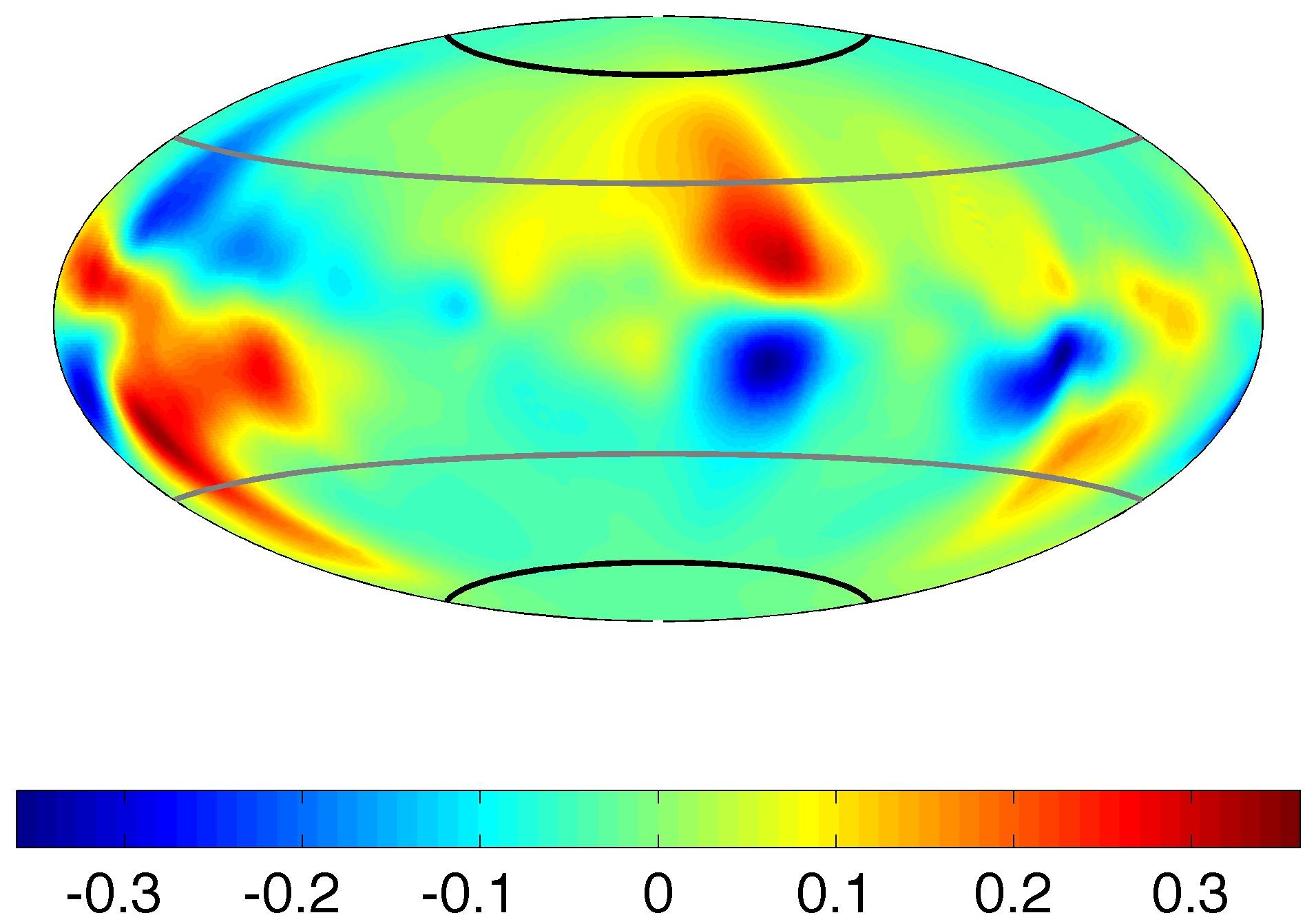}&
   \includegraphics[width=2in]{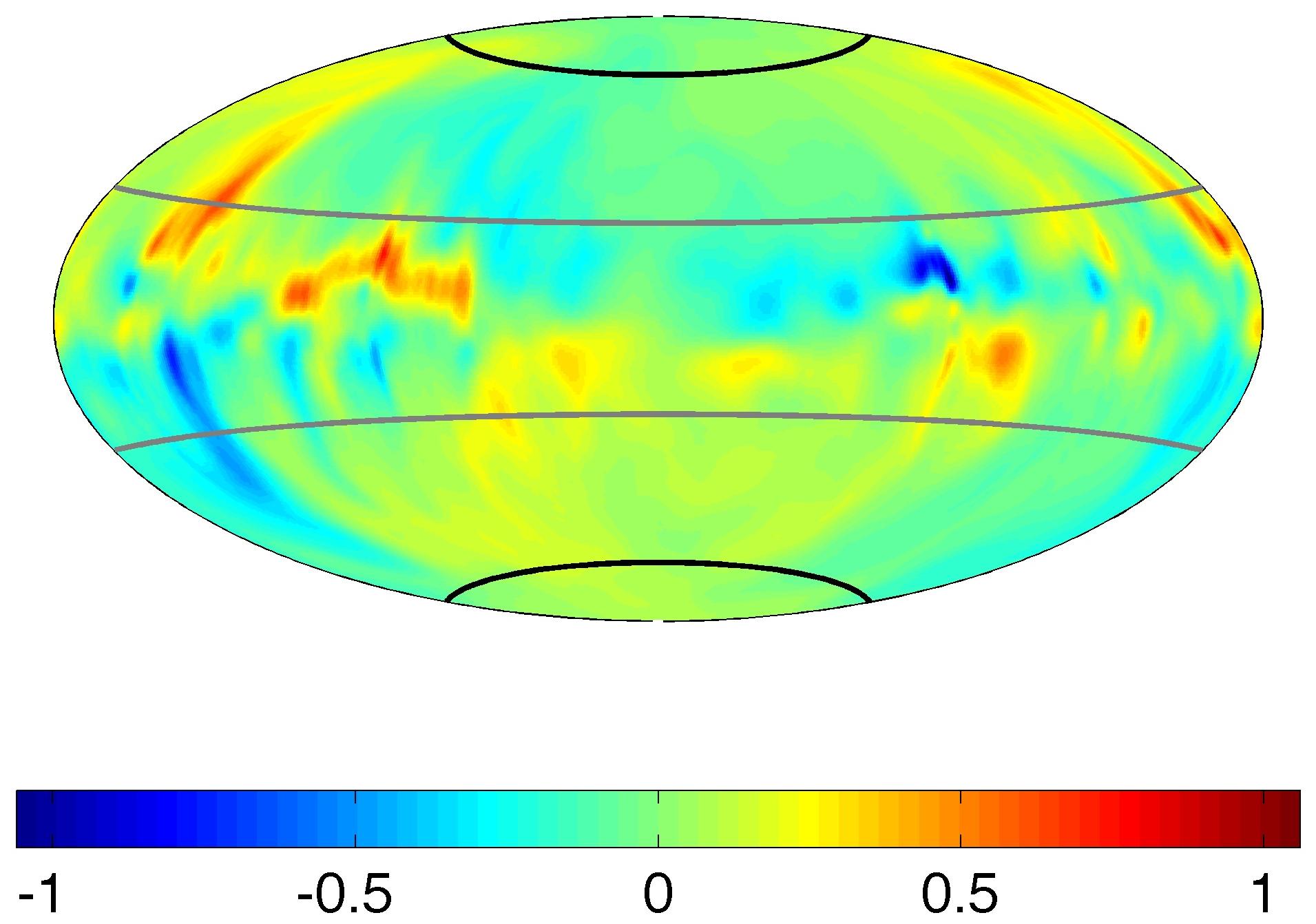}&
   \includegraphics[width=2in]{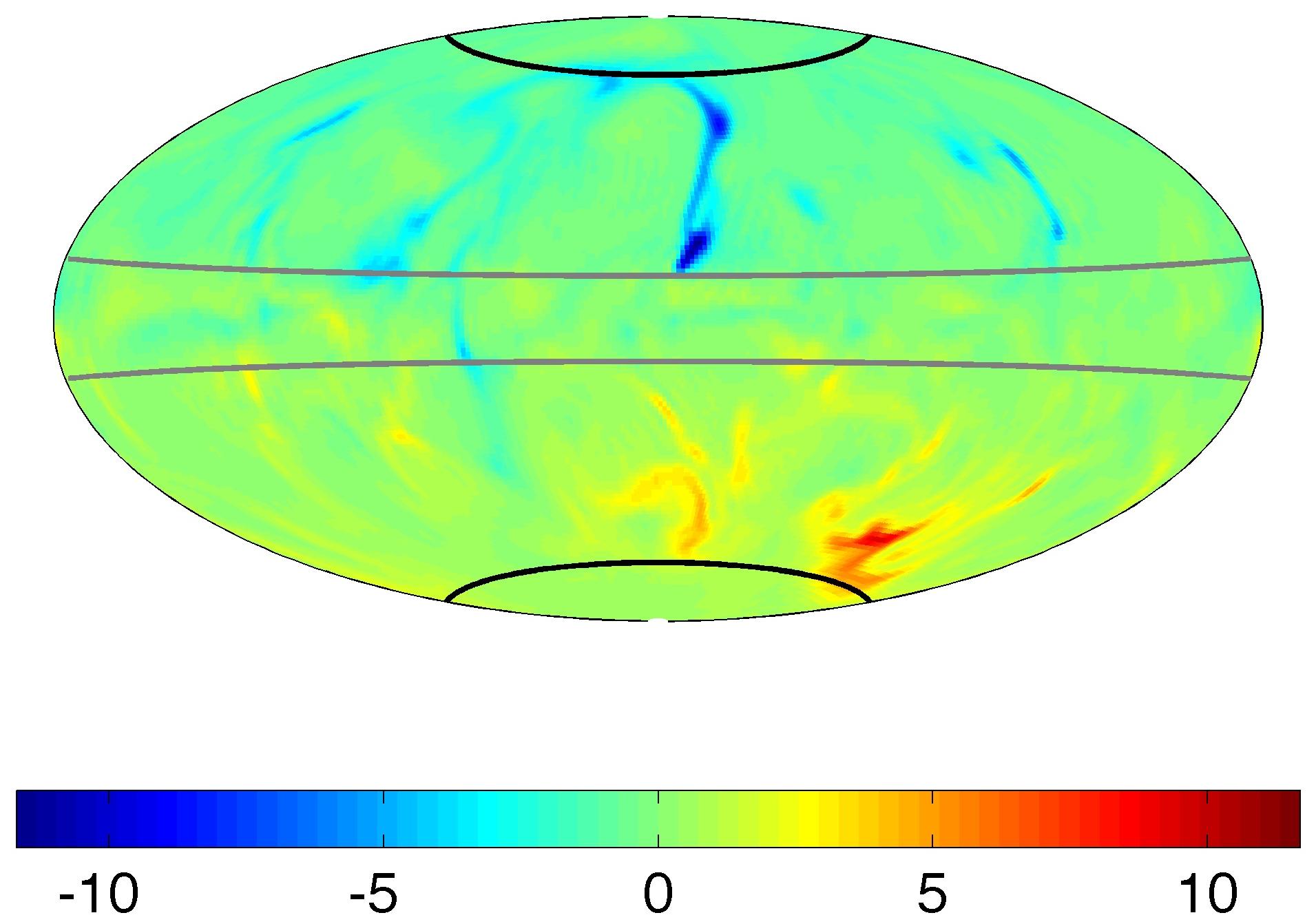}&
   \includegraphics[width=2in]{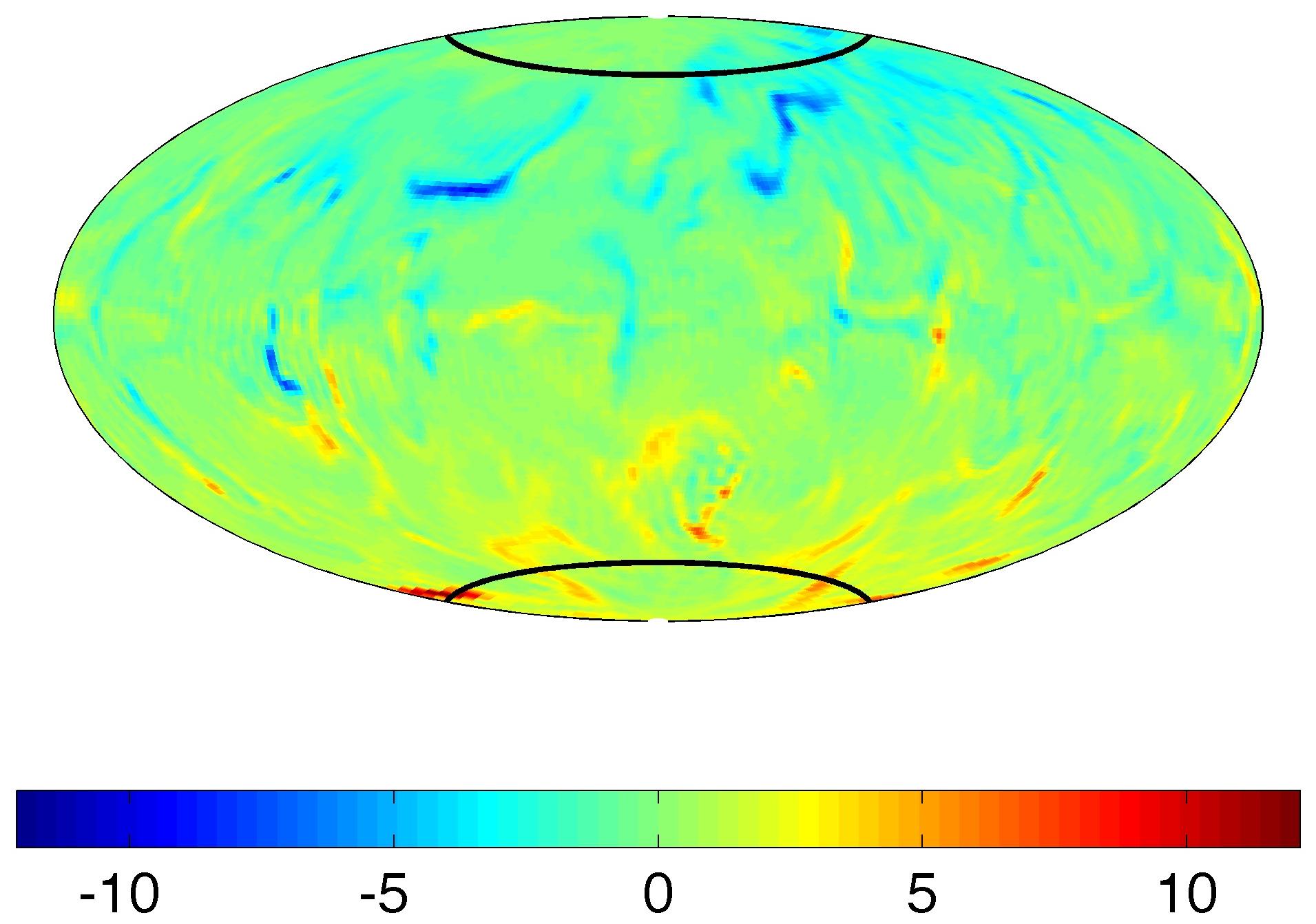}
   \\

   C&
   \includegraphics[width=2in]{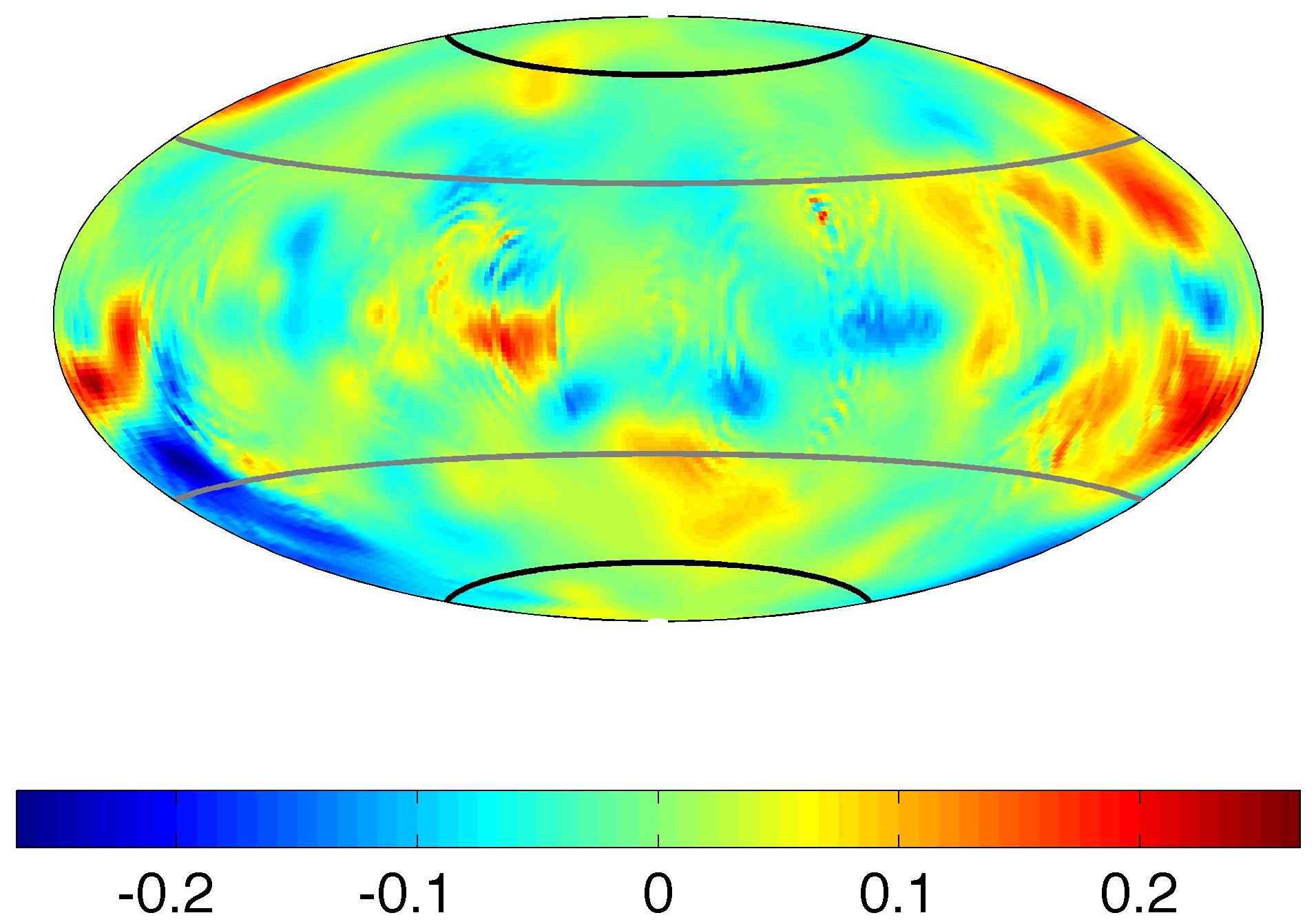}&
   \includegraphics[width=2in]{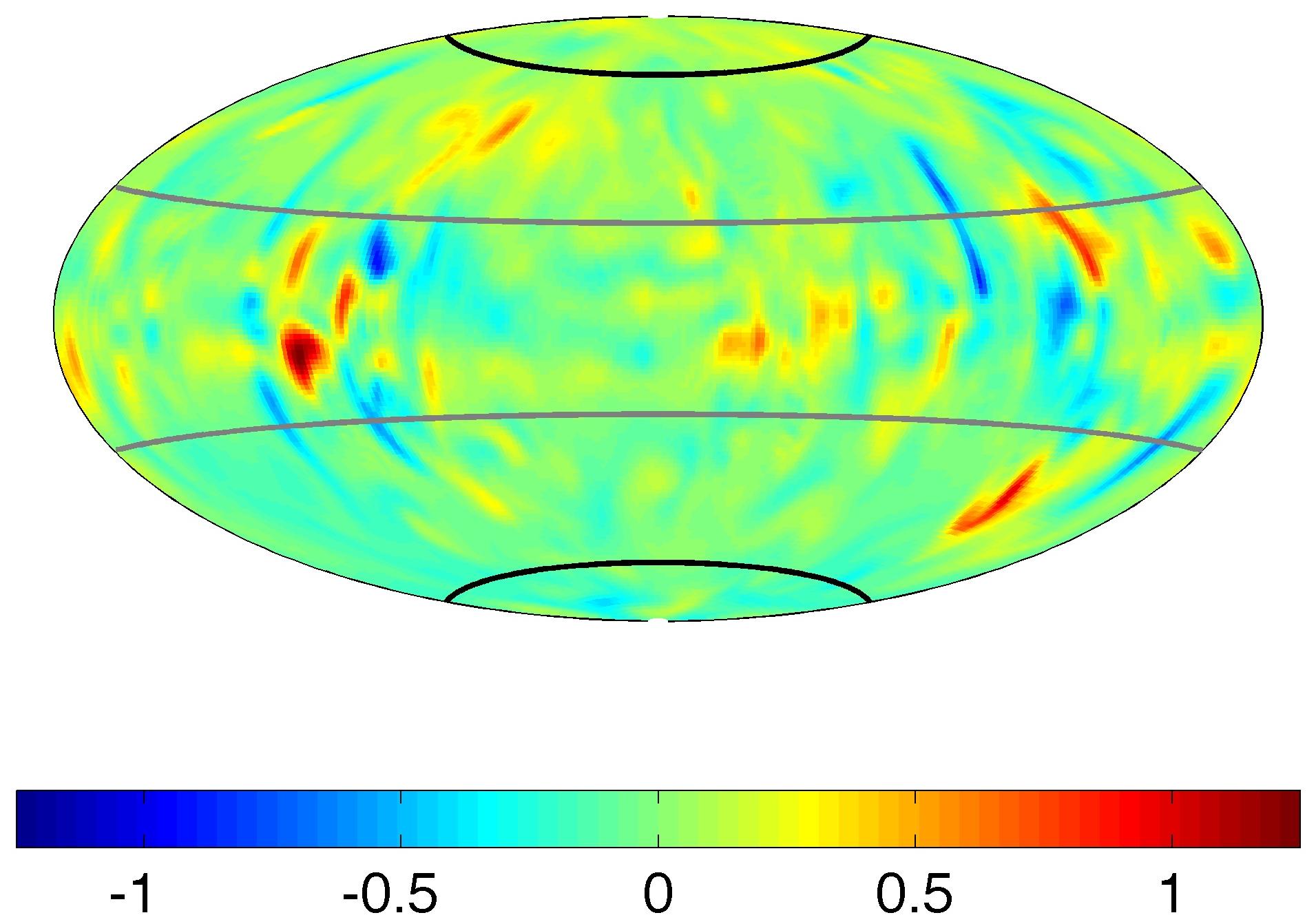}&
   \includegraphics[width=2in]{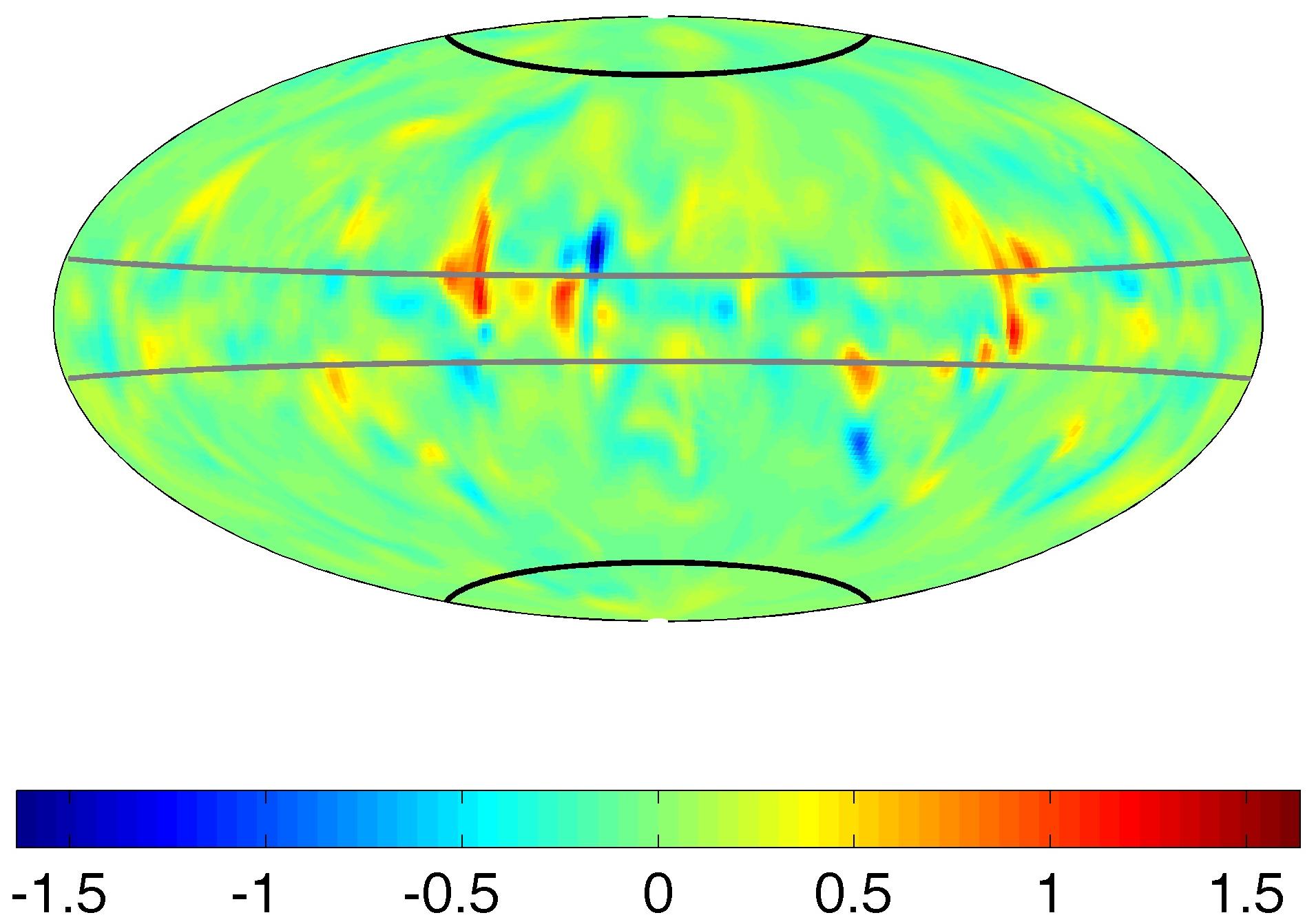}&
   \includegraphics[width=2in]{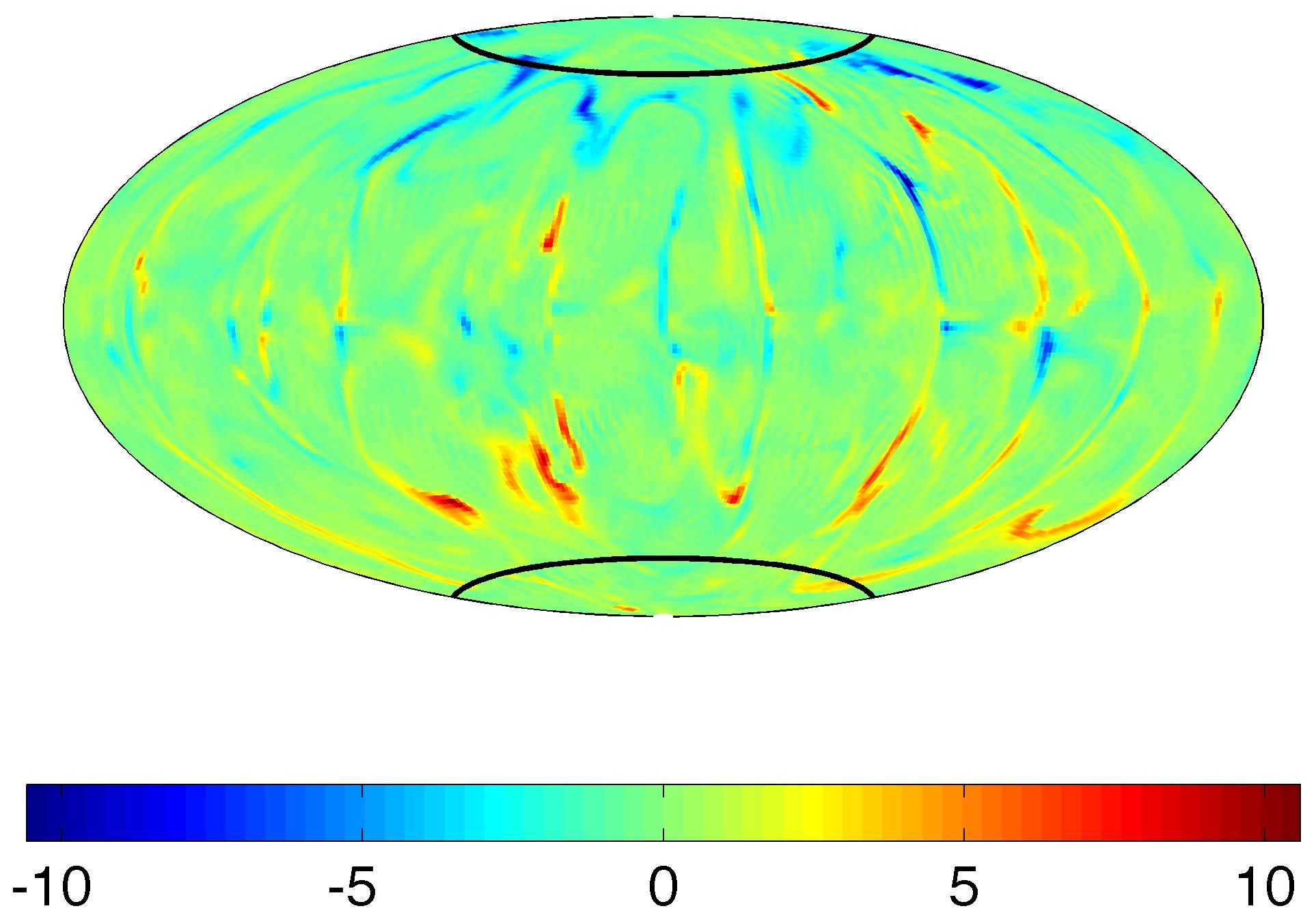}
      \\
  C'&
   \includegraphics[width=2in]{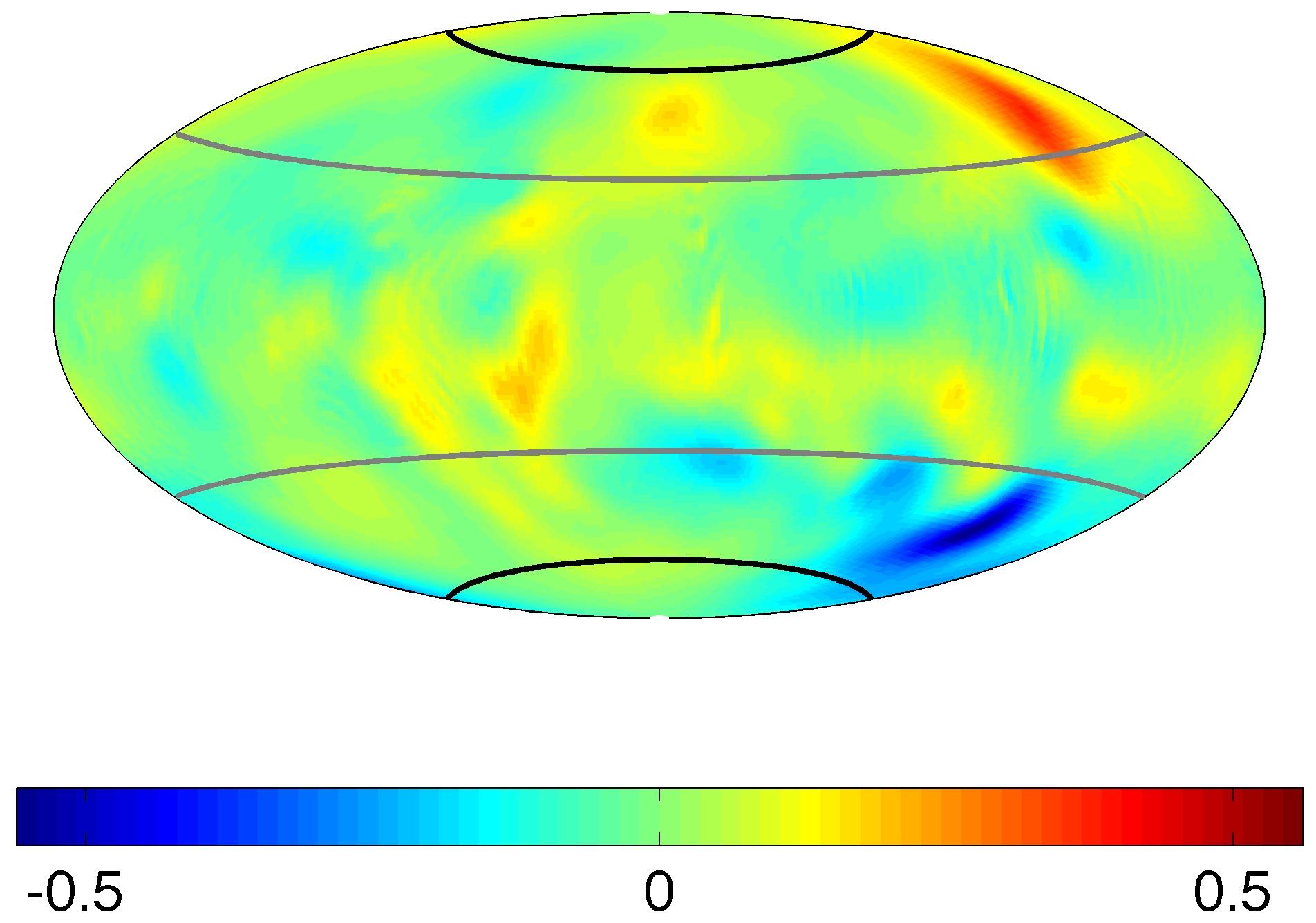}&
   \includegraphics[width=2in]{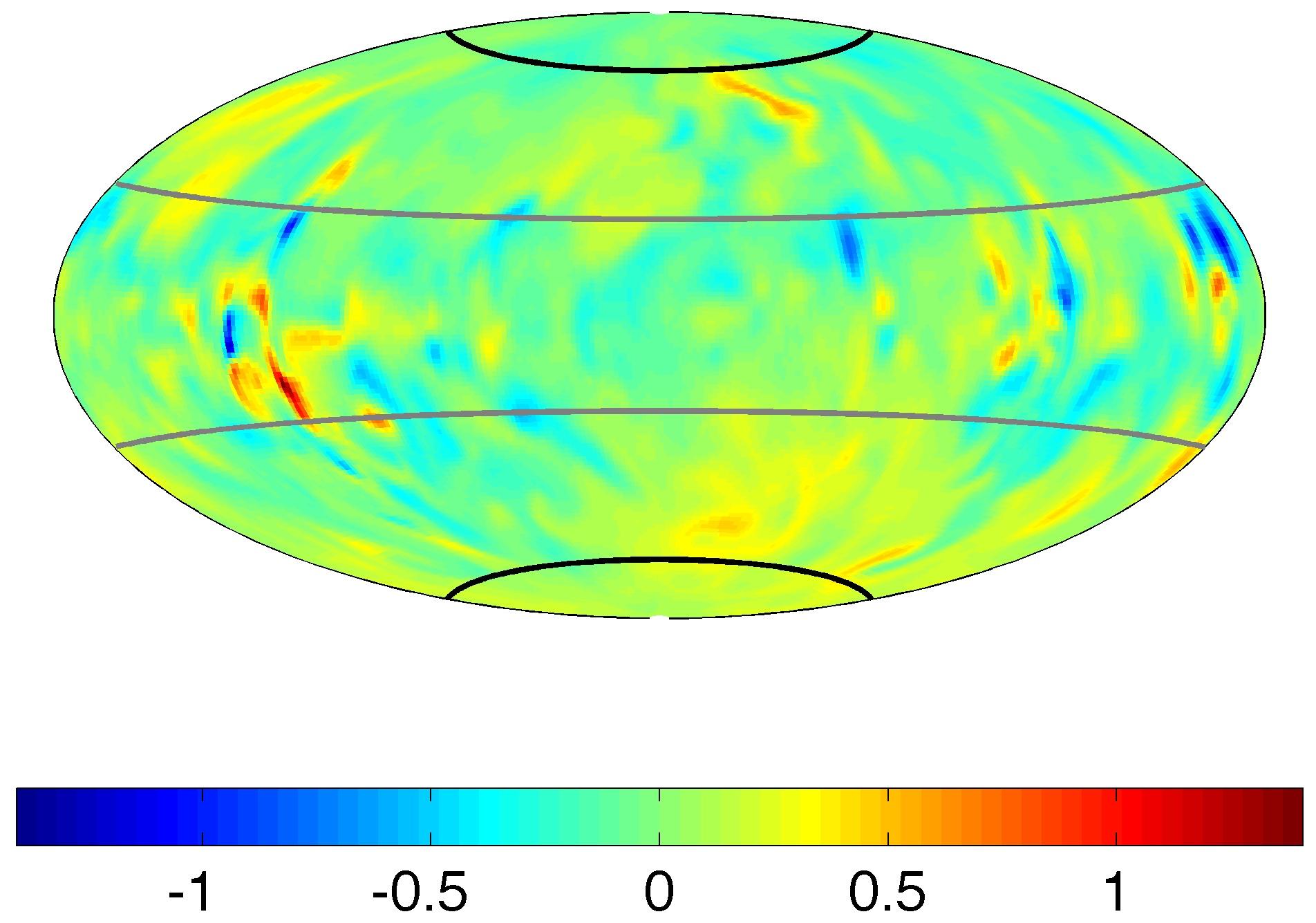}&
   \includegraphics[width=2in]{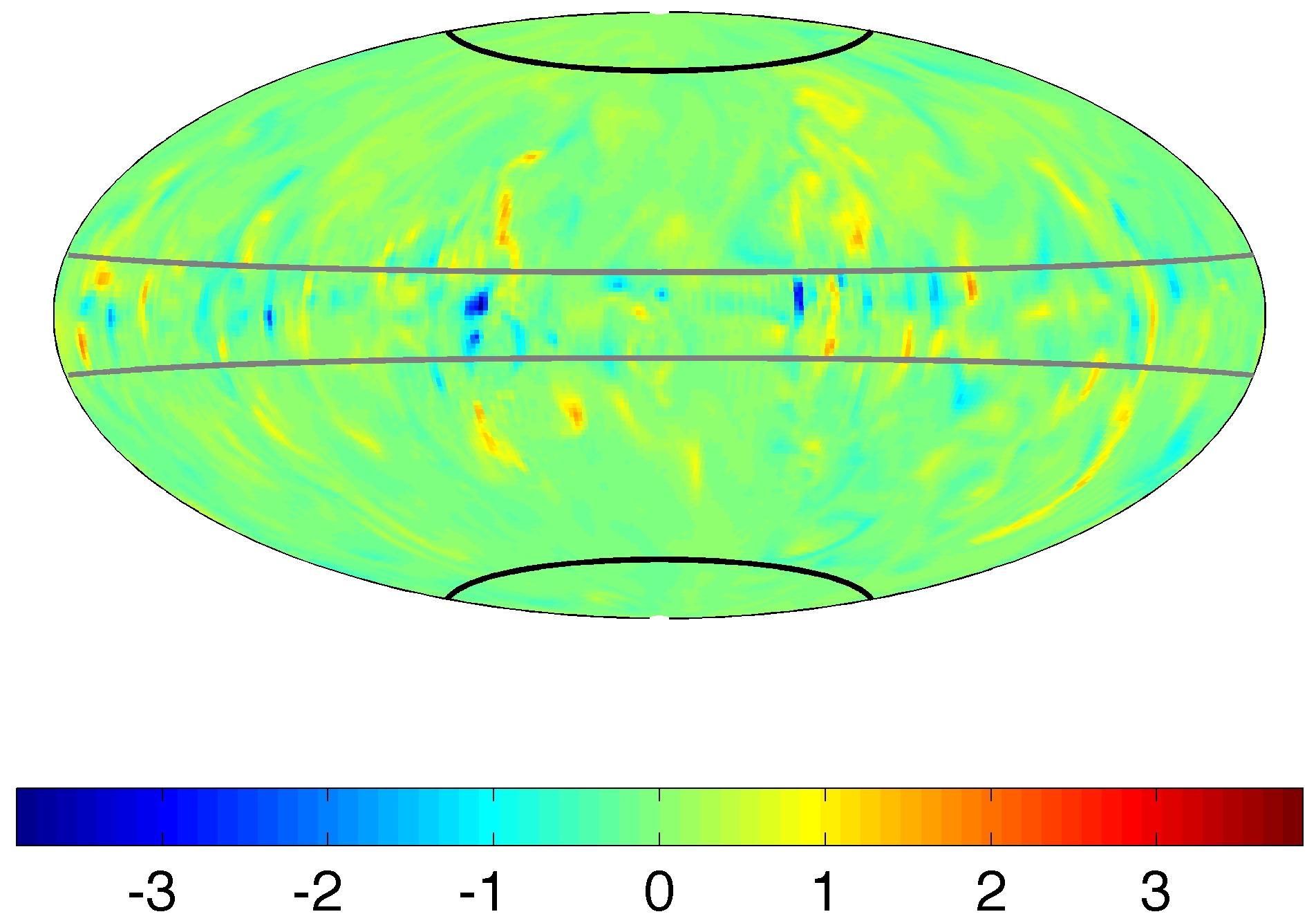}&
   \includegraphics[width=2in]{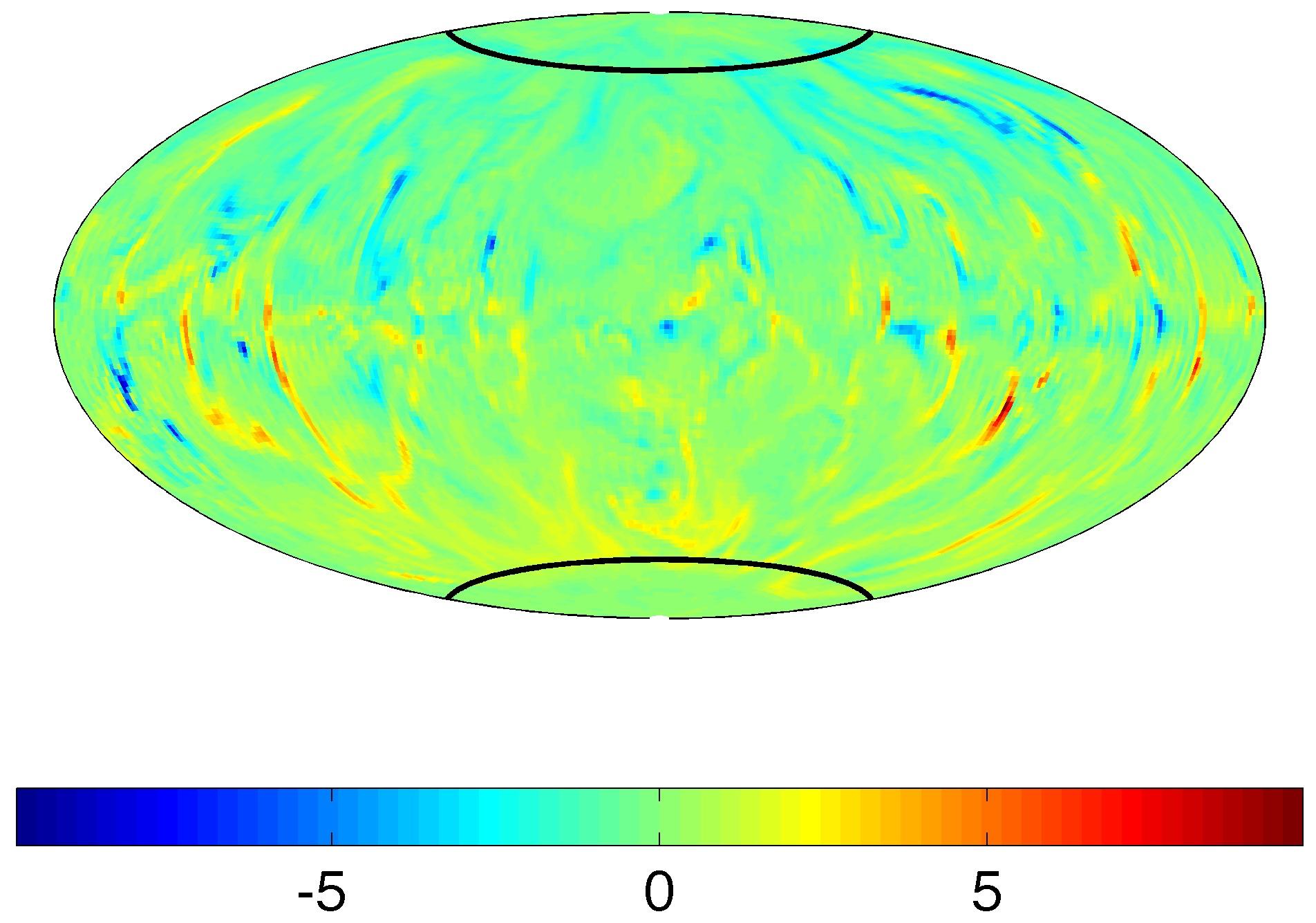}
   
   \end{tabular}
  \caption{
    Images of the radial magnetic field
    at the top of the simulated fluid are shown in a Hammer
    projection.  In each column snapshots
    for sets A, B, C and C', in columns left to right.  The rows are
    organized in values of $\chi_m$: 0.8, 0.9, 0.98 and $\infty$ from
    top to bottom.  Red (blue) lines correspond to positive (negative)
    radial field in units of $\sqrt{\rho\mu_0\lambda_i\Omega}$.  The
    projection of the tangent cylinder on the CMB is drawn here with a
    black solid line, and the projection of a tangent cylinder defined
    by the $r_m$ is drawn in grey.}
   \label{fig:Br_cmb}
   \label{lastfig}
\end{sidewaysfigure}

The radial component of the magnetic field at the top of the simulated
fluid is shown in figure~\ref{fig:Br_cmb}.

The symmetry of the magnetic field reflects the symmetry of the
internal flow. Similarly, as found by \citet{Heimpel2005a}, for
quasi-geostrophic flows (such as those in set A) the thermal plumes
generate Taylor columns in the flow that transport magnetic field
lines, resulting in a magnetic field morphology that correlates well
with the flow and the thermal structure.  For the weakly forced system
(set A), the case with a homogeneous electrical conductivity results
in a dipolar dominated dynamo with low non-axisymmetric
components. The variable conductivity cases, where the magnetic field
generated by the strong dynamo region has diffused through the low
electrical conductivity fluid, exhibit a dominant dipolar
field as well, which is highly axisymmetric (more so than the homogeneous
case).
In contrast, the strongly forced systems (sets B, C and C'), with a more
disorganized flow regime, result in non-axisymmetric fields with
significant higher degree components.

There is an identifiable transition in flow and magnetic field
morphology that corresponds to different values of the conductivity
profile radius ratio $\chi_m$. For cases with thick variable
conductivity layers (low $\chi_m$), thermal and flow fields resemble
those of non-magnetic convection (see figures 5 and 6). In contrast,
for relatively thin variable conductivity layers (high $\chi_m$)
result in flow and magnetic fields that closely resemble the
homogenous conductivity ($\chi_m= \infty$) cases. 
Such transition may be identified in Table~\ref{tab:averages}
as a large increase in the CMB Elsasser number, $\Lambda^{CMB}$,
for increasing $\chi_m$.

\begin{figure}[h!]
   \centering
   \includegraphics[width=5in]{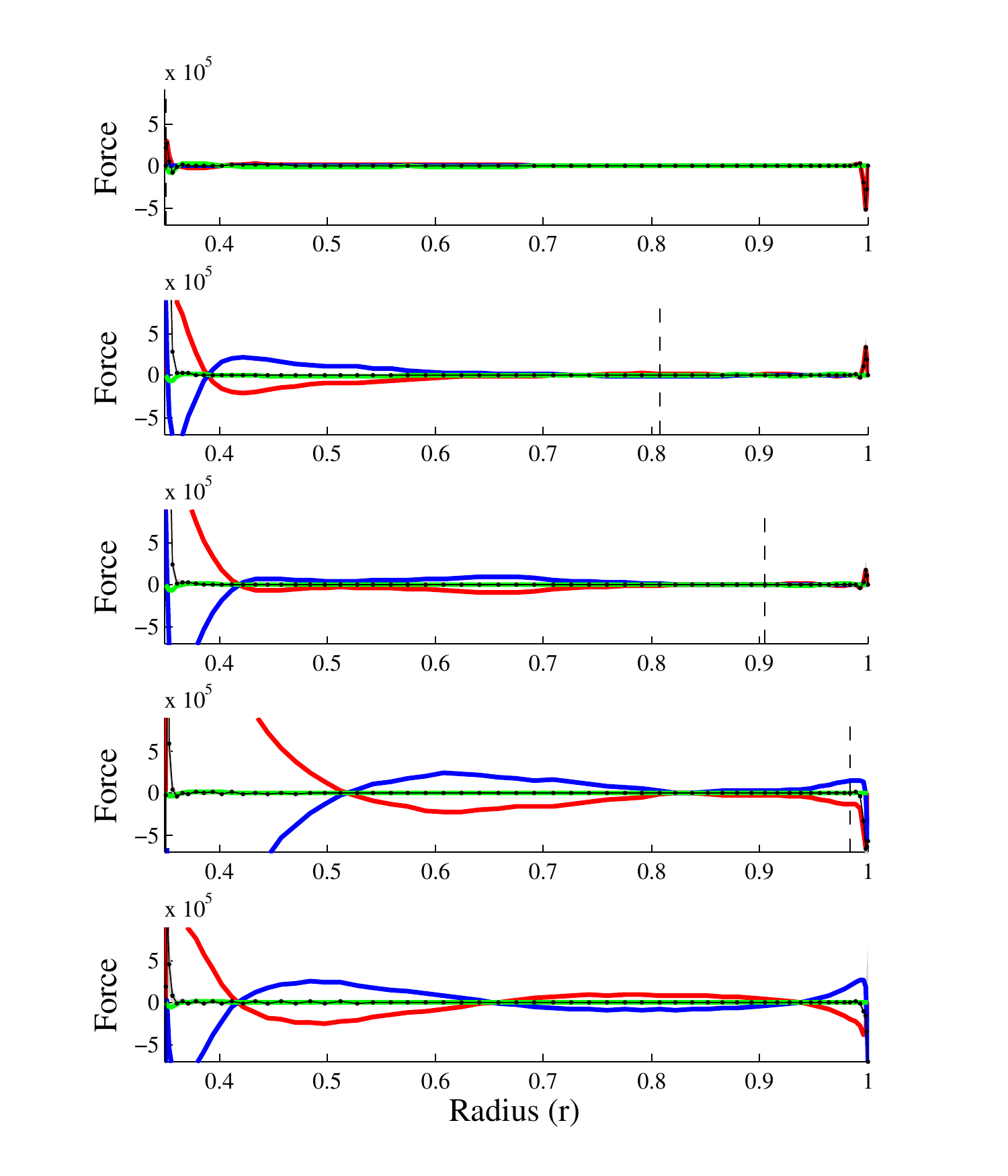}
   \caption{Time average of magnetic (blue) and Coriolis (red) forces
     as well as Reynolds stresses (green) over concentric cylinders at
     radii from $r_i$ to $r_o$. In panels top to bottom increasing
     values of $\chi_m=\emptyset, 0.8, 0.9, 0.89, \infty$ for set
     B. The sum of the Reynolds stresses, magnetic and Coriolis forces
     is shown in black dots and a solid black line with a grey shade
     showing the standard deviation of the time-average.  Each panel
     has a vertical dashed line marking $r_m$.  }
   \label{fig:forces33}
\end{figure}

\begin{figure}[h!]
   \centering
   \includegraphics[width=5in]{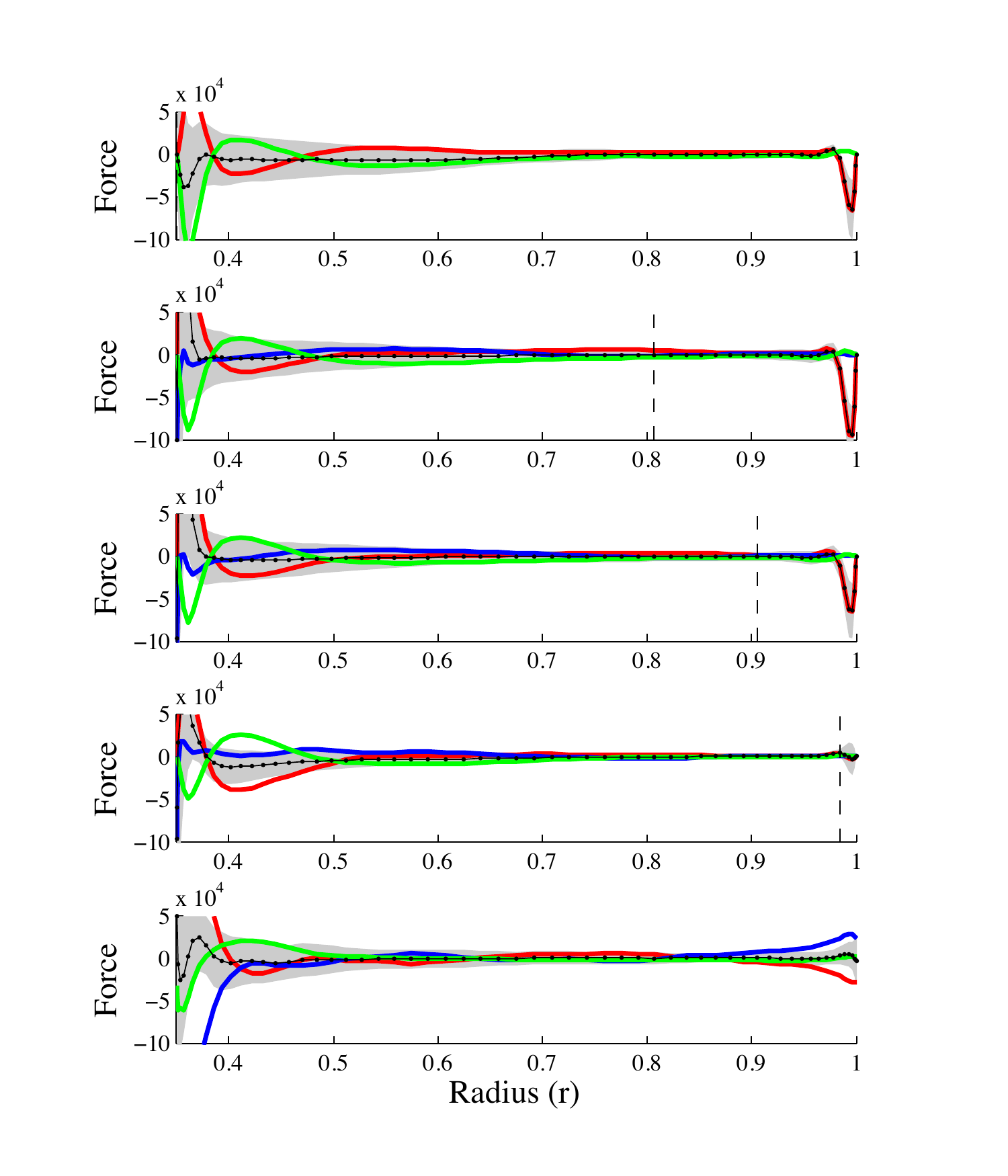}
   \caption{Similar to figure~\ref{fig:forces33} but for set C. From
     top to bottom increasing values of $\chi_m=\emptyset, 0.8, 0.9,
     0.98, \infty$. Reynolds stresses, magnetic and Coriolis forces
     are shown in green, blue and red respectively. The sum of them is
     shown in black dots joined by a solid black line.  }
   \label{fig:forces16}
\end{figure}

This transition in magnetic field morphology and magnitude
 corresponds as well to a change in the relative strength of
Coriolis and Lorentz forces.  Time averages of the Coriolis and
Lorentz forces as a function of radius for sets B and C are shown in
figures~\ref{fig:forces33} and \ref{fig:forces16}.  The volume
averages has been taken over cylinders of radius $r$ that have a
symmetry axis parallel to $\mathbf{\hat{z}}$.
In set B, runs where $\Lambda^{CMB}$ is large, the top of the core exhibits a
balance between Lorentz and Coriolis forces, with the exception of the
region included in the Ekman layer (i.e., the last three grid points
at the top of the simulated volume).
For set C, the Coriolis
is balanced by Lorentz in 
$\chi_m=\infty$.  For both sets of magnetic cases with low $\Lambda^{CMB}$, the
sum of Lorentz, Coriolis and Reynolds stresses is completely dominated
by the Coriolis force, which is balanced by viscous forces at the top
of the core.  In this set
for the case with $\chi_m=0.98$,
 it is hard to asses whether there is a  balance
between Coriolis and Lorentz forces or  whether the  Coriolis dominates.

\section{Discussion}\label{sec:diss}

We find that the geometry and temporal variability of the magnetic
field depends, to first order, on the spatial distribution of force
balances.  The decrease of the electrical conductivity at the top of the
simulated fluid results in significant changes in the resultant flow
regime. 

 \citet{Christensen2006b} reported an increase
in the wave number with increasing $Ra$ for non-magnetic convective
runs, in agreement with experimental results \citep{Aubert2001}. They
also point out that for magnetic cases the correlation between mean
wave number and Rayleigh number is incoherent, and they attribute it
to a difference in force balances present in the magnetic cases versus
the non-magnetic cases. 
In our models, for weakly forced cases, the prograde tilt of the Taylor columns
caused by the spherical boundary is allowed for the cases where the
Lorentz force is small at the boundaries (i.e., $\chi_m=0.8$ and
$\chi_m=0.9$ in set A). 
Based on the snapshots shown here, e.g., Figure~\ref{fig:Vort_eq},
there is
indication that the characteristic length scale of the flow depends
not only on the presence or absence of the Lorentz force but also on
its spatial distribution.  A reduction in the Lorentz force at the top
boundary results in an overall change in the flow for sets A and C, where the
convection becomes similar to that of the non-magnetic case.
In set B the flow behaviour  is different for high/low electrical conductivity regions
changing the characteristic length scale of the flow as a function of radius.

The models with free-slip boundaries (set C') show that strong
azimuthal flow in the non-magnetic case becomes weaker in
regions of considerable electrical conductivity (see figure~\ref{fig:Temp_eq}). This
result is expected since strong zonal flows, can develop for free slip
boundaries are damped by magnetic Lorentz forces.  The Lorentz forces
are proportional to the flow velocity and to the strength and volume
of the conductive layer.
In sets other than C', the resultant velocity is weaker. 
This is also expected, since strong zonal flows do not develop in no-slip
boundary systems.

\subsection{Boundary layers}\label{subsec:Boundary}

It has been suggested in a recent paper that rotating convection can
be controlled largely by the boundary layers \citep{King2009}.  They
compare the relation between: 1) the thermal boundary layer,
$\delta_k\sim\frac{D}{2\,N_u}$, where
$N_u=\frac{1}{4\pi\,r_i\,r_o}\frac{Q\,D}{\,\rho\,c\,\kappa\,\Delta T}$
is the Nusselt number, $c$ is the heat capacity, and $Q$ is the total
heat flow at the top boundary; and 2) the Ekman boundary layer
$\delta_E\sim D\,\sqrt{E}$.

For non-magnetic rotating convection viscous and Coriolis forces
govern the dynamics of the Ekman boundary layer.  
For convective dynamos 
there also exists
the magnetic Lorentz force. 
The non-dimensional magnetic parameter analogous to the Ekman
number 
is the Hartmann number 
$H =\left(\frac{B^2\,D^2}{\mu_0\,\lambda\,\rho\,\nu}\right)^{1/2}$ 
which is
defined as the ratio of Lorentz to viscous forces. Associated with the
Hartmann number is the Hartmann boundary layer.  Inside the Hartmann
layer viscous forces are important, while outside the layer magnetic
forces dominate. 
The Hartmann layer thickness is
$\delta_H=\frac{D}{|\mathbf{B}_\perp|}\,\sqrt{\mu_0\,\lambda\,\rho\,\nu}$,
where $\mathbf{B}_\perp$ is the magnetic field normal to the surface
\citep{Potherat2002}.
Written in terms of the non-dimensional parameters used here, 
\begin{equation}
\delta_H=D\,\sqrt{\frac{E}{\tilde{\sigma}_o\,\Lambda_r^{CMB}}},
\end{equation} 
where $\tilde{\sigma}_o$ is a measure of the normalized electrical conductivity in the layer.

\begin{figure}[h!]
   \centering
   \includegraphics[width=4in]{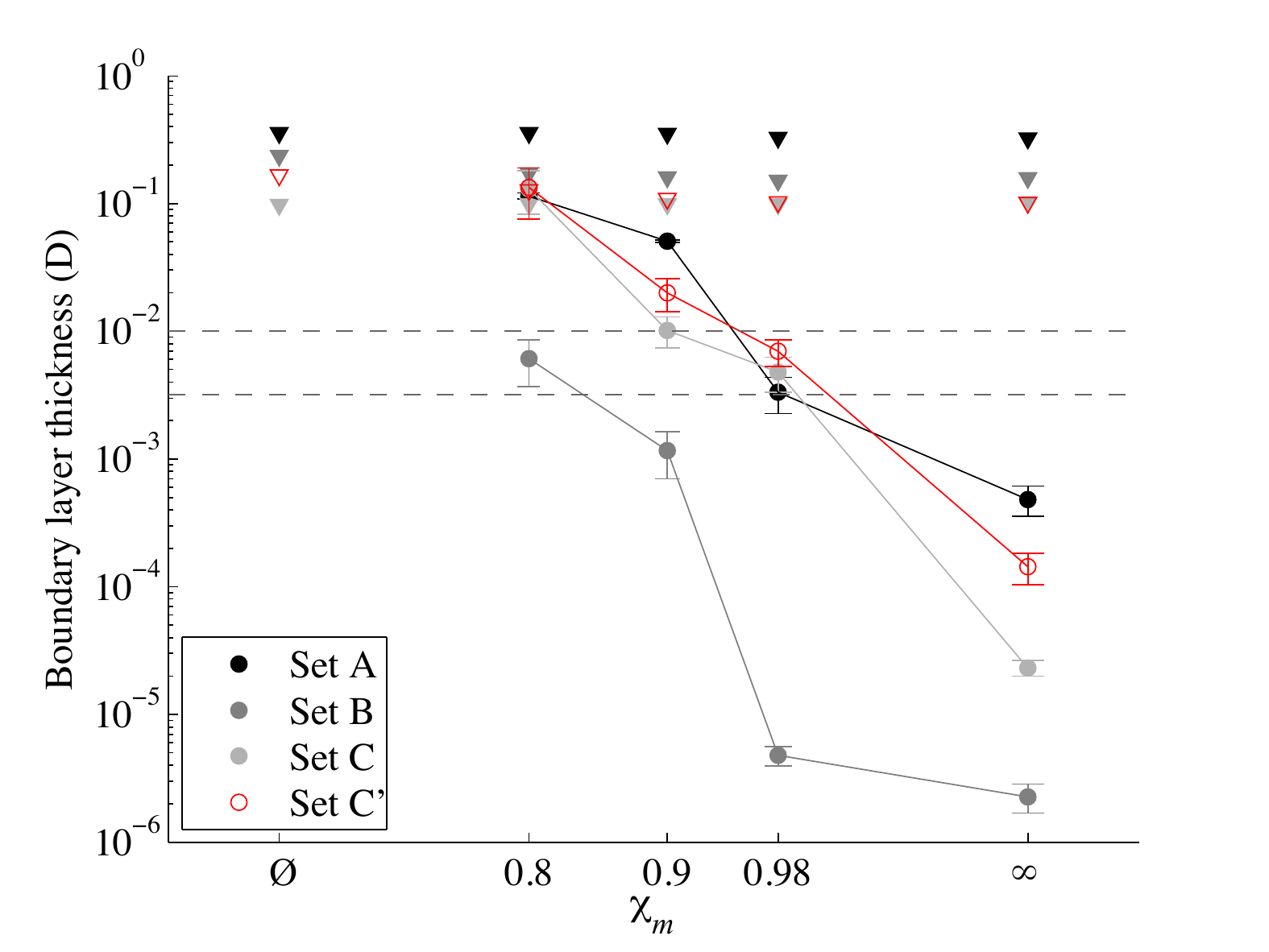} 
   \caption{This figure shows the time-averaged 
   boundary layer thickness in units of $D$ as a function of $\chi_m$.
   Triangles correspond to $\delta_k=\frac{1}{2\,N_u}$ (here the time 
   variance is small and is not plotted), and circles correspond to 
   $\delta_H=D\,\sqrt{\frac{E}{\tilde{\sigma}_o\,\Lambda_r^{CMB}}}$.
   Horizontal dashed lines for the Ekman boundary layer thickness
    are also plotted for sets A, C, and C'
   where $\delta_E=10^{-2}$ and for set B where $\delta_E=\sqrt{10^{-5}}$.
   }
   \label{fig:Deltas_vs_Xm}
\end{figure}

For this study the boundary layers are also affected by radially variable
electrical conductivity. Decreasing the conductivity
near the outer boundary decreases the Lorentz force. 
A decreased electrical conductivity allows for a
thicker boundary layer in which turbulent and/or molecular viscous
forces are important.
Figure~\ref{fig:Deltas_vs_Xm} shows the time averaged boundary layer 
thickness of  Hartmann, Ekman and thermal boundary layers
as a function of $\chi_m$.
In sets A, C and C' the Ekman boundary layer thickness is $\delta_E=10^{-2}$.
When Lorentz forces do not influence the flow 
$\delta_H>\delta_E$. This is case for $\chi_m=0.8$ in sets A, C and C' as well as 
and $\chi_m=0.9$ in sets A and C' 
(see also Table~\ref{tab:averages}).  In the case of set B where
$\delta_E\sim0.0032$, $\delta_H>\delta_E$ only for $\chi_m=0.8$, but
for $\chi_m=0.9$ $\delta_H=0.001160$ is only a factor of 3 times
smaller than $\delta_E$.  The transition in the flow behavior, found
to be a consequence of the force balances at the top of the core, is
correlated to the relation between the Ekman and Hartmann boundary
layers.

When pressure gradients and gravity forces
are balanced by the Coriolis force the flow is in a geostrophic balance.
If the Coriolis is balanced by Lorentz forces, 
the flow is said to be in a magnetostrophic balance.
Noting that we keep the thermal boundary layer 
constant for each set, and that $\delta_k$ is always large 
(see Table~\ref{tab:averages} and figure~\ref{fig:Deltas_vs_Xm}),
we found that the relative thickness of Hartmann and Ekman layers 
results in force balances at the top of the core allowing for Lorentz or Coriolis 
forces to control the flow, resulting in magnetostrophic ($\delta_H<\delta_E$)
or geostrophic ($\delta_E<\delta_H$) systems.

Inspection of the radial magnetic field (Figure~\ref{fig:Br_cmb}), reveals the
interplay between the Ekman and Hartmann boundary layers, and the
variable conductivity layers near the outer boundary for the various
cases. For set A, a relatively low $Ra$ results in relatively weak
magnetic field. Here the Hartmann layer thickness is similar to that
of the low conductivity layer.  For this set, we interpret the
presence of strong surface radial magnetic field patches near the
magnetic tangent cylinder to be indicative of the flow field, which
tends to be deflected by a Hartmann layer topography 
\citep{Potherat2002}.
For Set B, lower $E$ and higher
$Ra$, result in higher $\Lambda$ and a $\delta_{H}$ that is thinner
than the variable conductivity layer in the cases with $\chi_m = 0.8$
and $\chi_m = 0.9$. This results in high radial magnetic field patches
at the outer surface outside the magnetic tangent cylinder (i.e.,\ at
lower latitudes than the intersection of the tangent cylinder with the
outer surface).

In summary, for thick Hartmann layers, $\delta_H>\delta_E$,
the flow is unaffected by the presence of magnetic forces
and it may be controlled by the relation between 
Ekman and thermal boundary layers \citep{King2009}.
However for a thin layer, $\delta_H\lesssim\delta_E$, 
the Lorentz forces dominate 
outside the Hartmann boundary layer while inside both, thermal and Ekman boundary layers.
In our models, Ekman and Hartmann boundary layers 
define which force (Coriolis or Lorentz) is dominant just inside the
region in which viscous forces are significant, 
so that the system follows a geostrophic or a 
magnetostrophic regime.

For rotating-convection, \citet{King2009} argue that for stress-free
boundaries, there is a thermal Ekman layer which takes the place 
of the Ekman layer. Such layer would define the volume over which 
the rotation responds to density perturbations, modifying plume formation.
They argue that the Ekman layer is dynamically important in stress-free 
systems. 
We find a transition in set C' for $\Lambda^{CMB}$ to be between $\chi_m=0.9$
and $\chi_m=0.98$ This transition coincides with the change in the relation between
$\delta_H$ and the traditional $\delta_E=\sqrt{E}$ (see figure~\ref{fig:Deltas_vs_Xm}).
This is consistent with the relevance of $\delta_E$ in the dynamics with or without
stress-free boundary conditions.

\subsection{Planetary implications}

Computational limitations preclude the use of accurate 
diffusion coefficients in modeling the Earth and planets. 
Furthermore, since we have carried out only a few cases that show 
the correspondence of strong and a weak external magnetic fields 
to boundary layers with different viscosities and electrical conductivity 
profiles, we are not in a position to attempt to scale our present results 
to conditions in planetary interiors.  However,  we can come 
to some preliminary conclusions using estimated planetary values of 
the Ekman and Hartmann layer thicknesses. Considering 
turbulent molecular viscous diffusion we can estimate the earth-like 
Ekman numbers of $10^{-9}$ and $10^{-15}$, respectively, yielding Ekman 
layer thickness estimates of order 200 m and 0.2 m, respectively.  
Considering an Elsasser number at the top of Earth's core of 
0.3 \citep{Stevenson2003}, 
the Hartmann to Ekman layer thickness 
of Earth may be estimated to be $\delta_H/\delta_E\sim2$
where both, Ekman and Hartmann layers thickness are comparable,
and a magnetostrophic core is expected.

Observations of
Mercury's magnetic field by Mariner 10 and recent MESSENGER flybys are
consistent with a Hermean magnetic field that is weaker by about two
orders of magnitude, but roughly similar in morphology to that of
Earth (a moderately tilted axial dipole).
Because Mercury is small, and its rotation rate is quite low (its
sidereal day is 58.6 Earth days), its Ekman number is about 100 times
greater than that of Earth. The Elsasser number
scales like $\Lambda \propto B^2/\Omega$, Mercury's $\Lambda$ is less
than that of Earth by a factor of roughly 1/500. Since the Hartmann
layer thickness scales like $\delta_{H} \sim \delta_E/\sqrt{\Lambda}$ we
obtain estimates for a Hermean Hartmann to Ekman layer thickness 
to be $\delta_H/\delta_E\sim100$ where one expects 
a non-magnetostrophic flow. 

\section{Conclusions}\label{sec:con}

In our models, the dynamo-generated magnetic field morphology and
intensity are strongly affected by the relative strength of viscous,
Coriolis and Lorentz forces near the outer boundary. The relative
scale of Ekman and Hartmann boundary layer thicknesses is determined
by an electrical conductivity gradient. For uniform electrical
conductivity models Coriolis and Lorentz forces are typically of of
the same order, yielding a magnetostrophic balance.  Low electrically
conductivity near the top boundary can separate the
Ekman-Hartman layer into a thin Ekman layer and a thicker Hartmann
layer, resulting in changes of the detailed magnetic field morphology, and
a large-scale external magnetic field that is relatively weak (see
Table~\ref{tab:averages}).  Given the likelihood of a high
concentration of light element in Mercury's core, it is plausible that
a low electrical conductivity layer is present near the core mantle
boundary.  This means that the dynamics of the boundary layers
obtained in our models may be applicable to conditions in Mercury's
core. Our results imply that a low conductivity layer is consistent
with Mercury's weak observable magnetic field. However, while radially
variable electrical conductivity is the mechanism studied in this
paper, it is one of several models that can result in weak magnetic
fields. With the anticipated arrival of the MESSENGER spacecraft in
orbit, we will soon be in a position to use detailed mapping to
better constrain the relative contributions of the various dynamical
processes that generate Mercury's global magnetic field.

\section*{Acknowledgements}
The authors  thank S. C. Solomon and J. M.  Aurnou for helpful comments on the manuscript,
Erik King for enlightening discussions, and 
the two anonymous referees for comments and questions that broadened 
the focus and improved greatly the quality of the manuscript.
At DTM, NGP has been supported by the NASA MESSENGER project and by the NASA Planetary Geology and Geophysics Program, under contract NASW-00002 and grant NNX07AP50G, respectively.
Computational resources were provided by the Western Canada Research Grid (Westgird).

\appendix
\section{Symbols}\label{sec:symbols}

%

\begin{longtable}{|l|p{10cm}|l|} 
\caption{Symbols used throughout the document}\\    
\hline
Symbol    & Description & units\\
\hline
 \endhead
\hline \endfoot

      $a$				&		polynomial exponent 
      							of the electrical conductivity function	 &	\\      
      $\alpha$			&		thermal expansion coefficient	&	K$^{-1}$\\   
      $\mathbf{B}$		&		magnetic field induction
      							 vector				&		$(\rho\, \mu_0\,\lambda_i\,\Omega)^{1/2}$\\
      $c$				&		heat capacity			&		J (kg K)$^{-1}$\\
      $\chi$				&		shell radii ratio			&		\\
      $\chi_m$			&		electrically conductive 
      							volume radii ratio		&		\\
      $\delta_E$			&		Ekman boundary layer 
      							thickness				&		$D$\\      
      $\delta_H$			&		Hartmann boundary layer 
      							thickness				&		$D$\\ 
      $\delta_k$			&		thermal boundary layer 
      							thickness				&		$D$\\ 
      $E$				&		Ekman number			&		\\
      $g$				&		acceleration of gravity	&		$\nu^2\,D^{-3}$\\
      $g_o$				&		acceleration of gravity
      							at the CMB			&		$\nu^2\,D^{-3}$\\
      $H$				&		Hartmann number		&		\\ 
      $\kappa$			&		thermal diffusivity		&		m$^2$s$^{-1}$\\      
      $l_{max}$			&		maximum spherical 
      							harmonic degree		&		\\      
      $\Lambda$			&		model-normalized
      							Elsasser number		&		$\rho\, \mu_0\,\lambda_i\,\Omega$\\
      $\Lambda^{CMB}$	&		model-normalized
      							CMB Elsasser number	&		$\rho\, \mu_0\,\lambda_i\,\Omega$\\
      $\Lambda_r^{CMB}$	&		model-normalized
      							CMB radial field 
							Elsasser number	&		$\rho\, \mu_0\,\lambda_i\,\Omega$\\
      $\lambda$			&		fluid electrical 
      							diffusivity				&		m$^2$s$^{-1}$\\
      $\tilde{\lambda}$		&		normalized electrical 
      							diffusivity				&		$\lambda_i$\\
      $\lambda_i$		&		electrical diffusivity
      							at the ICB				&		m$^2$s$^{-1}$\\
      $M$				&		magnetic energy in the 
      							fluid core				&		\\
      $M^a$				&		axisymmetric magnetic energy in the 
      							fluid core				&		\\
      $M_o$				&		magnetic energy 
      							at the CMB			&		\\
      $M^a_o$			&		axisymmetric magnetic energy 
      							at the CMB			&		\\							     
      $\mu_0$			&		magnetic permeability
      							of vacuum				&		H m$^{-1}$\\      
      $Nu$				&		Nusselt number		&		\\      
      $\nu$				&		kinematic viscosity		&		m$^2$s$^{-1}$\\      
      $\mathbf{\Omega}$	&		Angular momentum		&		s$^{-1}$\\
      $P$				&		Pressure scalar		&		$\rho\,\nu\,\Omega$\\
      $\Phi$				&		toroidal potential		&		\\
      $\Psi$				&		poloidal potential		&		\\
      $Pm^*$			&		modified Magnetic 
      							Prandtl number			&		\\
      $Pr$				&		Prandtl number			&		\\
       $Q$				&		total heat flux at CMB	&		\\
       $Q^a$			&		ratio axisymmetric dipole to total 
       							magnetic energy		&		\\
       $Q^a_o$			&		ratio axisymmetric dipole to total 
       							magnetic energy at the CMB		&		\\
       $\mathbf{\hat{r}}$	&		Unit vector in the radial direction	&		\\
      $\rho$				&		Fluid density			&		kg m$^{-3}$\\
      $Ra$				&		Rayleigh number		&		\\
      $Ra_c$			&		critical Rayleigh number
      							for the onset of convection		&		\\
      $Re$				&		Reynolds number		&		\\
      $r_i$				&		Shell internal radius		&		$D$\\
      $r_m$				&		electrical conductivity 
      							transition radius		&		$D$\\
      $r_o$				&		Shell external radius		&		$D$\\
      $\sigma$			&		fluid electrical conductivity&		m$^{-2}$s\\
      $\tilde{\sigma}$		&		normalized electrical 
      							conductivity			&		$\sigma_i$\\
      $\sigma_i$			&		electrical conductivity
      							at the ICB				&		m$^{-2}$s\\
      $\tilde\sigma_m$		&		normalized electrical conductivity
      							at $r_m$				&		$\sigma_i$\\
      $\tilde\sigma_o$		&		normalized electrical conductivity
      							at the CMB			&		$\sigma_i$\\
      $T$				&		Temperature scalar		&		$\Delta T$\\
      $\mathbf{u}$		&		flow velocity vector		&		$\nu\, D^{-1}$\\
     $\mathbf{\hat{z}}$		&		Unit vector in the 
     							direction of the angular 
							momentum			&		
  
   \label{tab:symbols}	   \label{lasttable}  
   \end{longtable}

   \label{lastpage}

\bibliography{VC_preprint}
 \bibliographystyle{elsart-harv}

\end{document}

%% file: tab_1.txt
\begin{tabular}{|cccr@{$\pm$}lr@{$\pm$}lr@{$\pm$}lr@{$\pm$}lr@{$\pm$}lr@{$\pm$}lr@{$\pm$}l|}  
\hline 
 &	$Ra/Ra_{c}$ &  
$\chi_m$ &  
\multicolumn{2}{c}{$Re$}&  
\multicolumn{2}{c}{$\Lambda$}&  
\multicolumn{2}{c}{$\Lambda^{CMB}$}&  
\multicolumn{2}{c}{$Q^a_o$} & 
\multicolumn{2}{c}{$Q^a$} & 
\multicolumn{2}{c}{$\delta_H$} & 
\multicolumn{2}{c|}{$\delta_k$}  
\\  
\hline \hline 
\multirow{5}{*}{A} & 3.4 &  	 0.00&  	 133.5 & 9.5 &  	\multicolumn{2}{c}{} & 	\multicolumn{2}{c}{} & 	\multicolumn{2}{c}{} &  	\multicolumn{2}{c}{} &  	\multicolumn{2}{c}{}  &	 0.3575 & 0.0022   	\\  
 & 3.4 &  	 0.80&  	 121.7 & 6.1 &  	 3.8 & 0.2 &  	 0.029 & 0.001 &  	 0.78 & 0.01 &  	 0.0565 & 0.0036 &  	 0.114659 & 0.006442 &  	 0.3554 & 0.0014   	\\  
 & 3.4 &  	 0.90&  	 118.6 & 4.5 &  	 7.5 & 1.2 &  	 0.026 & 0.001 &  	 0.67 & 0.03 &  	 0.0257 & 0.0057 &  	 0.050601 & 0.001241 &  	 0.3497 & 0.0029   	\\  
 & 3.4 &  	 0.98&  	 123.2 & 7.0 &  	 23.0 & 4.2 &  	 0.250 & 0.100 &  	 0.61 & 0.10 &  	 0.0513 & 0.0126 &  	 0.003315 & 0.001058 &  	 0.3266 & 0.0117   	\\  
 & 3.4 &  	 $\infty$ &  	 109.6 & 9.5 &  	 36.4 & 6.8 &  	 1.055 & 0.272 &  	 0.60 & 0.06 &  	 0.1224 & 0.0195 &  	 0.000483 & 0.000129 &  	 0.3228 & 0.0163   	\\  
\hline \hline  
\multirow{3}{*}{B} & 7.8 &  	 0.00&  	 638.7 & 23.1 &  	\multicolumn{2}{c}{} & 	\multicolumn{2}{c}{} & 	\multicolumn{2}{c}{} &  	\multicolumn{2}{c}{} &  	\multicolumn{2}{c}{}  &	 0.2355 & 0.0023   	\\  
& 7.8 &  	 0.80&  	 542.6 & 34.2 &  	 67.6 & 8.1 &  	 0.120 & 0.043 &  	 0.05 & 0.07 &  	 0.0022 & 0.0020 &  	 0.006126 & 0.002463 &  	 0.1726 & 0.0155   	\\  
& 7.8 &  	 0.90&  	 505.8 & 32.1 &  	 99.1 & 11.9 &  	 0.403 & 0.134 &  	 0.01 & 0.01 &  	 0.0014 & 0.0009 &  	 0.001160 & 0.000463 &  	 0.1604 & 0.0135   	\\  
& 7.8 &  	 0.98&  	 360.4 & 27.6 &  	 387.4 & 56.0 &  	 20.269 & 4.161 &  	 0.43 & 0.07 &  	 0.1095 & 0.0170 &  	 0.000005 & 0.000001 &  	 0.1514 & 0.0102   	\\  
& 7.8 &  	 $\infty$ &  	 360.1 & 31.9 &  	 447.7 & 102.7 &  	 34.748 & 11.645 &  	 0.32 & 0.07 &  	 0.1105 & 0.0212 &  	 0.000002 & 0.000001 &  	 0.1572 & 0.0141   	\\  
\hline \hline 
\multirow{5}{*}{C} & 16.8 &  	 0.00&  	 814.1 & 31.4 &  	\multicolumn{2}{c}{} & 	\multicolumn{2}{c}{} & 	\multicolumn{2}{c}{} &  	\multicolumn{2}{c}{} &  	\multicolumn{2}{c}{}  &	 0.0974 & 0.0043   	\\  
& 16.8 &  	 0.80&  	 704.0 & 32.3 &  	 39.9 & 4.2 &  	 0.042 & 0.010 &  	 0.04 & 0.05 &  	 0.0014 & 0.0009 &  	 0.413985 & 0.155259 &  	 0.0990 & 0.0047   	\\  
& 16.8 &  	 0.90&  	 662.3 & 29.9 &  	 65.8 & 5.5 &  	 0.329 & 0.060 &  	 0.02 & 0.02 &  	 0.0016 & 0.0011 &  	 0.032156 & 0.008828 &  	 0.0984 & 0.0049   	\\  
& 16.8 &  	 0.98&  	 661.0 & 30.0 &  	 66.5 & 6.1 &  	 0.335 & 0.063 &  	 0.02 & 0.03 &  	 0.0016 & 0.0012 &  	 0.015181 & 0.004695 &  	 0.0984 & 0.0046   	\\  
& 16.8 &  	 $\infty$ &  	 506.8 & 27.2 &  	 336.3 & 30.2 &  	 22.479 & 2.637 &  	 0.13 & 0.02 &  	 0.0375 & 0.0059 &  	 0.000073 & 0.000010 &  	 0.1020 & 0.0054   	\\  
\hline \hline 
\multirow{5}{*}{C'} & 16.8 &  	 0.00&  	 1646.0 & 66.1 &  	\multicolumn{2}{c}{} & 	\multicolumn{2}{c}{} & 	\multicolumn{2}{c}{} &  	\multicolumn{2}{c}{} &  	\multicolumn{2}{c}{}  &	 0.1664 & 0.0261   	\\  
& 16.8 &  	 0.80&  	 1485.3 & 24.7 &  	 66.5 & 7.0 &  	 0.096 & 0.029 &  	 0.04 & 0.05 &  	 0.0018 & 0.0012 &  	 0.132790 & 0.057043 &  	 0.1277 & 0.0099   	\\  
& 16.8 &  	 0.90&  	 883.8 & 33.7 &  	 103.8 & 9.0 &  	 0.480 & 0.110 &  	 0.01 & 0.02 &  	 0.0015 & 0.0009 &  	 0.019941 & 0.005702 &  	 0.1084 & 0.0074   	\\  
& 16.8 &  	 0.98&  	 753.7 & 33.2 &  	 117.4 & 8.8 &  	 1.258 & 0.238 &  	 0.01 & 0.01 &  	 0.0015 & 0.0010 &  	 0.006946 & 0.001642 &  	 0.1023 & 0.0066   	\\  
& 16.8 &  	 $\infty$ &  	 627.4 & 33.2 &  	 276.7 & 35.7 &  	 9.576 & 2.099 &  	 0.20 & 0.03 &  	 0.0329 & 0.0074 &  	 0.000143 & 0.000040 &  	 0.1008 & 0.0060   	\\  
\hline  
\end{tabular} 

%% file: VC_preprint.bbl
\begin{thebibliography}{50}
\expandafter\ifx\csname natexlab\endcsname\relax\def\natexlab#1{#1}\fi
\expandafter\ifx\csname url\endcsname\relax
  \def\url#1{\texttt{#1}}\fi
\expandafter\ifx\csname urlprefix\endcsname\relax\def\urlprefix{URL }\fi

\bibitem[{Al-Shamali et~al.(2004)Al-Shamali, Heimpel, and
  Aurnou}]{Al-Shamali2004}
Al-Shamali, F., Heimpel, M.~H., Aurnou, J.~M., 2004. Varying the spherical
  shell geometry in rotating thermal convection. Geophys. Astrophys. Fluid Dyn.
  98, 153--169.

\bibitem[{Alexandrakis and Eaton(2007)}]{Alexandrakis2007}
Alexandrakis, C., Eaton, D.~W., 2007. Empirical transfer functions: Application
  to determination of outermost core velocity structure using {SmKS} phases.
  Geophysical Research Letters 34, L22317.

\bibitem[{Anufriev et~al.(2005)Anufriev, Jones, and Soward}]{Anufriev2005}
Anufriev, A., Jones, C., Soward, A., 2005. The {B}oussinesq and anelastic
  liquid approximations for convection in the {E}arth's core. Physics of the
  Earth and Planetary Interiors 152, 163--190.

\bibitem[{Aubert et~al.(2001)Aubert, Brito, Nataf, Cardin, and
  Masson}]{Aubert2001}
Aubert, J., Brito, D., Nataf, H., Cardin, P., Masson, J., 2001. A systematic
  experimental study of rapidly rotating spherical convection in water and
  liquid gallium. Physics of the Earth and Planetary Interiors 128~(1-4),
  51--74.

\bibitem[{Birch(1952)}]{Birch1952}
Birch, F., 1952. Elasticity and constitution of the {E}arth interior. J Geophys
  Res 57, 227--286.

\bibitem[{Braginsky(1993)}]{Braginsky1993}
Braginsky, S.~I., 1993. {MAC}-oscillations of the hidden ocean of the core. J
  Geomagn Geoelectr 45, 1517--1538.

\bibitem[{Braginsky(2007)}]{Braginsky2007}
Braginsky, S.~I., 2007. Formation of the stratified ocean of the core: A
  ternary alloy model. Earth and Planetary Science Letters 253, 507--512.

\bibitem[{Buffett et~al.(2000)Buffett, Garnero, and Jeanloz}]{Buffett2000}
Buffett, B., Garnero, E., Jeanloz, R., 2000. Sediments at the top of {E}arth's
  core. Science 290, 1338--1342.

\bibitem[{Caracas and Verstraete(2009)}]{Caracas2009}
Caracas, R., Verstraete, M., 2009. {Fe-Si} alloys in the lowermost mantle and
  the outer core. EosTrans. AGU 90~(22), Jt. Assem. Suppl. Abstract DI71A--07.

\bibitem[{Chandrasekhar(1961)}]{Chandrasekhar1961}
Chandrasekhar, S., 1961. Hydrodynamic and Hydromagnetic stability. Dover.

\bibitem[{Christensen et~al.(1998)Christensen, Olson, and
  Glatzmaier}]{Christensen1998}
Christensen, U., Olson, P., Glatzmaier, G., Jan 1998. A dynamo model
  interpretation of geomagnetic field structures. Geophysical Research Letters
  25~(10), 1565--1568.

\bibitem[{Christensen et~al.(1999)Christensen, Olson, and
  Glatzmaier}]{Christensen1999}
Christensen, U., Olson, P., Glatzmaier, G., 1999. Numerical modelling of the
  geodynamo: a systematic parameter study. Geophys J Int 138, 393--409.

\bibitem[{Christensen(2006)}]{Christensen2006a}
Christensen, U.~R., 2006. A deep dynamo generating {M}ercury's magnetic field.
  Nature 444, 1056--1058.

\bibitem[{Christensen and Aubert(2006)}]{Christensen2006b}
Christensen, U.~R., Aubert, J., 2006. Scaling properties of convection-driven
  dynamos in rotating spherical shells and aplication to planetary magnetic
  fields. Geophys. J. Int. 166, 97--114.

\bibitem[{Christensen and Wicht(2008)}]{Christensen2008}
Christensen, U.~R., Wicht, J., 2008. Models of field generation in partly
  stable plantary cores: Applications to {M}ercury and {S}aturn. Icarus 196,
  16--34.

\bibitem[{Clune et~al.(1999)Clune, Elliott, Miesch, Toomre, and
  Glatzmaier}]{Clune1999}
Clune, T., Elliott, J., Miesch, M., Toomre, J., Glatzmaier, G., Jan 1999.
  Computational aspects of a code to study rotating turbulent convection in
  spherical shells. Parallel Comput 25~(4), 361--380.

\bibitem[{Dormy et~al.(1998)Dormy, Cardin, and Jault}]{Dormy1998}
Dormy, E., Cardin, P., Jault, D., 1998. {MHD} flow in a slightly differentially
  rotating spherical shell, with conducting inner core. Earth and Planetary
  Science Letters.

\bibitem[{Evonuk and Glatzmaier(2004)}]{Evonuk2004}
Evonuk, M., Glatzmaier, G.~A., 2004. {2D} studies of various approximations
  used for modeling convection in giant planets. Geophysical and Astrophysical
  Fluid Dynamics 98, 241--255.

\bibitem[{Fearn and Loper(1981)}]{Fearn1981}
Fearn, D., Loper, D., Jan 1981. The effect of composition upon the
  stratification of the core. Geophys J Roy Astr S 65~(1), 258--258.

\bibitem[{Featherstone et~al.(2007)Featherstone, Browning, Brun, and
  Toomre}]{Featherstone2007}
Featherstone, N.~A., Browning, M.~K., Brun, A.~S., Toomre, J., 2007. Dynamo
  action in the presence of an imposed magnetic field. Astron. Nachr. 328,
  1126--1129.

\bibitem[{Glatzmaier(1984)}]{Glatzmaier1984}
Glatzmaier, G., 1984. Numerical simulations of stellar convective dynamos .1.
  the model and method. J Comput Phys 55, 461--484.

\bibitem[{Glatzmaier and Roberts(1995)}]{Glatzmaier1995a}
Glatzmaier, G., Roberts, P., 1995. A 3-dimensional self-consistent
  computer-simulation of a geomagnetic-field reversal. Nature 377, 203--209.

\bibitem[{{G{\'{o}}mez-P{\'{e}}rez}(2007)}]{Gomez2007a}
{G{\'{o}}mez-P{\'{e}}rez}, N., 2007. {Planetary magnetic fields in the solar
  system: A numerical study of dynamo models}. Ph.D. thesis, University of
  Alberta (Canada).

\bibitem[{G{\'{o}}mez-P{\'{e}}rez and Heimpel(2007)}]{Gomez2007b}
G{\'{o}}mez-P{\'{e}}rez, N., Heimpel, M.~H., Oct 2007. Numerical models of
  zonal flow dynamos: An application to the ice giants. Geophys. Astrophys.
  Fluid Dyn. 101~(5 {\&} 6), 371--388.

\bibitem[{Guillot(2005)}]{Guillot2005}
Guillot, T., 2005. The interiors of giant planets: Models and outstanding
  questions. Ann. Rev. Earth Planet. Sci. 33, 493--530.

\bibitem[{Hauck et~al.(2004)Hauck, Dombard, Phillips, and Solomon}]{Hauck2004}
Hauck, S., Dombard, A., Phillips, R., Solomon, S., 2004. Internal and tectonic
  evolution of {M}ercury. Earth and Planetary Science Letters 222, 713--728.

\bibitem[{Heimpel and Kabin(2008)}]{Heimpel2008}
Heimpel, M., Kabin, K., 2008. {M}ercury redux. Nature Geosci 1, 564--566.

\bibitem[{Heimpel et~al.(2005)Heimpel, Aurnou, Al-Shamali, and
  G{\'{o}mez}~P{\'{e}}rez}]{Heimpel2005a}
Heimpel, M.~H., Aurnou, J.~M., Al-Shamali, F.~M., G{\'{o}mez}~P{\'{e}}rez, N.,
  2005. A numerical study of dynamo action as a function of spherical shell
  geometry. Earth Planet. Sci. Lett. 236~(1-2), 542--557.

\bibitem[{Jacobs(1975)}]{Jacobs1975}
Jacobs, J.~A., 1975. The {E}arth's Core. Vol.~10. Academic Press, London.

\bibitem[{Kageyama et~al.(1993)Kageyama, Watanabe, and Sato}]{Kageyama1993}
Kageyama, A., Watanabe, K., Sato, T., 1993. Simulation study of a
  magnetohydrodynamic dynamo - convection in a rotating spherical-shell. Phys
  Fluids B-Plasma 5, 2793--2805.

\bibitem[{King et~al.(2009)King, Stellmach, Noir, Hansen, and
  Aurnou}]{King2009}
King, E.~M., Stellmach, S., Noir, J., Hansen, U., Aurnou, J.~M., 2009. Boundary
  layer control of rotating convection systems. Nature 457, 301--304.

\bibitem[{Kuang and Bloxham(1997)}]{Kuang1997}
Kuang, W., Bloxham, J., 1997. An {E}arth-like numerical dynamo model. Nature
  389, 371--374.

\bibitem[{Lister and Buffett(1998)}]{Lister1998}
Lister, J., Buffett, B., 1998. Stratification of the outer core at the
  core-mantle boundary. Physics of the Earth and Planetary Interiors 105,
  5--19.

\bibitem[{Margot et~al.(2007)Margot, Peale, Jurgens, Slade, and
  Holin}]{Margot2007}
Margot, J.~L., Peale, S.~J., Jurgens, R.~F., Slade, M.~A., Holin, I.~V., 2007.
  Large longitude libration of {M}ercury reveals a molten core. Science 316,
  710--714.

\bibitem[{McDonough(2003)}]{McDonough2003}
McDonough, W.~F., 2003. Compositional model for the {E}arth's core. In:
  Carlson, R.~W. (Ed.), The mantle and core. Elsevier-Pergamon, Oxford.

\bibitem[{Moffatt and Loper(1994)}]{Moffatt1994}
Moffatt, H., Loper, D., 1994. The magnetostrophic rise of a buoyant parcel in
  the {E}arth's core. Geophys J Int 117, 394--402.

\bibitem[{Nellis(2000)}]{Nellis2000}
Nellis, W., 2000. Metallization of fluid hydrogen at 140 {GPa} (1.4 {Mbar}):
  implications for {J}upiter. Planetary and Space Science.

\bibitem[{Poirier(1994)}]{Poirier1994}
Poirier, J., 1994. Light elements in the {E}arth's outer core: A critical
  review. Phys. Earth Planet. Int. 85, 319--337.

\bibitem[{Potherat et~al.({2002})Potherat, Sommeria, and Moreau}]{Potherat2002}
Potherat, A., Sommeria, J., Moreau, R., {2002}. {Effective boundary conditions
  for magnetohydrodynamic flows with thin Hartmann layers}. {Phys. Fluids}
  {14}~({1}), {403--410}.

\bibitem[{Sakuraba and Kono(1999)}]{Sakuraba1999}
Sakuraba, A., Kono, M., 1999. Effect of the inner core on the numerical
  solution of the magnetohydrodynamic dynamo. Physics of the Earth and
  Planetary Interiors 111, 105--121.

\bibitem[{Solomon et~al.(2008)Solomon, McNutt, Watters, Lawrence, Feldman, ,
  Head, Krimigis, Murchie, Phillips, Slavin, and Zuber}]{Solomon2008}
Solomon, S.~C., McNutt, R.~L., Watters, T., Lawrence, D.~J., Feldman, W.~C., ,
  Head, J.~W., Krimigis, S.~M., Murchie, S.~L., Phillips, R.~J., Slavin, J.~A.,
  Zuber, M.~T., 2008. Return to {M}ercury: A global perspective on
  {MESSENGER}'s first {M}ercury flyby. Science 321, 59--62.

\bibitem[{Stacey and Anderson(2001)}]{Stacey2001}
Stacey, F., Anderson, O., 2001. Electrical and thermal conductivities of
  {Fe-Ni-Si} alloy under core conditions. Physics of the Earth and Planetary
  Interiors 124, 153--162.

\bibitem[{Stacey and Loper(2007)}]{Stacey2007}
Stacey, F., Loper, D., 2007. A revised estimate of the conductivity of iron
  alloy at high pressure and implications for the core energy balance. Physics
  of the Earth and Planetary Interiors 161, 13--18.

\bibitem[{Stanley and Bloxham(2006)}]{Stanley2006}
Stanley, S., Bloxham, J., 2006. Numerical dynamo models of {U}ranus' and
  {N}eptune's magnetic fields. Icarus 184, 556--572.

\bibitem[{Stanley and Mohammadi(2008)}]{Stanley2008}
Stanley, S., Mohammadi, A., 2008. Effects of an outer thin stably stratified
  layer on planetary dynamos. Physics of the Earth and Planetary Interiors 168,
  179--190.

\bibitem[{Stevenson(2003)}]{Stevenson2003}
Stevenson, D.~J., 2003. Planetary magnetic fields. Earth Planet. Sci. Lett.
  208, 1--11.

\bibitem[{Stevenson(2008)}]{Stevenson2008}
Stevenson, D.~J., Jan 2008. Metallic helium in massive planets. P Natl Acad Sci
  Usa 105~(32), 11035--11036.

\bibitem[{Tanaka(2007)}]{Tanaka2007}
Tanaka, S., 2007. Possibility of a low {P-wave} velocity layer in the outermost
  core from global {SmKS} waveforms. Earth and Planetary Science Letters 259,
  486--499.

\bibitem[{Wicht(2002)}]{Wicht2002}
Wicht, J., 2002. Inner-core conductivity in numerical dynamo simulations.
  Physics of the Earth and Planetary Interiors 132, 281--302.

\bibitem[{Wood et~al.(2006)Wood, Walter, and Wade}]{Wood2006}
Wood, B., Walter, M., Wade, J., 2006. Accretion of the {E}arth and segregation
  of its core. Nature 441, 825--833.

\end{thebibliography}
